\documentclass{aa}

\usepackage{txfonts}

\DeclareRobustCommand{\VAN}[3]{#2}
\let\VANthebibliography\thebibliography
\def\thebibliography{\DeclareRobustCommand{\VAN}[3]{##3}\VANthebibliography}

\usepackage{graphicx}  
\usepackage{amsmath}  
\usepackage{amssymb}  
\usepackage{natbib,twoopt}
\usepackage[breaklinks=true,colorlinks = true,linkcolor = blue,urlcolor  = blue,citecolor = blue,anchorcolor = blue]{hyperref} 
\usepackage{threeparttable}
\usepackage{subfig}
\usepackage{pdflscape}
\usepackage{multirow}

\newcommand{\nhtwo}{$N_{\textrm{H}_2}$}
\newcommand{\tdust}{$T_{\textrm{dust}}$}
\newcommand{\tP}{\texttt{Transphere}}
\newcommand{\asec}{$^{\prime\prime}$}
\newcommand{\amin}{$^{\prime}$}
\newcommand{\menv}{$M_{\textrm{env}}$}
\newcommand{\lbol}{$L_{\textrm{bol}}$}
\newcommand{\nkau}{$n$($r=1000$~AU)}
\newcommand{\rout}{$r_{\textrm{out}}$}
\newcommand{\micron}{$\mu$m}
\defcitealias{motte_earliest_2007}{MBS07}

\begin{document}

   \title{Linking high- and low-mass star formation}

   \subtitle{Observation-based continuum modelling and physical conditions}

   \author{R. L. Pitts\inst{1}\fnmsep\thanks{ORCiD: 0000-0002-7937-4931}, L. E. Kristensen\inst{1}, J. K. J\o{}rgensen\inst{1}
          \and S. J. van der Walt
          }

   \institute{Niels Bohr Institute, Centre for Star \& Planet Formation, University of   Copenhagen, \O{}ster Voldgade 5-7, 1350 Copenhagen K, Denmark\\
              \email{rebecca.pitts@nbi.ku.dk}}

   \date{Received September 28, 2021; accepted October 20, 2021}

\abstract
{Astronomers have yet to establish whether high-mass protostars form from high-mass prestellar cores, similar to their lower-mass counterparts, or from lower-mass fragments at the heart of a pre-protostellar cluster undergoing large-scale collapse. Part of the uncertainty is due to a shortage of envelope structure data on protostars of a few tens of solar masses, where we expect to see a transition from intermediate-mass star formation to the high-mass process.}
{We sought to derive the masses, luminosities, and envelope density profiles for eight sources in Cygnus-X, whose mass estimates in the literature placed them in the sampling gap. Combining these sources with similarly evolved sources in the literature enabled us to perform a meta-analysis of protostellar envelope parameters over six decades in source luminosity.}
{We performed spectral energy distribution (SED) fitting on archival broadband photometric continuum data from 1.2 to 850~\micron to derive bolometric luminosities for our eight sources plus initial mass and radius estimates for modelling density and temperature profiles with the radiative-transfer package \tP.}
{The envelope masses, densities at 1000 AU, outer envelope radii, and density power law indices as functions of bolometric luminosity all follow established trends in the literature spanning six decades in luminosity. Most of our sources occupy an intermediate to moderately high range of masses and luminosities, which helps to more firmly establish the continuity between low- and high-mass star formation mechanisms. Our density power law indices are consistent with observed values in the literature, which show no discernible trends with luminosity, and have a mean $p=-1.4\pm0.4$. However, our sub-sample, with a mean power law index of $-1.1\pm0.3$, is slightly flatter than would be expected for spherical envelopes in free fall ($p=-1.5$).}
{We attribute flattened density profiles for our eight sources to one or more of the following: ongoing accretion from their natal filaments, convolution of sources with neighbours or the larger filament, spherical averaging of asymmetric features (for example fragments), or inflation of the envelope by a moderate far-ultraviolet (FUV) field. Finally, we show that the trends in all of the envelope parameters for high-mass protostars are statistically indistinguishable from trends in the same variables for low-mass protostars.}


\maketitle


\section{Introduction}\label{sec:intro}
Astronomers typically classify stars into two or three mass regimes for a number of reasons that can generally by summarised as their expected courses of evolution and environmental influences. High-mass stars are widely agreed to include any star of $M\gtrsim8$~$M_{\odot}$---the minimum initial mass of a supernova progenitor \citep[][and references therein]{heger2003,smartt2009}---and stars of lower mass sometimes are divided into intermediate-mass ($1.5$~$M_{\odot}<M<8$~$M_{\odot}$), and low-mass ($M\lesssim1.5$~$M_{\odot}$) depending on whether or not the stars are small and cool enough to sustain a convection in their outer layers \citep[][\S2.2]{stellarstructure}. Low-mass stars make up the overwhelming majority of stars in the Galaxy, and along with intermediate-mass stars, they are the hosts of planetary systems. High-mass stars are rare and may not have planets, but they produce most of a galaxy's light, and are major drivers of the physical and chemical evolution of the interstellar medium (ISM) via fierce winds, ionising radiation, and supernovae. Most stars of all masses form in clusters \citep{lada03}, but high-mass stars do so most exclusively. This, on top of their rarity and short lifetimes, makes high-mass stars difficult to study in the process of formation; high-mass star-forming regions (SFRs) are characteristically distant, heavily extincted, and very crowded.

The formation process for high-mass stars has long been a subject of heated debate, with the main source of contention being whether high-mass stars begin as high-mass cores ($\sim0.1$~pc, that is they are core-fed) or if they are made high-mass by starting in a particularly dense environment (that is they are clump-fed, where clumps are parsec-scale condensations). The former paradigm is encapsulated in the core accretion or turbulent core model \citep{mckee2003}, essentially a scaled up, accelerated analogue to how low- and intermediate-mass stars are understood to form. The latter paradigm started with the so-called competitive accretion model proposed by \citealt{bonnell2001,bonnell2004}, where similar-sized condensations can gain more mass if they happen to start near the centre of the cloud's potential. This has since been largely superseded by the hierarchical collapse model \citep{vazquez2009,vazquez2019,padoan2020}, where the large-scale cloud collapses into filaments and nodes (or ridges and hubs, respectively) and concentrates gas into parsec-scale clumps. Therein, low- to intermediate-mass cores that are born from turbulent fragmentation \citep{padoan2002} can rapidly accrete a massive envelope. High-mass prestellar core candidates exist \citep[see for example][]{russeil2010,tackenberg2012,csengeri2017,tige2017,motte_hmsf_review}, but none have been confirmed, and similar candidates from earlier studies have subsequently been found to contain deeply embedded protostars. On the other hand, high-mass prestellar clumps and high-mass clumps containing protostellar cores of a wide range of masses are abundant in the literature \citep[][to name a few]{bontemps_fragmentation_2010,svoboda2016,sanhueza2019,kong2021,pitts2021}. This evidence would seem to favour the clump-fed scenario, but only one high-mass prestellar core would need to be confirmed to prove the core-fed scenario is also viable.

Another potential way to distinguish the above scenarios is to look at their predictions for the distribution of matter in protostellar envelopes. The radial mass density distributions for protostellar envelopes of all masses are typically modelled as a power law of the form $\rho(r)\propto r^p$, where the index $p$ is between 0 and $-2$. Interpretations may vary depending on whether one subscribes to the historically favoured inside-out collapse model of \citealt{shu77}, or the outside-in mode that hierarchical collapse appears to produce \citep{gomez2021}. Under the former scenario, a slope of -2 is indicative of an isothermal or Bonnor-Ebert sphere, -1.5 indicates an isolated sphere in free-fall, and -1 is the slope for the density distribution of gas in virial equilibrium to where the inside-out collapse has not yet propagated \citep{logotrope97}. In the hierarchical collapse scenario, solving the continuity equation for a spherical envelope reveals $p=-2$ to be an attractor towards which any other value of $p$ will evolve \citep{vazquez2019}. Low- to intermediate-mass protostars, tending as they do to form in less dense environments, should have envelopes well-described by a free-falling spherical envelope model. If we start with that assumption and if the core accretion model of massive SF is true, massive protostars should have similar values of $p$. If a hierarchical collapse model suits massive stars better, one of three possibilities may be observed:
\begin{itemize}
    \item Evolution towards $p=-2$ will be quicker, so most observed massive protostars will have approximately that value of $p$.
    \item Turbulent fragmentation will cause the spherical envelope approximation to break down, and $p$ will be flatter than observed for lower-mass stars as different cores pull the mass in different directions.
    \item The ridge-hub morphology of the larger cloud in the hierarchical collapse scenario will lead to accretion flows that strongly violate the assumption of spherical symmetry, which may lead to either flatter values of $p$ or an increased dispersion in $p$.
\end{itemize}
Thus in most cases, if the transition from intermediate to high-mass stars is accompanied by a transition to formation by hierarchical collapse, we should observe a population-wide change in the distribution of $p$ and any parameters derived from the density profile (for example mass).

To test the above prediction, we needed constraints provided by detailed models of the density (and temperature) distributions over a large sample of protostars of all masses, but especially in the poorly-sampled region between about 10 and 100~$M_{\odot}$ ($\sim50$--1000~$L_{\odot}$). Based on the relative scales of the stellar initial mass function (IMF) and the similar-shaped core mass function (CMF), it is expected that between 15 and 50\%, typically around 30\%, of the mass of a core will end up in the star \citep{alves2007,jes2007a,bene2018,konyves2020}, which means that the sampling gap spans the range of core masses with the potential to form either intermediate- or high-mass stars. Maps of the high-mass SFR Cygnus-X ($d\sim1.4$~kpc, \citealt{rygl_parallax_2012}) by \citealt{motte_earliest_2007}, hereafter abbreviated as \citetalias{motte_earliest_2007} revealed a wealth of protostellar cores with masses overlapping the high end of the desired mass range. Further, subsequent studies have since resolved some of these cores into fragments with masses suggestive of intermediate-mass star progenitors \citep[see for example][]{bontemps_fragmentation_2010,duarte_cabral_2013}, although continuum data from \emph{Herschel} do not resolve these fragments. We selected ten of these sources for the following analysis in this article, and in anticipation of future chemical modelling.

In \S\ref{sec:meth}, we discuss how we obtained and pre-processed our continuum data to get fluxes for SED fitting; how we used SED fitting to estimate parameters, like envelope mass and luminosity, for bench-marking 1D envelope models; and how we then used those SED fits and derived parameters to model the temperature and density structures of our sources. In \S\ref{sec:res} we compile and compare the results from SED-fitting and 1D envelope modelling, and compare both sets of results to values that other studies derived for the same objects. In \S\ref{sec:disc}, we place our results in the context of individual source morphologies, and of comparable data on as many resolved protostellar sources as we could find spanning seven decades in protostellar luminosity, to identify trends and determine if they differ between low- and high-mass protostars. We conclude with a summary of our results in \S\ref{sec:conc}

\section{Methods}\label{sec:meth}
As discussed in \S\ref{sec:intro}, our science goal was to establish whether or not different physics govern the evolution of high- and low-mass protostellar envelopes by comparing trends in their structural parameters at roughly the same evolutionary phase, but over a wide range of masses and luminosities. Our technical goals include filling the sampling gap in the literature at intermediate masses and luminosities, which we were well-positioned to do thanks to concurrent work on 10 moderately-massive protostellar sources in Cygnus-X. However, extracting envelope parameters from available data is a multi-stage process; the temperature and density profiles of protostellar envelopes cannot be directly observed over much of a protostellar envelope. Resolution limits aside, the dominant gas and lightest molecule in any protostellar envelope, H$_2$, does not have a permanent dipole moment and therefore cannot emit at temperatures below about 500~K. Other tracers are needed to determine the density distribution and detect the protostellar emission as it emerges, reprocessed, from the cold envelope. Our tracer of choice is thermal dust emission, which has fairly large random uncertainties in its abundance relative to H$_2$ (as explained in \S\ref{ssec:seds}) but is robust to a much wider range of temperatures than most molecular tracers.

The procedure described in the following subsections has three main phases: photometry, SED-fitting, and spherical envelope modelling. The necessity of photometry becomes immediately evident in the phases after it, but the results of SED-fitting and spherical envelope modelling may at first glance look redundant. There are several reasons why we do both. Spherical envelope modelling incorporates more detailed physics through ray-tracing, and is able to recover radial structure that SED-fitting collapses to a point. However, the parameter space for spherical envelope modelling is much too large to optimise over without prior constraints. The parameters from SED fitting and their uncertainties provide those constraints.
Dust SED fitting is also the first line of study for many pre- and protostellar objects, and is a way to derive mass-weighted average parameters that can be compared in lieu of radially resolved models of the envelope structure, if such models are not feasible with the available data. As a long-term goal, we want to make both sets of data available so that similar future comparisons of low and high-mass protostellar sources can include our data, regardless of whether or not the structure of individual protostellar envelopes can be resolved.

\subsection{Observational data}\label{ssec:obs}
\begin{figure*}
    \centering
    \includegraphics[width=\textwidth]{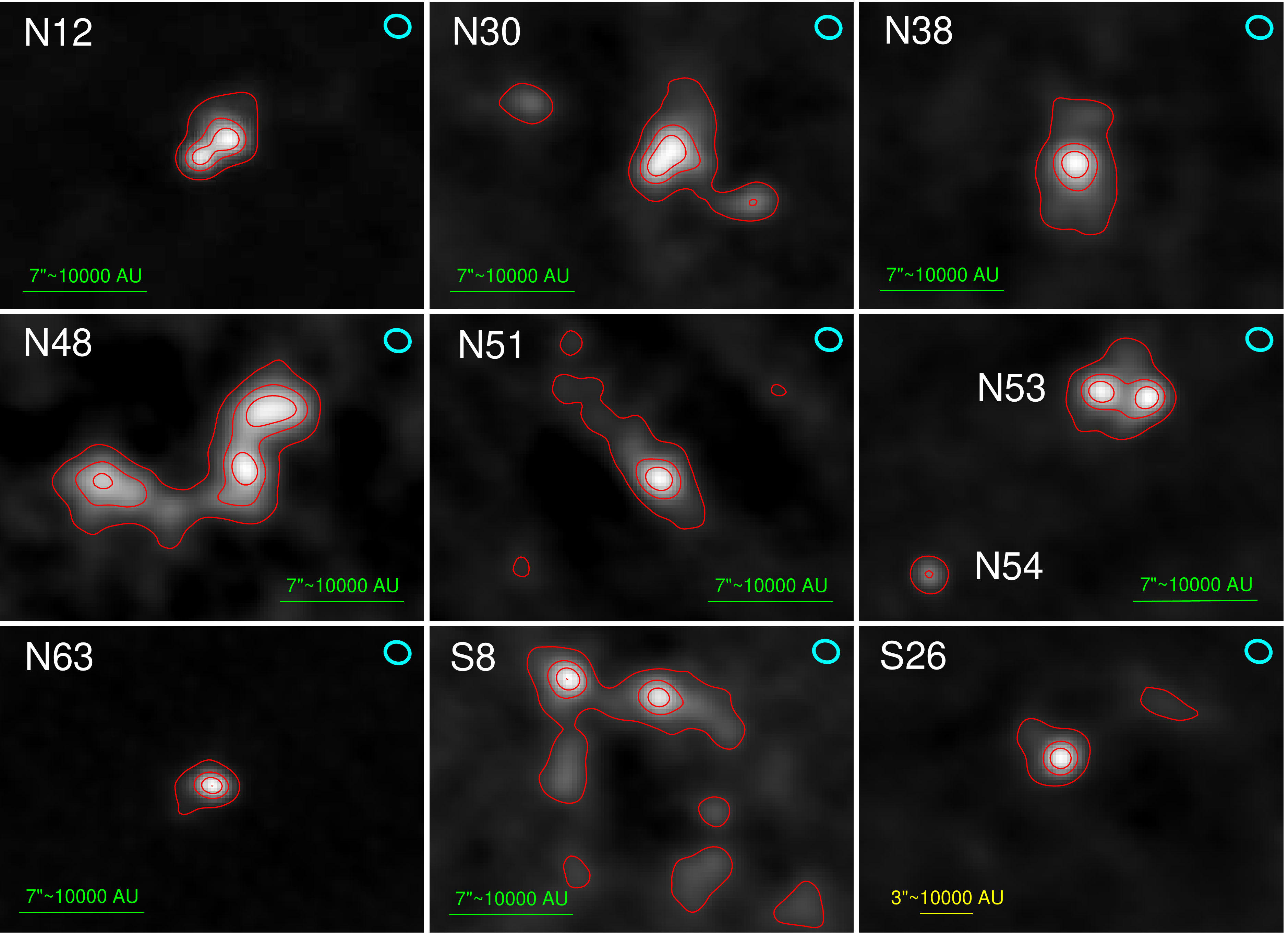}
    \caption{SMA 868~\micron\ (345 GHz) images of all 10 PILS-Cygnus sources (N53 and N54 are both featured in the middle right panel) contoured in red at 10\%, 40\%, and 70\% of the maximum flux. The cyan ellipses in the upper right corners show the FWHMa of both axes and position angle of the synthesised SMA beam, and the line segments alone the bottoms of each panel show how long 10000 AU and its corresponding angular length appears at the adopted distance to each object. We note the change of scale for S26. All images are normalised to their peak flux densities, which are given in Table~\ref{tab:coords}.}
    \label{fig:smaimgs}
\end{figure*}

The objects we contribute to this metastudy come from the PILS-Cygnus survey (PI: Lars E. Kristensen), for which 10 protostellar envelopes in Cygnus-X were selected based on brightness in continuum, SiO 2--1, and H$_2$O $2_{02}-1_{11}$ emission \citep{motte_earliest_2007,mottram_water_2014}. These were mapped with the SubMillimeter Array (SMA) telescope on Mauna Kea in the continuum and a variety of molecular species. The SMA continuum observations, taken in June and November of 2017, covered a 32~GHz-wide band centred at 345~GHz (868~$\mu$m) in both compact and extended array configurations for a combined spatial resolution of $0\farcs7$ and sensitivity of 0.15 Jy~km~s$^{-1}$ (see van der Walt et al. 2021, to appear in A\&A, for detailed description of data reduction). Figure~\ref{fig:smaimgs} shows the 868~\micron\ (345 GHz) continuum images of all 10 sources
, contoured at 10\%, 40\%, and 70\% of the maximum flux in each image, with the synthesised SMA beam dimensions shown in the upper right of each panel and the image scales along the bottom of each panel. These images were not used directly in dust SED fitting because much of the outer envelope emission was resolved out. They were, however, subject to Gaussian decomposition to separately determine the number of core fragments inside each envelope as a check against \citealt{bontemps_fragmentation_2010} and to compare core multiplicity to structural envelope parameters in search of trends.

\begin{figure*}
    \centering
    \begin{tabular}{cc}
      \includegraphics[width=0.56\textwidth]{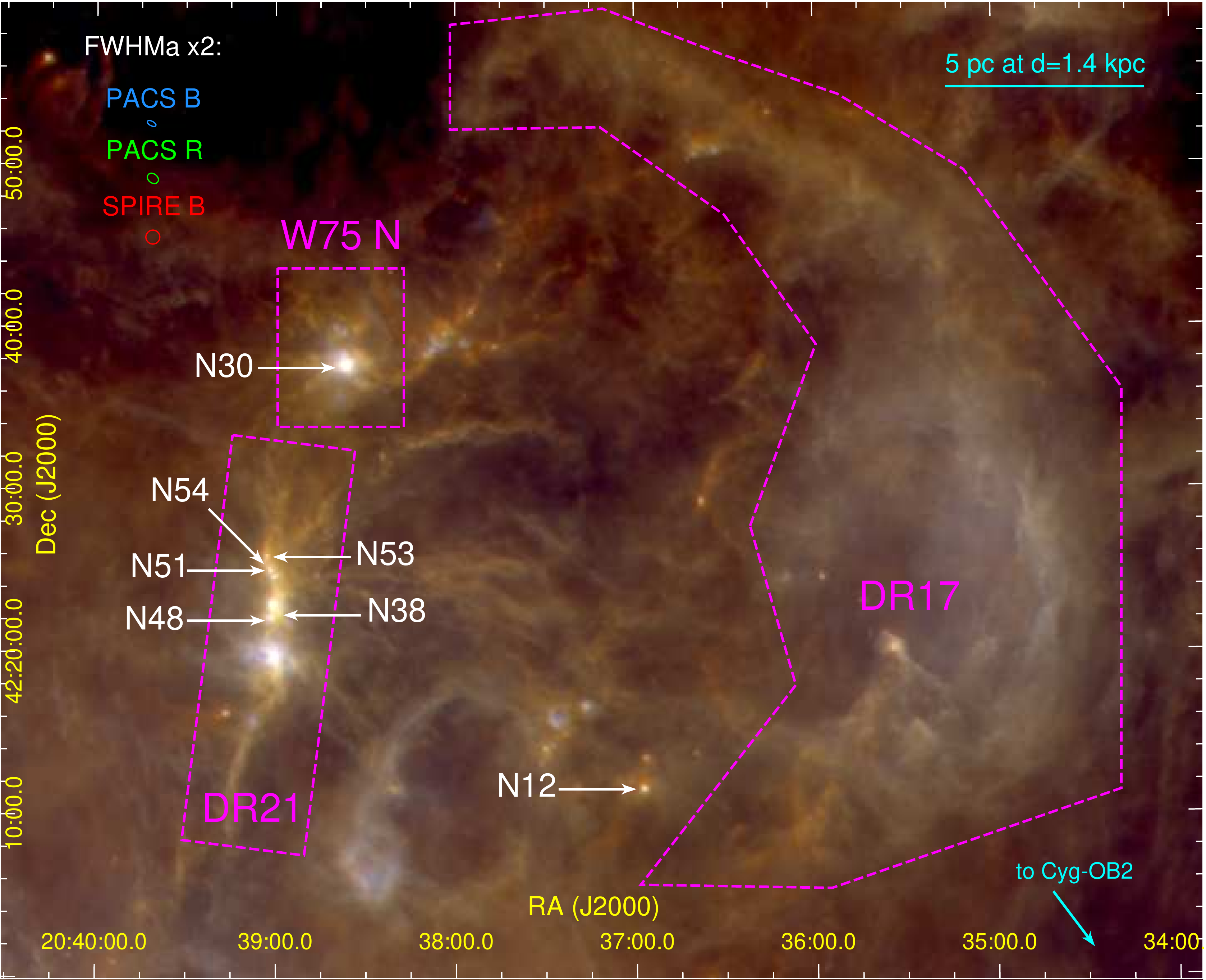}   & \includegraphics[width=0.42\textwidth]{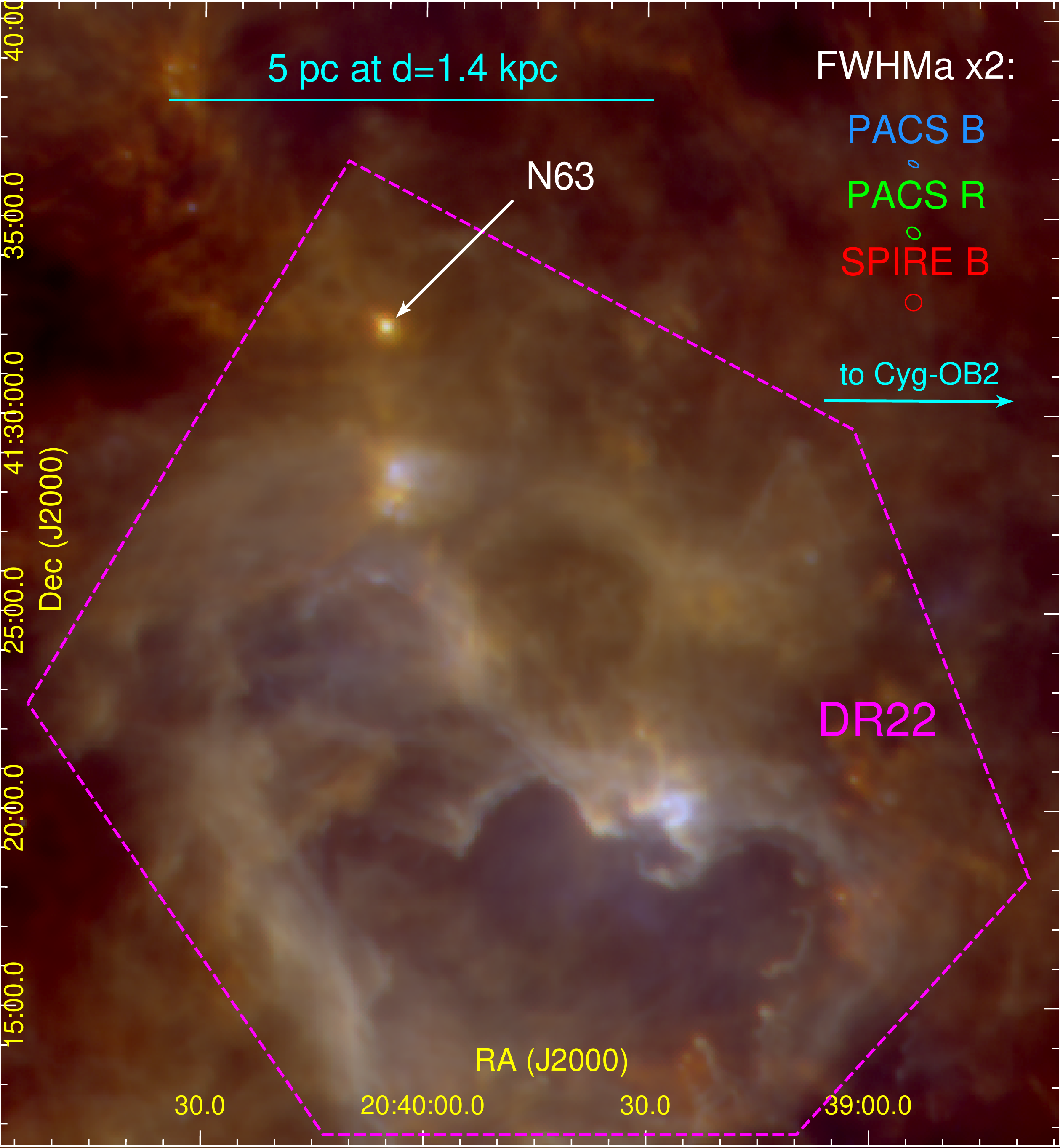} \\ 
       \includegraphics[width=0.56\textwidth]{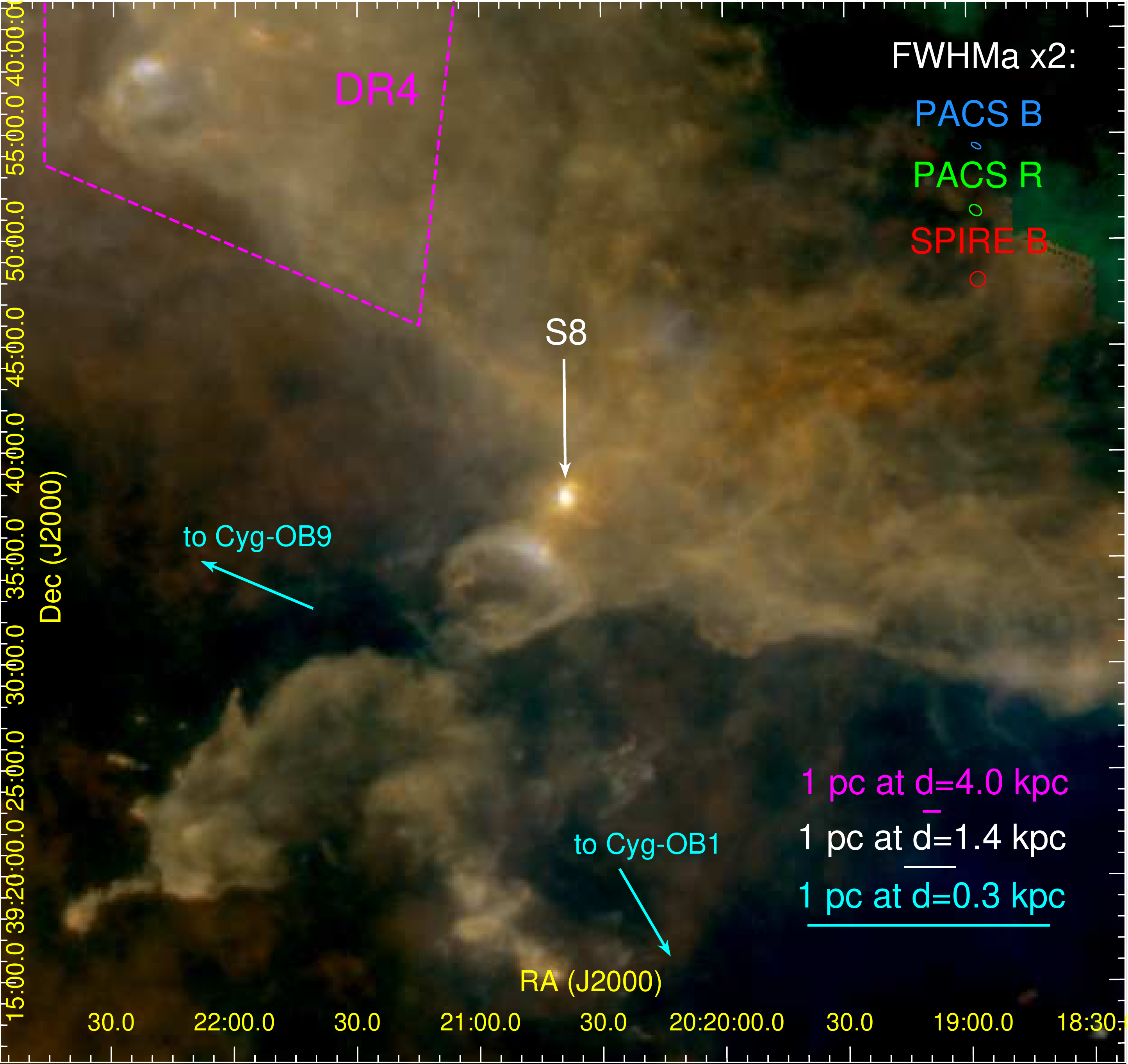}  & \includegraphics[width=0.42\textwidth]{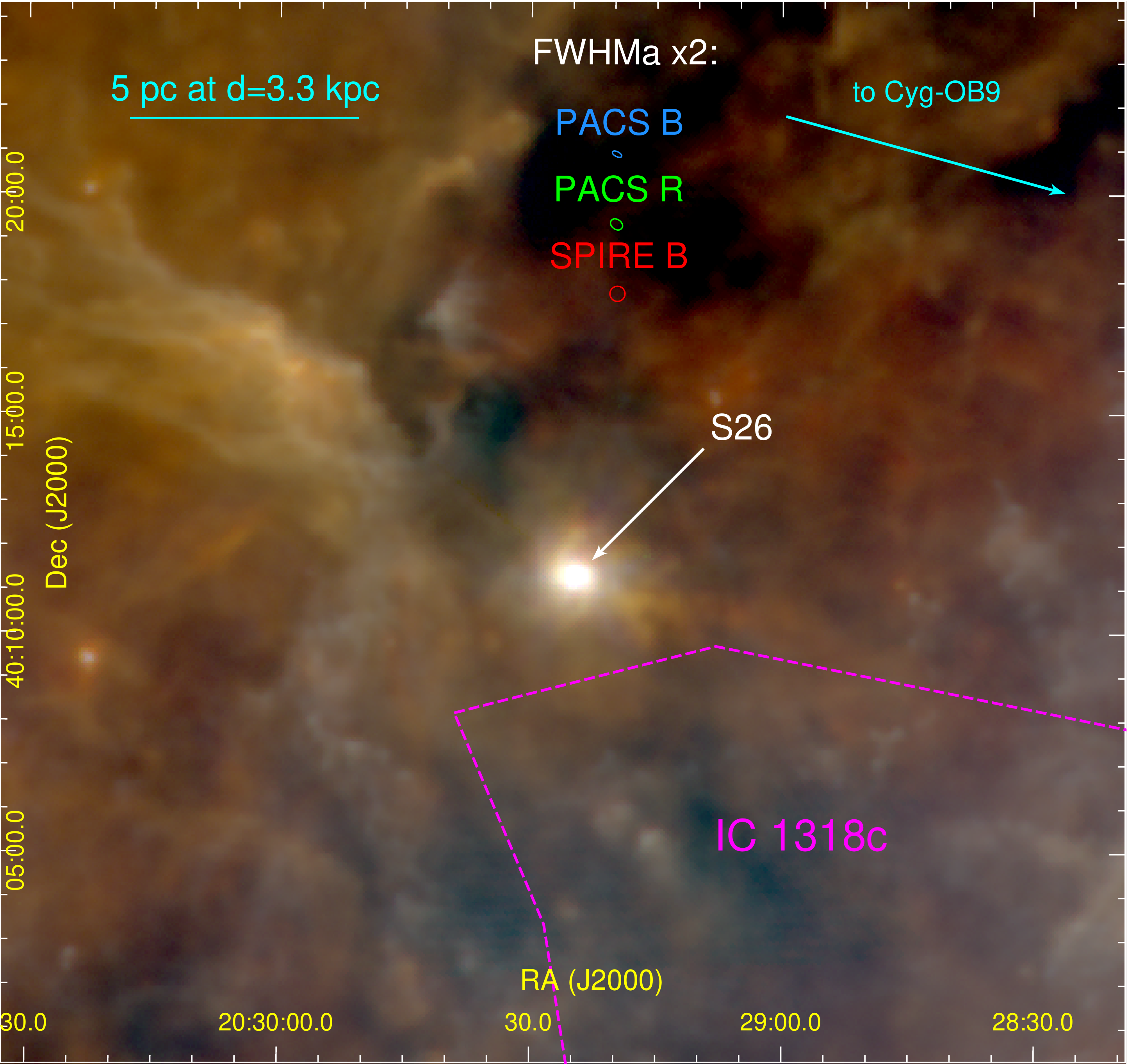}
    \end{tabular}
    \caption{\emph{Herschel} three-colour cut-outs showing the PILS-Cygnus sources in their natal environments at 70, 160, and 250~\micron, with the DR objects labelled. \emph{Herschel} beams are enlarged by two-fold for visibility.}
    \label{fig:cutouts}
\end{figure*}

Table~\ref{tab:coords} gives the sky coordinates and designations per \citetalias{motte_earliest_2007} for all 10 objects, and Fig.~\ref{fig:cutouts} shows \emph{Herschel} RGB cutouts around all of the sources at 70, 160, and 250~\micron, with the sources and nearby objects from the DR catalogue \citep{drcat1966} labelled. Of those 10 sources, nine could be identified in archival mid-infrared (MIR) to submillimetre (submm) broadband continuum data, and eight could be separated from the extended filamentary emission and/or other sources in the DR~21 Ridge. N38 could not be conclusively identified at any wavelength, and N54 was too faint beyond 160~\micron\ to construct a believable SED.

In most calculations, we adopted the average distance to Cygnus-X of 1.4~kpc from \citealt{rygl_parallax_2012}, except for S26, for which \citealt{rygl_parallax_2012} finds a distance of 3.3~kpc. 
We used more precise values from \citealt{rygl_parallax_2012} for individual objects, where available, to estimate the projected distances from Cygnus OB2 (hereafter abbreviated Cyg~OB2). However, distance estimates to the centre of Cyg~OB2 range from about 1.4~kpc \citep{hanson03} to about 1.7~kpc \citep{massey_1991,berlanas2019}, and the recent work by \citealt{berlanas2019} using \emph{Gaia} parallaxes suggest both estimates may be correct because Cyg-OB2 may be two OB associations, one each at 1.4 and 1.7~kpc, projected on top of each other. This revelation deeply complicates any interpretation of the PILS-Cygnus source morphologies based on deprojected distances to Cyg~OB2. For now, we only report the projected distances in Table~\ref{tab:coords}, which should be interpreted as lower limits on the absolute separations from Cyg~OB2.

\begin{table*}
    \centering \caption{Table of PILS-Cygnus source coordinates}\label{tab:coords}
\begin{threeparttable}
    \begin{tabular}{l l l l l l l l}\hline
        Source\tnote{(a)} & RA\tnote{(b)} & Dec\tnote{(b)} & Associated Objects & D$_{\odot}$ (pc)\tnote{(c)} & D$_{CygOB2}^{proj}$ (pc)\tnote{(d)} & Fragments\tnote{(e)} & $S^{max}_{868\mu m}$ (Jy/beam)\tnote{(f)}\\ \hline
        N12	& 20:36:57.68	& 42:11:30.8 & J20365781+4211303 & $1400\pm100$\tnote{(g)} & $31\pm2$ & 2+ & 0.50\\
        N30	& 20:38:36.45	& 42:37:33.8 & W75N(B) & $1300\pm70$ & $43^{+2}_{-3}$ & 4 & 1.29\\
        N38\tnote{(h)}	& 20:38:59.26	& 42:22:28.6 & DR 21(OH) W & $1500^{+80}_{-70}$ & $47\pm2$ & 1 & 0.62\\
        N48	& 20:39:01.46	& 42:22:05.9 & DR 21(OH) S & $1500^{+80}_{-70}$ & $47\pm2$ & 5+ & 0.25\\
        N51	& 20:39:01.97	& 42:24:59.2 & G081.7522+00.5906 & $1500^{+80}_{-70}$ & $48^{+2}_{-3}$ & 1+ & 0.42\\
        N53	& 20:39:03.13	& 42:25:52.5 & W75S~FIR3 & $1500^{+80}_{-70}$ & $48\pm2$ & 2+  & 0.41\\
        N54\tnote{(h)}	& 20:39:04.03	& 42:25:41.1 &  & $1500^{+80}_{-70}$ & $48\pm2$ & 1 & 0.27$^*$\\
        N63	& 20:40:05.50	& 41:32:12.6 & BGPS G081.174-00.100 & $1400\pm100$\tnote{(g)} & $42^{+3}_{-2}$ & 2 & 1.66\\
        S8	& 20:20:39.29	& 39:37:54.2 & Mol 121 & $1400^{+2600}_{-1100}$\tnote{(gi)} & $>80$\tnote{(i)} & 2+ & 0.30\\
        S26	& 20:29:24.87	& 40:11:19.3 & RAFGL 2591 VLA 3 & $3330\pm110$ & $\cdots$\tnote{(j)} & 1+ & 0.60\\ \hline
    \end{tabular}
    \begin{tablenotes}
    \item[a] Object names from \citetalias{motte_earliest_2007}.
    \item[b] Coordinates corrected using \emph{Spitzer} images overlaid with SMA contours and ellipses defined by FWHMa from \protect\citetalias{motte_earliest_2007}.
    \item[c] Heliocentric distances and uncertainties are from \citealt{rygl_parallax_2012} and references therein unless otherwise noted.
    \item[d] Projected from Cyg~OB2 calculated using D$_{\odot}$.
    \item[e] Number of core components visible in SMA images (868~\micron, $0\farcs7$ resolution). A plus (+) indicates that weak and/or filamentary emission suggest there may be additional fragments.
    \item[f] Maximum 868~\micron\ flux density among all core fragments. ($^*$Peak flux density of N54 measured from image centred on N54.)
    \item[g] \citealt{rygl_parallax_2012} did not explicitly measure the parallaxes at the locations of these sources, so we adopted the average and uncertainty over Cygnus~X.
    \item[h] SEDs could not be reconstructed for these sources. See text for details.
    \item[i] The distance to S8 is poorly constrained; the given uncertainties reflect the full range of distance estimates in the literature (see review by \citealt{varri2010}, appendix A37), and the smallest possible projected distance can by calculated by assuming both S8 and Cyg~OB2 are at the latter's minimum distance from us ($\sim1.3$~kpc).
    \item[j] The projected distance between S26 and Cyg~OB2 is negligible compared to their line-of-sight separation, $\gtrsim1.57$~kpc.
    \end{tablenotes}
\end{threeparttable}
\end{table*}

We concentrated mainly on the FIR/submm continuum data from \emph{Herschel}/PACS \citep{poglitsch_photodetector_2010} and \emph{Herschel}/SPIRE \citep{griffin_herschel-spire_2010} as part of the Hi-GAL survey \citep{higal}, and JCMT/SCUBA(-2) \citep{francesco_scuba_2008,holland_scuba-2_2013} for SED computation. 
To compute bolometeric luminosities, we supplemented these data with photometry from 2MASS \citep{2mass}, \emph{Spitzer}/IRAC \citep{fazio_infrared_2004, beerer_aspitzerview_2010}, \emph{Spitzer}/MIPS \citep{rieke_multiband_2004}, MSX \citep{egan_msx_2003}, and upper limits from IRAS \citep{neugebauer_infrared_1984}. We performed most of the photometry ourselves; for MSX and IRAS, whose large uncertainties limited their influence upon the fit, we used point-source catalogue fluxes. Table~\ref{tab:bands} shows the instrument and filter specifications for every data set we used for any purpose, photometry or otherwise, including the corresponding wavelengths, resolutions, and pixel scales. All fluxes and their associated uncertainties are listed in Appendix~\ref{app:flux}, Table~\ref{tab:fluxtab}.

\begin{table}
    \centering
    \caption{Instrument and filter specifications for all filters used.} \label{tab:bands}
    \begin{threeparttable}
    \begin{tabular}{l l l l}\hline
        Band & $\lambda$ (\micron) & Resolution & Pixel Size \\ \hline
        2MASS-J &	1.235 & 4\asec & 2\asec\\
        2MASS-H &	1.662 & 4\asec & 2\asec\\
        2MASS-Ks &	2.159 & 4\asec & 2\asec\\
        IRAC-1 &	3.6 & 1\farcs66 & 0\farcs6\tnote{(a)}\\
        IRAC-2 &	4.5 & 1\farcs72 & 0\farcs6\tnote{(a)}\\
        IRAC-3 &	5.8 & 1\farcs88 & 0\farcs6\tnote{(a)}\\
        IRAC-4 &	8.0 & 1\farcs98 & 0\farcs6\tnote{(a)}\\
        MSX-A &	8.28 & 18\farcs3 & 6\asec \\
        MSX-C &	12.13 & 18\farcs3 & 6\asec \\
        MSX-D &	14.65 & 18\farcs3 & 6\asec \\
        MSX-E &	21.4 & 18\farcs3 & 6\asec \\
        MIPS-24 &	24 & 6\asec & 2\farcs45\\
        IRAS-25 &	25 & 0\farcm5 & $\cdots$\tnote{(b)} \\
        IRAS-60 &	60 & 1\amin & $\cdots$\tnote{(b)} \\
        PACS-B &	70 & 5\farcs86~$\times$~12\farcs16 & 3\farcs2\\
        PACS-R &	160 & 11\farcs64~$\times$~15\farcs65 & 3\farcs2\\
        SPIRE-B &	250 & 17\farcs9 & 6\asec \\
        SPIRE-G &	350 & 24\farcs2 & 10\asec \\
        SCUBA-B\tnote{(c)} &	450 & 11\asec $/$ 7\farcs9  & 3\asec $/$ 2\asec \\
        SPIRE-R &	500 & 35\farcs4 & 14\asec \\
        SCUBA-R\tnote{(c)} &	850 &  19\farcs5 $/$ 13\asec & 6\asec $/$ 3\asec \\ 
         \hline
    \end{tabular}
    \begin{tablenotes}
    \item[a]Images of S8 had 0\farcs86 pixels.
    \item[b]Catalogued fluxes were used as upper limits.
    \item[c]Numbers to the left of the slash are for SCUBA Legacy data, which we used for photometry. Numbers to the right are for SCUBA-2 data, which we used for density profile fitting but not photometry because too much flux was lost to secondary beam components.
    \end{tablenotes}
    \end{threeparttable}
\end{table}

For IRAC and 2MASS data, in which the sources were largely resolved, we performed aperture photometry using the \texttt{Astropy}-adjacent \texttt{Photutils} package with some adaptations for the highly crowded surroundings. Instead of background annuli, which could not be configured to avoid all neighbouring sources, we took the average of two background apertures of identical size to the target aperture, placed in uncontaminated areas on opposite sides of the target. Uncertainties in the fluxes listed in Appendix~\ref{app:flux} are the statistical uncertainties in the differences between the target aperture flux and the background aperture flux with the uncertainties in each aperture flux propagated through. Systematic uncertainties due to the presence of filamentary emission or absorption are likely to be larger. S8 in particular has a complex extended morphology, like a comma embedded in a Y-shape, that made it difficult to decide what should be included or excluded from the source. The automatic photometry results were remarkably discontinuous with the rest of the SED, so we redid the photometry for it manually in SAO DS9, following the same procedures and using the 90$\%$ contour levels as a guide for the aperture size.

\subsection{Deconvolution of Sources}\label{ssec:decon}
Because many of the sources are unresolved and separated from each other or nearby objects by less than the beam FWHM of most \emph{Herschel} bands, we developed a simple Gaussian deconvolution and fitting routine with all parameters free to vary within strict limits determined by eye (for example the coordinates of the centre of each Gaussian were allowed to fall anywhere within an area the size of each image's resolution element, centred on the coordinates given in \citealt{motte_earliest_2007}). It fits up to four components and a flat background to cutouts with dimensions tailored to enclose no more than four targets, but also extend at least 10$\sigma$ away from any source in at least two directions. The fluxes used to compute the SEDs and all derived parameters use the best-fit Gaussian integral of the deconvolved, background-less target flux.

Since the objects often did not separate cleanly at the longest wavelengths, and filamentary contamination remained an issue, RMS deviations from the observations were typically used in lieu of the unrealistically small calculated fitting uncertainties. We tried to estimate the uncertainties by distributing the RMS deviations amongst the modelled sources according to their fluxes, but this was not possible with N30, N38, and N54. With N30, it was unclear whether a one- or two-component Gaussian was preferable at short wavelengths, and at long wavelengths, most of the flux that should have been assigned to N30 was being assigned to a minor object identified as N31 in \citetalias{motte_earliest_2007}. MIR images from Spitzer suggest N31 is either a filamentary spur or some other distortion of N30, so we opted for the one-component model in the hopes that the inclusion of the spur and the relatively high fitted background would roughly balance out. In FIR/submm images where N38 would have been resolved if it were visible, the nearest source was 26\asec\ away, a much greater distance than the position uncertainties for any of the aforementioned observing instruments and only a couple arcseconds shy of the distance between N38 and N48. 
N54's location suggested that it would be blended with N53 at all FIR/submm wavelengths, but at the wavelengths in which their centres should have been more than a FWHM apart (ergo, able to be deconvolved), N54 is entirely swamped by both N53 and another source on the opposite side that was not part of our sample, N52. The total flux from N54 was within the margins of error for N53 and N52 at PACS wavelengths, and at longer wavelengths the fit was sometimes negative before we explicitly restricted the Gaussian models to positive amplitudes. Like N38, the fits from N54 could not be trusted. At 70 and 160~\micron, blending with N54 is not expected to contribute more flux to N53 than is accounted for by its large error margins. The same seems likely at longer wavelengths because N54 is quite blue in RGB images using 250, 160, and 70~\micron\ data, but our data do not allow confirmation.

\subsection{SED-Fitting}\label{ssec:seds}
The SED fitting routine used for the deconvolved data is described in detail in \citealt{pitts_seds_2019}. The program uses \texttt{MPFIT} \citep{mpfit} to derive an initial dust temperature ($T_{\textrm{dust}}$) and molecular hydrogen column density ($N_{\textrm{H}_2}$), interpolates on grids of colour correction factors\footnote{Colour correction functionality is only available for $1.5\leq\beta\leq2.0$ for \emph{Herschel} observations. Corrected fluxes are typically different from the uncorrected values by less than the typical 10\% margin of uncertainty. Not colour correcting sources with temperatures >10~K systematically depresses the SPIRE fluxes by $\lesssim10\%$, but this is more than counteracted by the tendency of our sources to be increasingly contaminated at longer wavelengths.}, propagates the estimated uncertainty in the colour correction factor to the input uncertainty, and recomputes the SED from the colour-corrected data. The program normally assumes a single modified blackbody, or greybody, component along the line of sight following the prescription of \citealt{hildebran83}. Up to three additional components of the same functional form may be added, provided they have sufficiently distinct peak temperatures. The greybody equation takes the form
\begin{equation}
    I_{\nu}\approx B_{\nu}(T_{\mathrm{dust}})\left( 1 - \mathrm{exp}\left(-\left( \frac{\nu}{\nu_0} \right)^{\beta} \frac{\kappa_0N_{\mathrm{H}_2}}{\gamma} \mu m_{\mathrm{H}} \right) \right)
\end{equation}\label{eq:mbb}
where $B_{\nu}$(\tdust) is the Planck function, $\beta$ is the dust emissivity index (typically with $1\lesssim\beta\lesssim2$), $\kappa_0$ is the dust opacity at reference frequency $\nu_0$, $\gamma$ is the gas-to-dust ratio (assumed to be 100), $\mu$ is the mean molecular weight per hydrogen molecule (2.8; see Appendix A of \citealt{kauf08}), and $m_{\mathrm{H}}$ is the mass of a hydrogen atom. For each object, we took the best-fitting value of $\kappa_0$ at 350~$\mu$m ($\nu_0=856$ GHz) for grains with thin ice mantles from \citealt{ossenkopf_dust_1994}. Note that Eq.~\ref{eq:mbb} does not assume optically thin dust emission, because for most PILS-Cygnus sources, optical depth at 70~$\mu$m is not much less than 1.

Ordinarily \tdust\ is somewhat degenerate with $\beta$ and \nhtwo\ is degenerate with both $\gamma$ and $\kappa_0$, so only $T_{\textrm{dust}}$ and $N_{\textrm{H}_2}$ were initially left as free parameters. However, we found that we could get $\beta$ to converge by disabling colour correction and fitting only data at wavelengths longer than either 70 or 160~\micron\, depending on whether or not the source was warmer than about 30~K. For N30, N51, and S26, which are also prominent at shorter wavelengths, we first fit only data at $\lambda\gtrsim70$~\micron\ to derive $\beta$, then fixed $\beta$ and fit a sum of greybodies with the understanding that the warmer components would only be first-order approximations of the luminosity. Since the uncertainty in $\beta$ dominates over all other factors, we estimated the uncertainties in the temperature and column density parameters for the warm and/or hot components by varying $\beta$ within its uncertainty margins from the single-component fit, and taking the range of output values as the margins of uncertainty for the other parameters. We emphasise that the warmer temperature components are probably not physical, and that the tendency of such components to separate around 10~\micron\ is an artefact of the silicate absorption feature. Except for the purpose of calculating bolometric luminosities, we are only interested in the parameters of the cold thermal dust SED, and warmer gas accounts for $\ll1\%$ of the total mass.

We also note that some sources---N12, N51, N53, and N63---have had emission at 70~\micron\ excluded or assigned to a different temperature component of the SED fit. In these cases, the 70~\micron\ emission may be inconsistent with the rest of the cold(est) dust SED component due to contamination by nonthermal emission from very small grains (see, for example \citealt{desert90,bernard_2008} for more details). That the rest of the sources have 70~\micron\ emission consistent with the SED fit to data at longer wavelengths does not indicate a lack of such contamination. It only indicates either that the amount of contamination is at the level of noise, or that the sum of thermal and nonthermal contributions at 70~\micron\ lie within the expected range of values possible for a single cold greybody SED given the uncertainty in $\beta$.

As usual, $\gamma$ and $\kappa_0$ are thought to be uncertain by at least a factor of about two and 2.5, respectively \citep[][in particular suggest the gas-to-dust-ratio may drop by more than a factor of 3 in cold clumps/cores]{ossenkopf_dust_1994,beckwith90,motte_earliest_2007,reach}, so the uncertainties in $T_{\textrm{dust}}$ and $N_{\textrm{H}_2}$ and all derived quantities refer to those contributed by the observational data, with the factor of two to three left implicit. This code was originally intended for parsec-scale prestellar clumps, not centrally-heated cores, so the temperature and density results may be best interpreted primarily as initial values for \tP. To compute the masses and luminosities from SED-fitting, we used the 3$\sigma$ mean radii from \citetalias{motte_earliest_2007} corrected by the distances from \citealt{rygl_parallax_2012}.

Finally, we remark that this SED fitting routine does not model the broad 9.7~\micron\ silicate absorption feature or the prominent emission lines attributed to polycyclic aromatic hydrocarbons (PAHs) between 3 and 20~\micron. Modelling any of these features would have changed the total luminosity by much less than its margins of uncertainty, and no other parameter besides the total luminosity depends significantly on accurately characterising the emission at $\lambda<50$~\micron.

\subsection{Modelling with \tP}\label{ssec:tp}
\texttt{Transphere} \citep{dullemond_tp_2002} can, among other things, calculate the dust radiative transfer in 1D through a spherical envelope using user-specified envelope masses, luminosities, and density power law indices, and output radial temperature profiles and SED fits at a reference radius. We excluded data at $\lambda<50$~\micron\ from fitting because at those wavelengths, the emission is dominated by the deep interior which is likely to be in a disk, ergo the approximation of spherical symmetry breaks down \citep[Appendix C]{jes_2007,lommen_2008,wish2}. The important user-supplied parameters are listed, and their effects on the fitted SED and radial temperature profiles described, as follows:

\textbf{Stellar luminosity $L_{\textrm{TP}}$} controls the amplitude of the SED and the position of the peak in wavelength-space via the usual $L_{\textrm{TP}}\propto T^4$ and Wien's Displacement Law. Increasing $L_{\textrm{TP}}$ also boosts the temperature profile at all radii. Instead of $L_{\textrm{bol}}$, we used only the luminosity calculated from \emph{Herschel} data and longer wavelengths as initial values of $L_{\textrm{TP}}$ because of the aforementioned breakdown of spherical symmetry at shorter wavelengths.

\textbf{Stellar surface temperature $T_{\star}$} has no noticeable effects on the envelope SED if it is initialised within a range that allows the temperature profile to converge (typically between 3000 and 10000~K, endpoints not always inclusive). Indeed, we should not expect stellar surface temperature to matter because, for a protostar embedded in a spherical envelope, all surface emission should be reprocessed by dust in the envelope before it escapes. Increasing $T_{\star}$ does slightly steepen the radial temperature profile at small ($r\lesssim100$~AU) radii, but this is not relevant to our studies because the assumption of spherical symmetry breaks down at these radii. We fixed $T_{\star}=5000$~K for all PILS-Cygnus sources.

\textbf{Envelope mass $M_{\textrm{env}}$} has a rather complex effect on the SED. Increasing $M_{\textrm{env}}$, all else equal, pushes the SED towards longer wavelengths and makes the SED both slightly taller and narrower due to increased reprocessing of photons from the central source. The radial temperature profile also steepens with increasing $M_{\textrm{env}}$, all else equal, reaching higher temperatures at small radii and lower temperatures at large radii (greater than a few~$\times10^3$~AU). This change in temperature profile reflects the radiative trapping effects of the increased opacity required at small radii \citep{hartmann98}. Initial envelope mass values were set to the masses calculated from dust SED fitting and varied iteratively with luminosity.

\textbf{Power-law index $p$} does affect the envelope SED, but the variation is within the data uncertainties for the entire plausible range of $p$. Increasing the magnitude of $p$ (making it more negative) with all else equal will steepen the radial temperature profile for radii less than a few~$\times10^3$~AU, but temperature profiles for different values of $p$ converge at larger radii. We initialised $p$ to $-1.8$ for all PILS-Cygnus sources, and then adjusted it iteratively.

\textbf{Inner envelope radius $r_{\textrm{in}}$} dictates how close to the protostar we truncate the envelope, and determines whether or how much of the inner envelope is sampled with more finely-spaced bins in radius to resolve optically thick gas. By default the program switches to a finer radial grid at 50~AU; temperature profiles failed to converge for any value of $r_{\textrm{in}}$ within about 20~AU of this radius, and contained one or more discontinuities for $r_{\textrm{in}}$ between about 80 and 200~AU. We fixed $r_{\textrm{in}}$ at 20~AU to avoid this issue and err on the side of including material that a more sophisticated model might place in an inner disk.

\textbf{Outer envelope radius $r_{\textrm{out}}$} has a similar effect on the SED as $M_{\textrm{env}}$, but not identical: increasing $r_{\textrm{out}}$ makes the SED narrower and very slightly taller, but does not significantly change the wavelength of the peak if the envelope mass and all other variables are held constant. On the radial temperature profile, varying $r_{\textrm{out}}$ has the effect of changing the concavity in log-log space. We initialised $r_{\textrm{out}}$ to 20000~AU and fit it iteratively along with $p$.
    
\textbf{Interstellar Radiation Field (ISRF) strength} affects the height of the SED at $\lambda\lesssim50$~\micron\ and the temperature profile in the outermost few~$\times10^3$ AU (out of 1 to a few~$\times10^4$~AU). Increasing the ISRF strength can boost the SED at $\lambda\lesssim50$~\micron, but never enough to explain the observed fluxes at those wavelengths without convergence errors. We fixed the ISRF strength at $1\,G_0$ with $G_0$ in \citealt{habing68} units.

To find the optimal envelope masses, luminosities, and density profiles of the PILS-Cygnus sources, we took an iterative $\chi^2$-minimisation approach using models computed with \tP. First, we computed a large grid of model SEDs over a wide range of masses, luminosities, and density power law indices for a handful of possible outer envelope radii ($r_{\textrm{out}}$). Second, we minimised $\chi^2$ between the model SEDs and data with $\lambda>50$~\micron\ by interpolating on a grid of $\chi^2$ as a function of luminosity ($L_{\textrm{TP}}$) and envelope mass ($M_{\textrm{env}}$) only, given a fixed $p$. Third, we created smaller grids of model intensity images for each object, given $L_{\textrm{TP}}$ and $M_{\textrm{env}}$, for all reasonable values of $p$ and $r_{\textrm{out}}$. Fourth, we interpolated on a grid of $\chi^2$ as a function of $p$ and $r_{\textrm{out}}$ to minimise $\chi^2$ between the model images and close-cropped cutouts from the observational data. Finally, we took the resulting $p$ and $r_{\textrm{out}}$, and repeated the second through fourth steps until $L_{\textrm{TP}}$,  $M_{\textrm{env}}$, $p$, and $r_{\textrm{out}}$ converged. Convergence typically occurred by the third iteration.

Data at $\lambda\lesssim30$~\micron\ were excluded from the fit because these data mostly trace warm (\tdust$\gtrsim100$~K) dust and gas close to the protostar where it is likely to have collapsed into a disk. The disk region around a protostar is not well-approximated by a 1D spherical model (see, for example \citealt{jes_2002} or Appendix~C of \citealt{wish2}). Accordingly, because we have not attempted to model and subtract off a disk component, the envelope masses from \tP\ modelling are likely to be upper limits, especially for more evolved sources with brighter MIR emission components. As discussed in Appendix~C of \citealt{wish2}, not subtracting the warm inner disk effectively boosts the steepness of a protostar's flux profile by convolving the (marginally) resolved outer envelope with the unresolved inner disk. This in turn is expected to inflate the power-law index of the density profile, so those too are expected to be somewhat overestimated.

The MIR to submm data we have cannot be used to determine the ISRF independently, so we did not include it in our grid model. From trial and error, we did find that N12 is slightly better fit when some protostellar luminosity is exchanged for a stronger ISRF, since the ISRF selectively boosts the part of the SED to the short-wavelength-side of the peak. Indeed, most PILS-Cygnus sources are expected to be exposed to a strong ISRF due to the proximity ($\lesssim$300~pc) of various OB associations, especially Cygnus-OB2 (see cyan arrows in Fig.~\ref{fig:cutouts} pointing towards OB clusters). However, in these cases, the ISRF could not be boosted enough at the expense of intrinsic luminosity to fit data at $\lambda\lesssim30$~\micron\ without convergence errors in \tP, most likely because any gas that is far enough out and low enough in density to feel the effects of the ISRF is too rarefied to account for any significant fraction of the mass or emission. Moreover, the widths of the dust filaments and pillars in Fig.~\ref{fig:cutouts} suggest all the sources may be well-shielded out to radii well beyond \rout, even N12. This would imply that most or all of the MIR emission that could not be fit with \tP\ is coming from a hot inner disk heated by the protostar itself, rather than any external emission from tens to hundreds of pc away.

\section{Results and analysis}\label{sec:res}
In this section we discuss the results of both SED-fitting and modelling with \tP, starting with the fitting parameters in \S\ref{ssec:params}. We compare the results for each type of analysis, and discuss potential sources of discrepancies in \S\ref{ssec:methcomp}. Then compare both sets of results to literature covering one or more of our eight targets, identifying which set of our results is more appropriate for comparison and justifying any significant inconsistencies (\S\ref{sssec:lit}). Finally, we place our sources in the context of literature on similarly-young protostars across a wide range of masses and luminosities, and show how we determined the significance (or lack thereof) of apparent differences in trends between different mass regimes (\S\ref{sssec:mcmc}).

\subsection{SED-Fitting and \tP\ Modelling Parameters}\label{ssec:params}
Table~\ref{tab:partab} lists the single or multi-component greybody SED-fitting parameters for all sources. Half of the sources were fit with a single component, denoted as the ``cold'' component in sources where multiple components could be fit. The columns, from left to right, are: source name, the 3$\sigma$ mean radius from \citetalias{motte_earliest_2007} corrected for distance measurements from \citealt{rygl_parallax_2012} ($r_{1.2\textrm{mm}}$), dust temperature of the cold ($T_{\textrm{dust}}<60$~K) temperature component, H$_2$ column density in the cold dust component ($N[\textrm{H}_{2}]$), the dust emissivity index $\beta$, dust temperature and (H+)H$_2$ column density of any additional fitting components ($T_{\textrm{dust},\;i>1}$ and $N_{i>1}[\textrm{(H+)H}_2]$, respectively) listed in order of increasing \tdust. The warmer temperature components rely on limited data, so they should be treated with caution, especially those peaking in the poorly-sampled region between 25 and 70~\micron. 

\begin{table*}
    \centering
    \caption{Modified blackbody fitting parameters for PILS-Cygnus sources.}    \label{tab:partab}
    \begin{threeparttable}
    \begin{tabular}{l l l l l l l}
    \hline
        Source & $r_{1.2\textrm{mm}}$ (AU) & $T_{\textrm{dust}}$ (K) & $N[\textrm{H}_{2}]$ (cm$^{-2}$) & $\beta$ & $T_{\textrm{dust},\;i>1}$ (K)\tnote{(a)} & $N_{i>1}[\textrm{(H+)H}_2]$ (cm$^{-2}$)\tnote{(a)}\\ \hline
        
        N12\tnote{(a)} & $1.9\times10^4$ & $22\pm2$ & $9.5\pm2.7\times10^{22}$ & $1.2\pm0.2$ & $\cdots$ & $\cdots$\\
        
        N30 & $2.3\times10^4$ & $37\pm1$ & $7.8\pm0.4\times10^{22}$ & $1.3\pm0.1$ & $396.4\pm0.1$ & $5.0\pm0.5\times10^{15}$\\
        
        N48 & $2.6\times10^4$ & $40\pm10$ & $7\pm1\times10^{22}$ & $0.6\pm0.4$ & $\cdots$ & $\cdots$\\
        
        N51 & $2.6\times10^4$ & $22\pm1$ & $9\pm2\times10^{22}$ & $1.7\pm0.6$ & $68\pm3$, & $<6.3\times10^{20}$,\\
        
         &  & &  & & $350\pm20$ & $<10^{16}$\\
        
        N53 & $2.2\times10^4$ & $18\pm5$ & $1.2\pm1.1\times10^{23}$ & $1.1\pm0.7$ & $\cdots$ & $\cdots$\\
        
        N63 & $1.2\times10^4$ & $25\pm4$ & $1.2\pm0.4\times10^{23}$ & $0.7\pm0.3$ & $\cdots$ & $\cdots$\\
        
        S8  & $2.9\times10^4$ & $38\pm1$ & $3.0\pm0.2\times10^{22}$ & $1.1\pm0.2$ & $98\pm5$, & $9\pm2\times10^{19}$,\\
        
          &   &   &   &   & $360\pm20$, & $1.0\pm0.3\times10^{16}$,\\
        
           &   &   &   &   & $920\pm40$ & $2.5\pm0.5\times10^{13}$\\
        
        S26 & $5.3\times10^4$ & $45\pm2$ & $4.9\pm0.3\times10^{22}$ & $1.2\pm0.1$ & $140\pm20$, & $1.0\pm0.1\times10^{20}$,\\
        
           &   &   &   &   & $433\pm5$ & $4\pm1\times10^{17}$\\        
    \hline
    \end{tabular}
    \begin{tablenotes}
    \item[a]$T_{\textrm{dust},\;i>1}$ and $N_{i>1}[\textrm{(H+)H}_2]$ are respectively the temperatures and column densities of additional greybody components.
    \item[b]These are the adopted fitting parameters for N12, in which PACS-70~\micron\ data are excluded. When the 70~\micron data are included, \tdust, \nhtwo, and $\beta$ become $31\pm2$~K, $4.3\pm0.6\times10^{22}$~cm$^{-2}$, and $0.6\pm0.1$, respectively.
    \end{tablenotes}
    \end{threeparttable}
\end{table*}

\begin{figure*}
    \centering
    \includegraphics[width=\textwidth]{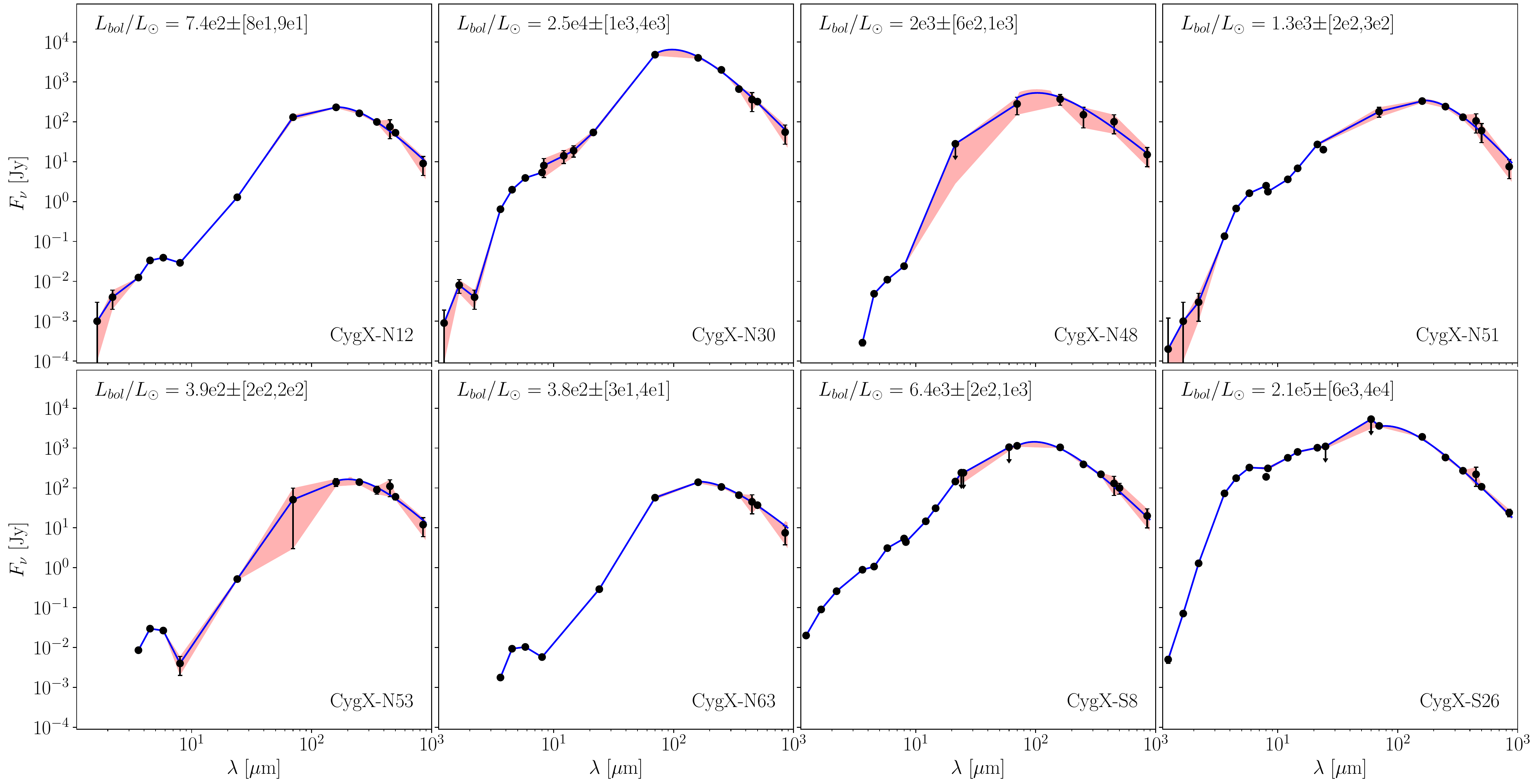}
    \caption{Full SEDs for all sources, which were fitted, interpolated, and integrated by trapezoidal sums to calculate the bolometric luminosities. The curves running through data at $\lambda\geq70$~\micron\ are the FIR/submm component of the SED fits discussed in \S\ref{ssec:seds}, while the rest of the data has been log-linearly interpolated in lieu of integrating over the less-physical warmer SED components.
    For N30 and N51, no SED components passed through the 70~\micron\ data point, so the SED fit was only used for $\lambda\geq160$~\micron\ while the 70~\micron\ data were linearly interpolated like the shorter wavelength data. The light red filled curves show the margins of uncertainty as constrained by the flux errors, and were likewise integrated to estimate the total uncertainty in luminosity.}
    \label{fig:splines}
\end{figure*}

Table~\ref{tab:tptab} gives the FIR luminosities and envelope masses derived from greybody fitting, the bolometric luminosity derived from spline-fitting the data in Table~\ref{tab:fluxtab}, and the fitting parameters for \tP\ modelling. The columns, from left to right, are: source name, bolometric luminosity from spline interpolation of data from 1.22 to 850~$\mu$m ($L_{\textrm{bol}}$; see also Fig.~\ref{fig:splines}), luminosity of the FIR or cold dust component derived from greybody fitting ($L_{\textrm{FIR}}$), total gas mass derived from \nhtwo\ within the 3-$\sigma$ source area given the source dimensions in \citetalias{motte_earliest_2007} at 1.2~mm ($M_{\textrm{c}}$), protostellar luminosity from \tP\ modelling ($L_{\textrm{TP}}$), protostellar envelope mass from \tP\ modelling ($M_{\textrm{env}}$), the fitted power law index of the \tP\ model's density profile ($p$), and the deconvolved fitted radius in the 450~\micron\ SCUBA-2 data ($r_{\textrm{out}}$). Because of the wavelength restrictions on \tP\ modelling, $L_{\textrm{TP}}$ from greybody fitting is more directly comparable to $L_{\textrm{FIR}}$ than $L_{\textrm{bol}}$. Note also that $M_{\textrm{c}}$ is not directly comparable to $M_{\textrm{env}}$ because it incorporates the warmer temperature components and makes different assumptions about the dimensions of each source. All temperature components warmer than the coldest component combined contribute $<1\%$ of the total mass of the object, well within the uncertainties in mass, so $M_{\textrm{env}}$ were not systematically smaller.

We used a range of possible $r_{\textrm{out}}$, defined from the minimum and maximum measured radii of flux contours at 10\% of the background-subtracted maximum flux in the 450~\micron\ images, before deconvolution with the 450~\micron\ SCUBA-2 primary beam, to minimise $\chi^2$ in $p$ and $r_{\textrm{out}}$ relative to the model radial flux profile. For more symmetric sources (flagged 'A' in Table~\ref{tab:tptab}) we radially averaged flux profiles from the 450~\micron\ SCUBA-2 data, while for sources with bright filamentary protrusions (flagged 'B'), we selected a radial cut free of such structures. However, $r_{\textrm{out}}$ turned out to be at least somewhat degenerate with all of the other \tP\ fitting parameters. The resulting \rout\ fits, rounded to the nearest 5000~AU due to precision limits and then deconvolved by subtracting the $1.64\sigma$ 450~\micron\ SCUBA-2 beam radius in quadrature\footnote{The radius of a Gaussian at 10\% of the maximum is about $1.64\sigma$. For the SCUBA-2 beam at 450~\micron, with a FWHM of 7.9\asec, the $1.64\sigma$ radius is about 5.5\asec.}, produced better matches to the dust SEDs than a simple average of the measured minimum and maximum half-widths at 10\% of the maximum flux. However, the uncertainties on the resulting \rout\ values were ill-defined---for small sources like N48, refining \rout\ by 1000~AU produced noticeable improvements, but for most sources we could only tell that the uncertainties were at least 20\% of \rout. The uncertainties reported in Table~\ref{tab:tptab} are the greater of either $\pm0.2r_{\textrm{out}}$, or the deconvolved differences between the fitted \rout\ and the upper and lower measured limits used to constrain the fitting. For $\chi^2$ minimisation in envelope mass and luminosity, it was enough to adopt the nearest multiple of $10^4$~AU and refine the $r_{\textrm{out}}$ by eye using the SED (see \S\ref{ssec:tp}). 
The uncertainties on the masses and FIR luminosities from greybody fitting are propagated from the uncertainties in the parameters from Table~\ref{tab:partab}. Uncertainties for quantities derived from \tP\ modelling are only approximations from the $\sim95\%$ confidence levels on the maps of $\chi^2$.

\begin{table*}
    \centering
    \caption{Parameters for PILS-Cygnus sources derived from SED fitting and \tP\ modelling.}\label{tab:tptab}
    \begin{threeparttable}

    \begin{tabular}{l l l l l l l l l }
    \hline
        Source & $L_{\textrm{bol}}$ ($L_{\odot}$) &$L_{\textrm{FIR}}$ ($L_{\odot}$) & $M_{\textrm{c}}$ ($M_{\odot}$) & $L_{\textrm{TP}}$ ($L_{\odot}$) & $M_{\textrm{env}}$ ($M_{\odot}$) & $p$ & flag($p$)\tnote{(a)} & $r_{\textrm{out}}$ (AU)\\ \hline
        
        N12 & $740^{+80}_{-90}$ & $330-490$\tnote{(b)} & $22-54$\tnote{(b)} & $360^{+40}_{-50}$ & $130\pm30$ & $-1.2^{+0.3}_{-0.1}$ & B & $2.4\pm0.5\times10^4$ \\
        
        N30 & $2.5^{+0.1}_{-0.3}\times10^4$ & $1.5\pm0.2\times10^4$ & $220\pm20$ & $1.4\pm0.3\times10^4$ & $190\pm50$ & $-1.0\pm0.1$ & A & $2.9^{+0.6}_{-0.7}\times10^4$\\
        
        N48 &$2000\pm700$ & $1000\pm300$ & $30\pm20$ & $900\pm200$ & $50\pm10$ & $\gtrsim-0.5$ & B & $1.4^{+0.8}_{-0.3}\times10^4$\\
        
        N51 & $1300\pm200$ & $400-700^\dagger$ & $130\pm30$ & $750\pm70$ & $90\pm10$ & $-1.1\pm0.1$ & B & $2.4^{+0.5}_{-0.8}\times10^4$\\
        
        N53 & $400\pm200$ & $270\pm80$ & $120\pm90$ & $170^{+20}_{-30}$ & $240^{+40}_{-20}$ & $-1.0\pm0.2$ & B & $1.3\pm0.5\times10^4$\\
        
        N63 & $380^{+30}_{-40}$ & $220\pm70$ & $39\pm13$ & $310^{+40}_{-80}$ & $80^{+30}_{-10}$ & $-1.2^{+0.3}_{-0.1}$ & A & $2.4^{+0.5}_{-0.9}\times10^4$ \\
        
        S8  & $6400^{+200}_{-1000}$ & $3500\pm300$ & $63\pm6$ & $3300\pm500$ & $90\pm10$ & $<-1.6$ & B & $3.9^{+0.9}_{-0.8}\times10^4$\\
        
        S26\tnote{(c)} & $2.10^{+0.06}_{-0.4}\times10^5$ & $5.1\pm0.7\times10^4$ & $300\pm30$ & $6.0\pm0.2\times10^4$ &$530^{+110}_{-150}$ & $-1.3\pm0.1$ & A & $8\pm2\times10^4$\\
    \hline
    \end{tabular}
    \begin{tablenotes}
    \item[a]Flag($p$) indicates whether we estimated $p$ using the radially-averaged cut (denoted A), or the ``best'' (lowest-$\chi^2$) cut (denoted B).
    \item[b]These ranges express the range of values plus or minus uncertainties that we get depending on whether 70~\micron\ data are included in the greybody SED fit. See also the note about N12 under Table~\ref{tab:partab}.
    \item[c]S26 is not only more than twice as far away as the other sources, but is also thought to contain at least one outflow seen pole-on \protect\citep{trinidad_2003}.
    \end{tablenotes}
    \end{threeparttable}
\end{table*}

N12 in particular was almost as well-fit with PACS-70~\micron\ data as without, but the physical conditions derived in each case are dramatically different. Including PACS-70~\micron\ data increases \tdust\ to $31\pm2$~K, and reduces $\beta$ to $0.6\pm0.1$ and \nhtwo\ to $4.3\pm0.6\times10^{22}$~cm$^{-2}$. This fit is slightly poorer, but still within the uncertainties of the data. 
For consistency with the adopted parameters from \tP\ modelling, however, we adopt the lower-\tdust, higher \nhtwo\ (higher mass, lower luminosity) results.

\subsection{Comparison of Methods}\label{ssec:methcomp}
\begin{figure}
    \centering
    \includegraphics[width=\columnwidth]{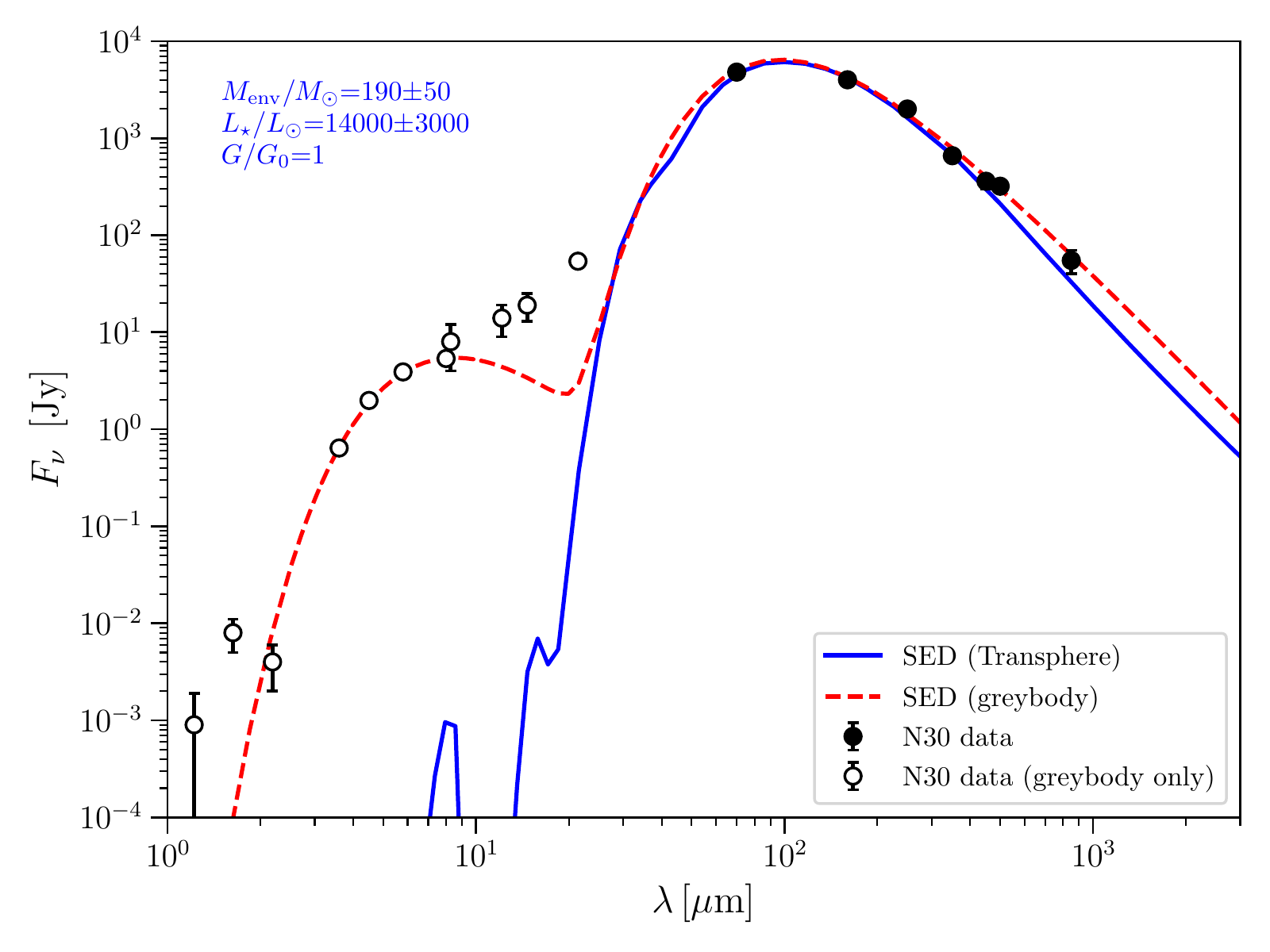}
    \caption{SEDs of N30. The dashed red line is the modified greybody fit to the data, shown as black and grey markers with error bars. The solid blue line is the best fit \tP\ SED to only the data shown in black. The grey points at $\lambda<70\,\mu$m were excluded from the fit because the local ISRF is poorly constrained, and we were generally unable to set it large enough to fit the short-wavelength data without crashing \tP.}
    \label{fig:n30sed}
\end{figure}

\begin{figure*}
    \centering
    \subfloat[A.]{\includegraphics[width=0.49\textwidth]{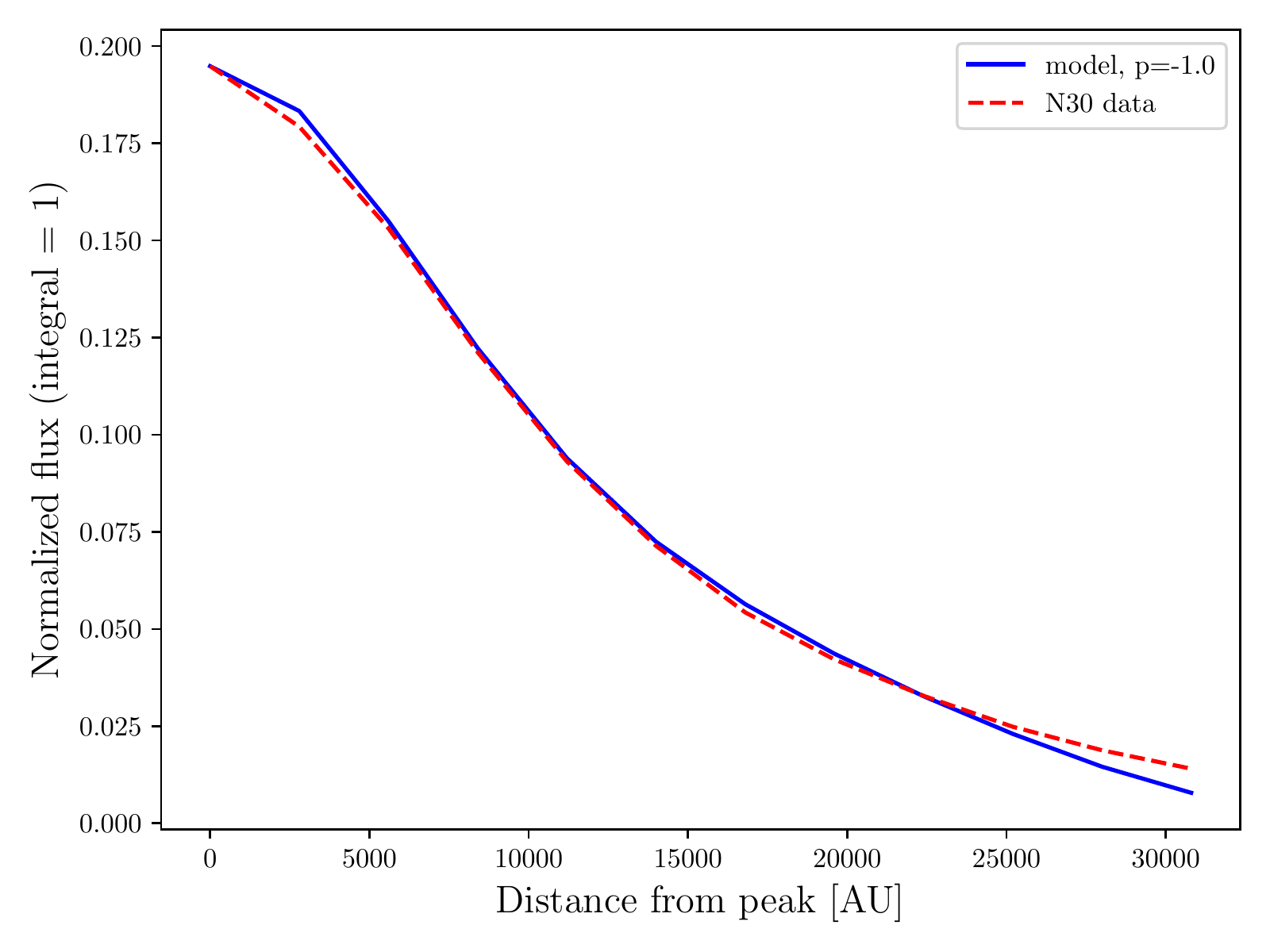}}\quad
    \subfloat[B.]{\includegraphics[width=0.49\textwidth]{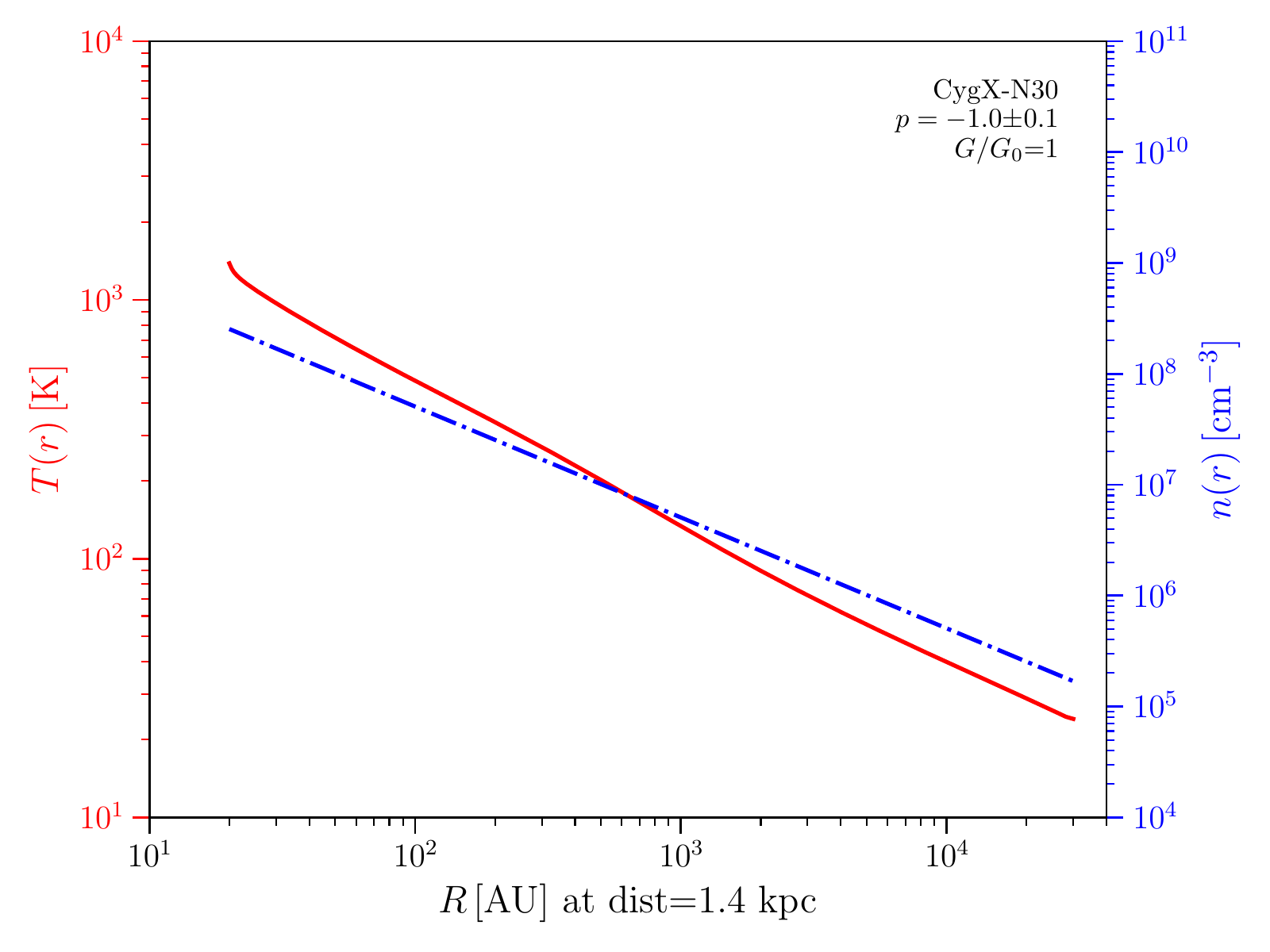}}
    \caption{\textbf{A.} Radial emission profile comparison between the data (red dashed line) and the \tP\ model (blue solid line), both normalised by their respective maxima. \textbf{B.} Resulting model temperature (red solid line and red axis) and density (blue dash-dotted line and blue axis) profiles.}
    \label{fig:n30profs}
\end{figure*}

Fig.~\ref{fig:n30sed} shows the SEDs, both from greybody fitting and \tP, of N30, and Fig.~\ref{fig:n30profs} shows N30's radial flux, temperature, and density profiles given the parameters in Table~\ref{tab:tptab}. Similar figures for all other objects can be found in Appendix~\ref{app:figs1} and \ref{app:figs2} . 
It is unclear from either SED fit whether the 70~\micron\ data should have been excluded: both fits tracked each other closely at that wavelength for 5 out of eight sources (the objects with discrepancies between the two SED models at 70~\micron\ are N48, N51, and N63). It is noteworthy that for the five lowest-\tdust\ sources, the $\beta$ index of a (fixed-\tdust) SED fit to just the 450 and 850~\micron\ data would be steeper than the measured $\beta$ based primarily on the fits to Herschel data. We have considered a variety of explanations, none of which appear to work by themselves. It is highly likely that the shallow $\beta$ coming predominantly from the \emph{Herschel} data is due to our increasing inability to separate emission from the cores and that from their surrounding filaments. If that were the only factor, however, the SCUBA data should be either shallower still or entirely below the rest of the dust SED.

Half of our sources have mutually consistent $M_{\textrm{c}}$ and $M_{\textrm{env}}$, but all eight have mutually consistent $L_{\textrm{FIR}}$ and $L_{\textrm{TP}}$. The discrepancies between the envelope masses from each method can be at least partly accounted for by noting that $M_{\textrm{env}} \propto r_{\textrm{out}}^{3-p}$: for six of the eight sources, $r_{\textrm{out}}$ as measured at 450~\micron\ is substantially larger than the same at 1.2~mm (denoted $r_{1.2\textrm{mm}}$ in Table~\ref{tab:partab}) despite the sources supposedly being resolved at both 450~\micron\ and 1.2~mm. We suspect this is because the 1.2~mm data resolve out some of the more extended structure that would be included at 450~\micron, and it was often less clear what to exclude at the shorter wavelength. The two sources that were measured to be smaller at 450~\micron, N48 and N53, are in crowded high-background areas, and may be over-subtracted at 450~\micron\ or blended at 1.2~mm.

Some of the discrepancy may also come from the power-law indices for each method and how they relate mathematically (if not physically). Physically, there is little reason to expect that $\beta$ and $p$ should be related, and \tP\ does not fit or incorporate $\beta$ explicitly, but mathematical effects of $\beta$ and $p$ on the SED are similar: the smaller their absolute values, the wider the SED. When comparing radial profiles from \tP\ modelling to the radial profiles of the sources at 450~\micron, neither smoothing nor Gaussian deconvolution of sources could be used without altering the derived density profile, so we derived $p$ from the ``best'' (lowest-$\chi^2$ relative to the model) background-subtracted radial cut in high-contamination areas, and used the radially-averaged $p$ in nominally isolated sources. Even in relatively isolated sources, $p$ for the average and ``best'' radial cuts could be different by up to 50$\%$. For greybody fitting, because the $\beta$ is an average over a model of the whole source, $\beta$ often more closely resembles the radially-averaged $|p|$. In fact, under certain conditions (namely optically thin dust emission, for $T>10$~K, in the Rayleigh-Jeans approximation), $p$ and $\beta$ are physically related because of how intensity can be defined as a function of projected distance from the centre of a centrally-heated source. Under these assumptions, if the density and temperature profiles can be described as power laws of indices $p$ and $q$ respectively, then the intensity as a function of projected radial distance is also a power law with an index of $1-(p+q)$ \citep{scoville_kwan_1976,beltran_2002}. Then in the optically thin Rayleigh-Jeans limit, it can be shown analytically that the temperature power-law index is related to the dust emissivity index via $q=2/(4+\beta)$ \citep{chandler_1989,kenyon_1993}.

We had convolved the SCUBA legacy data\footnote{The SCUBA-2 data have a better resolution and smaller uncertainties, but the secondary beam component contains a substantially larger fraction of the total flux (40\% at 450~\micron, \citealt{holland_scuba-2_2013}) and is larger in radius.} to the radius of the secondary or ``error'' beam, which has a FWHM 40\asec, and then attempted the same Gaussian deconvolution routine as with the \emph{Herschel} data. Because these data were much more aggressively background-subtracted than \emph{Herschel} data, we allowed the fitted flat background to accommodate negative average background levels so we could add it to the peak amplitude of the Gaussian fit. We still integrated over a Gaussian with zero constant background level. Nevertheless, we know that negative average background levels are more of an issue at 450~\micron\ than 850~\micron, so data at the shorter wavelength may have still been boosted too much despite removing the flat background from integration. That makes it less clear whether $\beta$ from the SCUBA data alone would be more reliable than the values of $\beta$ we derived, which are dominated by \emph{Herschel} data. For most objects that were isolated or separable, the results seemed reasonable to within the 50\%\ uncertainty of the 450~\micron\ data as compared to the 350 and 500~\micron\ \emph{Herschel}/SPIRE data. We also checked the results against simple aperture photometry with SAO DS9 for both SCUBA and SCUBA-2 data, and with \emph{Herschel} images where the clumps could be isolated. For S8 alone, we redid the photometry manually, cutting out the bright spiral-arm-like feature identified in \citetalias{motte_earliest_2007} as S7. Unfortunately, PSF fitting of the SCUBA data was never an option since, at the native SCUBA and SCUBA-2 resolutions, all of the objects are resolved along at least one axis.

Given that the \tP\ models incorporate more physics, 
the masses from those models are probably more accurate than the masses from multi-component greybody fitting, of which some components are not necessarily physical. The luminosities could only be fully accounted for by spline-fitting the SED as shown in Fig.~\ref{fig:splines}; until the degeneracy between intrinsic luminosity and ISRF strength is broken, we will have to evaluate which luminosity is more consistent with other results on a case-by-case basis. Thus, wherever we adopt values for chemical modelling purposes, we prefer the \tP\ model values for everything except the luminosity, for which $L_{\textrm{bol}}$ is preferred given the relative lack of underlying assumptions.

\subsection{Comparison with Literature}\label{ssec:litcomp}
\subsubsection{Other Studies of Cygnus-X}\label{sssec:lit}
Among literature where the same objects are analysed at comparable size scales (as opposed to separately evaluating components of each object identified at mm wavelengths), there is precious little consistency between different groups using the same methods, let alone between different methods. Most of the scatter stems from the lack of consensus on how to define the sizes of these objects, upon which the masses in particular sensitively depend. In some cases the discrepancies are further exacerbated by the use of different power-law indices $\beta$ and $p$, either because they were fixed parameters \citep[as with, for example][]{duarte_cabral_2013,bgps_xiii,cao_2019,glostar} or because they were fitted using a different technique \citep[for example][]{palau14}. We find that in general, the results of greybody fitting agree slightly more often with literature than the results of \tP\ modelling, in large part because most other groups also performed observational SED-fitting of some kind, while very few attempted to model density profiles.

After correcting for their older distance measurements, we found that 6 out of 8 sources had at least one mass measurement consistent with the respective mass estimates in \citetalias{motte_earliest_2007}. For N30 and N48, neither $M_{\textrm{c}}$ nor $M_{\textrm{env}}$ were as large as the envelope masses computed by \citetalias{motte_earliest_2007}, though the discrepancy with N30 was small. The rest of the sources seem to agree more closely with the \citetalias{motte_earliest_2007} masses when they have lower \tdust, since the temperatures we found were typically higher than \citetalias{motte_earliest_2007} assume. All of this makes sense given the methods they used to determine mass and luminosity---they had no resolved data between 21 and 350~\micron, so they derived the mass ranges using only their 1.2 mm fluxes and assumed temperatures between 15 and 25~K. Their assumed temperatures were usually lower than we found, and the opacity coefficient they use folds in an assumption that $\beta=1.8$ \citep{ossenkopf_dust_1994}, which we found to be too steep. For a given temperature, a slightly steeper $\beta$ significantly increases any fitted column or surface density parameters.

The most direct and comprehensive comparison we can make to our greybody fitting results is with \citealt{cao_2019}, who included all of the PILS-Cygnus targets except for S26 in their sample. They decomposed the emission into different spatial scales, fit single-component SEDs pixel-by-pixel to only data with $\lambda>70$~\micron\, assuming $\beta=2$ and optically thin emission, and excluded data that were too far from the resulting curve (which usually included the 70~\micron\ data). We use the same angular size measurements, but our method of deconvolving and fitting the targets is much more basic. One would expect that they would derive lower \tdust\ and luminosities, and our masses would be similar to or larger than their estimates. Indeed, we systematically derive higher temperatures for five of the seven sources largely because of our inclusion of the 70~\micron\ data. However, we found that our SED fits yielded significantly lower masses for all objects except for N51 and N53, which were within the uncertainties. Instead, their luminosities matched our greybody-derived luminosities within the uncertainties for four objects, and for a further two (N51 and N53), the luminosities we derived with \tP\ matched instead. Only N48 had a dramatically discrepant luminosity, and none of its other derived parameters matched, either--not unexpected given N48's large uncertainties and our aggressive decontamination efforts. We believe most of the mass discrepancies are related to our different values of $\beta$, as discussed above with respect to \citetalias{motte_earliest_2007}; increasing the temperature decreases the column density required to match the amplitude of the SED further.

Only one other group attempted density profile fitting to derive $p$ for more than one object in our sample. \citealt{palau14} simultaneously performed the density profile fitting and SED-fitting with a single temperature component fit to all data with $\lambda>50$~\micron, including free-free cm-wave radiation where it was available. Their single-component SED fits were usually offset towards higher temperatures relative to both their data and our coldest SED components for the same objects, and it is unclear why that would happen with their wavelength restrictions. Our \tP\ models used the same lower wavelength limit, and while the \tP\ models were often narrower than the greybody fits, the peak positions of the two fits were seldom misaligned. The fitted values of $p$ from \citealt{palau14} are also systematically greater in magnitude, as are their $\beta$ values except in the case of S26. These results would make sense if the narrower SED fits of \citealt{palau14} were because their fits systematically ignored the 70~\micron\ \emph{Herschel} data, but except in the case of S26, their model SEDs appear to fit the 70~\micron\ data and run parallel to but offset from the longer-wavelength data. 

\subsubsection{Comparison to Protostars of All Sizes}\label{sssec:mcmc}
\begin{figure*}
    \centering
    \includegraphics[width=\textwidth]{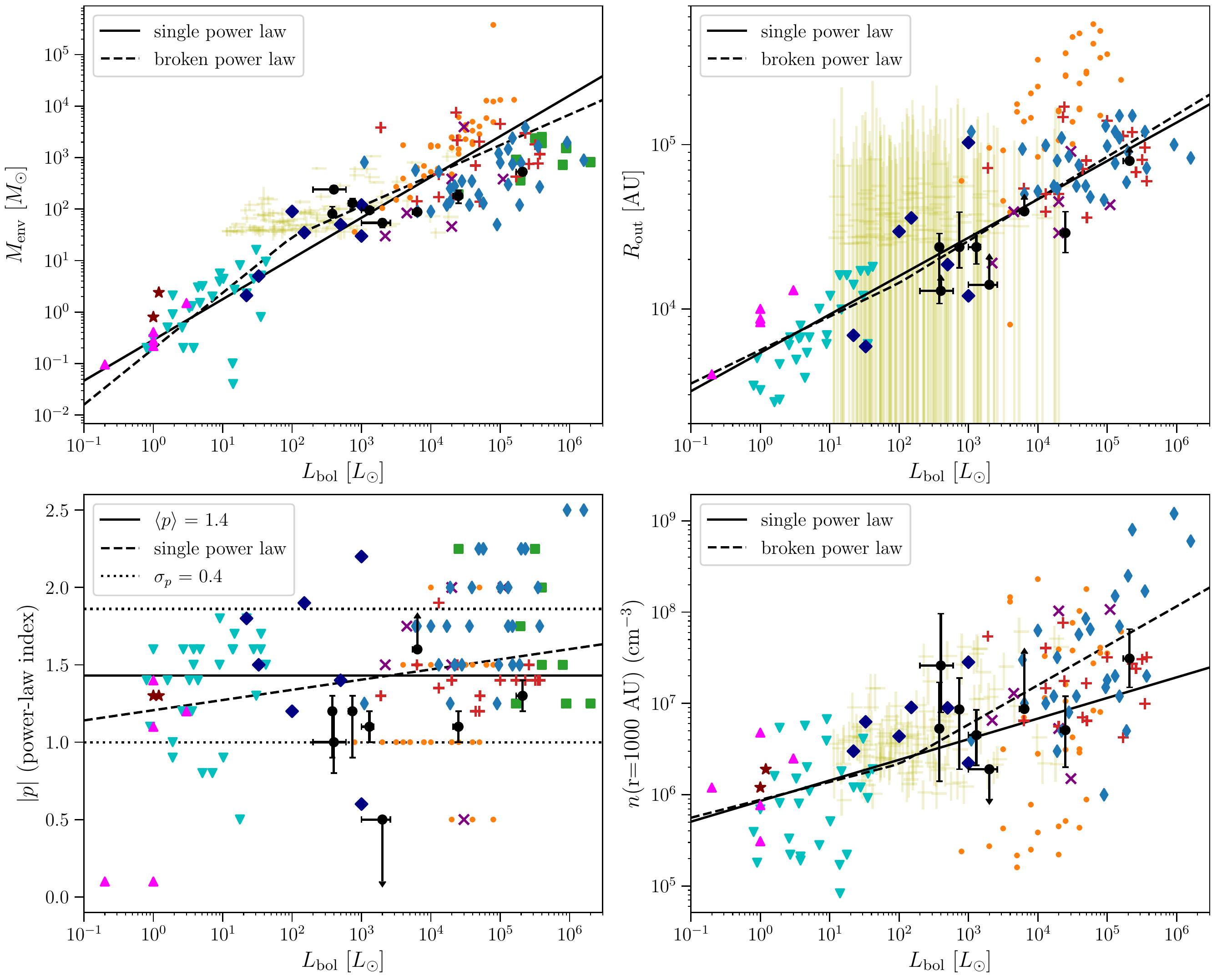}
    \caption{Comparison of PILS-Cygnus Sources (large black circles with error bars) to objects in the literature from \citealt{tak2000} (purple x's), \citealt{jes_2002} (magenta upward-pointing triangles), \citealt{mueller02} (thin blue diamonds), \citealt{shirley02} (maroon stars), \citealt{hatch03} (green squares), \citealt{williams05} (small orange circles), \citealt{crimier_2009,crimier_2010,crimier_2010b} (wide navy diamonds), \citealt{wish2} (cyan downward-pointing triangles), \citealt{tak2013} (red plus signs), and \citealt{cao_2019} (pale gold error bars). Solid and dashed black lines represent single and broken power law fits to the data, respectively.}
    \label{fig:crim}
\end{figure*}

Following the example of \citealt{crimier_2010}, in Fig.~\ref{fig:crim} we have plotted the \tP-derived envelope masses, $\chi^2$-fitted 450-\micron\ outer envelope radii, density power-law index, and gas number density at $r=1000$~AU against bolometric luminosity for all of the PILS-Cygnus sources (black circles with error bars in Fig.~\ref{fig:crim}). For comparison, we include similar results for about 180 other resolved protostellar sources ranging from $<0.1$ to $>10^4\;M_{\odot}$ from \citealt{tak2000}, \citealt{jes_2002}, \citealt{mueller02}, \citealt{shirley02}, \citealt{hatch03}, \citealt{williams05}, \citealt{crimier_2009,crimier_2010,crimier_2010b}, \citealt{wish2}, and \citealt{tak2013}. In an attempt to further fill out the sampling shortage between $\sim50$ and 1000~$L_{\odot}$, we also tried to include possible ranges of values from the extensive but resolution-limited survey by \citealt{cao_2019} (olive points with large error bars), excluding prestellar sources and sources whose deconvolved sizes were smaller than the \emph{Herschel}/PACS 160~\micron\ beam, and using the mean and sample standard deviation (1.4 and 0.4, respectively) of the density power law index of the rest of the data to estimate the particle densities at 1000~AU. The resulting distribution of data from \citealt{cao_2019} appears to have a slope that is flatter than the aggregate distribution of the other sources, but that is likely an artefact of the sharp lower limit in radius and mass, so we treat these data with caution and exclude them from curve fitting.

Concentrations of data at specific power law indices are artefacts of limited precision in their respective surveys; for example, \citealt{williams05} rounded their power law indices to the nearest multiple of 0.5. Uncertainties for data other than ours are not shown on the plot, but in all cases, we either adopted the values stated in the literature or assume uncertainties based on contextual information in the original literature (for example if no luminosity uncertainty was given, we adopted the largest percentage uncertainty in the flux) for curve fitting purposes. A machine-readable table of all values used for plotting will be included with the online publication.

For the plots of \menv\ versus \lbol, \rout\ versus \lbol, and \nkau\ versus \lbol, we minimised the negative log likelihoods for single and broken power law fits (solid and dashed black lines in Fig.~\ref{fig:crim}, respectively), modelled their respective posterior distributions using \texttt{emcee} \citep{emcee}, and compared the sum of squared residuals for each fit. For each single power law fit, we solved for:
\begin{itemize}
    \item the power law exponent,
    \item the normalisation at $L_{\textrm{bol}}=1\,L_{\odot}$, and 
    \item ln$\epsilon$, a nuisance parameter that assumes any residual sources of scatter that are not accounted for by the input uncertainties in the other parameters are constant in \lbol.
\end{itemize}
 The broken power law fit is similar to the single power law fit except that the normalisation is solved for at the power-law break point in $L_{\textrm{bol}}$ instead of at $1\,L_{\odot}$. We somewhat arbitrarily set the break point at $L_{\textrm{bol}}=60\,L_{\odot}$ because the fits failed to converge when that parameter was free, and that value is the approximate centre (in the log scale) of the gap in reliable data for intermediate luminosity protostars. As expected given the amplitude of scatter around both sets of fits, we found that in all cases, residual sums of squares for the broken power law and single power law differed by $<10\%$ ($<2$\% for \menv\ and \rout\ vs. \lbol), smaller than the uncertainties for nearly all available data. Thus, we provide the following single power law results as our final fits:
\begin{equation}\label{eq:mvl}
        \mathrm{log}\left(\frac{M_{\mathrm{env}}}{M_{\odot}}\right) = 0.79^{+0.01}_{-0.02}\mathrm{log}\left(\frac{L_{\mathrm{bol}}}{L_{\odot}}\right) + \mathrm{log}(0.30^{+0.07}_{-0.06})
    \end{equation}
\begin{equation}\label{eq:rvl}
        \mathrm{log}\left(\frac{R_{\mathrm{out}}}{\mathrm{AU}}\right) = 0.23\pm0.02\mathrm{log}\left(\frac{L_{\mathrm{bol}}}{L_{\odot}}\right) + \mathrm{log}(5.4^{+0.7}_{-0.6}\times10^3)
    \end{equation} 
\begin{equation}\label{eq:nvl}
        \mathrm{log}\left(\frac{n_{1000\mathrm{AU}}}{\mathrm{cm}^{-3}}\right) = 0.2^{+0.3}_{-0.2}\mathrm{log}\left(\frac{L_{\mathrm{bol}}}{L_{\odot}}\right) + 5.9\pm0.8.
    \end{equation} 
For Eqs.~\ref{eq:mvl} and \ref{eq:rvl}, the values of ln$\epsilon$ were $-0.08^{+0.04}_{-0.11}$ and $-1.0\pm0.2$, respectively. Because the uncertainties for most values of \nkau\ are large, the corresponding nuisance parameter became so small ($\sim10^{-17}$) that removing it improved the fit. Corner plots of the posterior distributions of all parameters in these fits are available in Appendix~\ref{app:mcmc}.

For the plot of $|p|$ against \lbol, we followed a similar procedure as for the other three plots to fit a log-linear function in \lbol\ with a nuisance parameter ln$\epsilon$, and to model the posterior distributions. Appendix~\ref{app:mcmc} also contains the corresponding corner plots for these fitting parameters. We compared this fit (dashed black line in the bottom left panel of Fig.~\ref{fig:crim}) to the mean and standard deviation of $|p|$, 1.4 and 0.4 respectively (solid and dotted black lines in the bottom left panel of Fig.~\ref{fig:crim}), by performing a Student's t-Test with the null hypothesis that the slope of the line is indistinguishable from zero. We acknowledge that the uncertainties in $|p|$ for our data and those from the literature are not the same for all data, but they do not appear to depend systematically on \lbol\ and do not vary much over the full range of \lbol---typical uncertainties are about 0.3. Thus, we believe the Student t-Test should at least be indicative enough to verify the insignificance of the slope suggested by the fit remaining within a standard deviation of the mean over the full \lbol\ domain. The two-tailed p-value (not to be confused with the density power law index $|p|$) we calculated from the t-statistic was 0.87. If our use of this test is valid, then it is statistically impossible to distinguish the distribution of our and others' $|p|$ indices from a Gaussian with a mean of 1.4 and a standard deviation of 0.4. We also note that all values of $|p|>2$ have uncertainties large enough that they are consistent with $|p|=2$, and that having $|p|>2$ is probably not physical, so the apparent upward trend in $|p|$ with \lbol\ is even less likely to be real than it looks.

\section{Discussion} \label{sec:disc}
In this section, we start by discussing the results and implications for the eight sources in Cygnus-X, especially their density profiles in the context of their individual shapes and immediate environmental conditions. Then we zoom out to discuss our meta-analysis of envelope parameters in the literature, with our eight sources included. There, we focus on the statistical significance of the scaling relationships derived, that is, whether the relationships for low- and high-mass (luminosity) stars are statistically distinguishable or consistent with one power law per parameter over the full range of sampled protostellar luminosities.

\subsection{Profiles of Cygnus-X sources}\label{ssec:disc8}
All of our sources have strong emission peaks at wavelengths of order $10^2$~\micron, and less emission towards shorter wavelengths, as shown in Fig.~\ref{fig:splines}. In the protostellar spectral classification scheme of \citealt{lada87} and \citealt{andre93}, Class 0 and I sources have flux densities increasing with wavelength from NIR to submm wavelengths, whereas Class II and III sources have flux densities decreasing with wavelength over the same range. The decline in flux density towards shorter wavelengths for all of our sources identifies them as Class 0 or I. The original criteria from \citealt{andre93} for Class 0 sources are not directly readable from an SED, and under the strictest interpretation of the revised criteria in \citealt{barsony94}, all of our sources would be Class I. However, non-detection at $\lambda<10$~\micron\ with \emph{Spitzer} is very different from a non-detection in the late 1980s to early 1990s, when the most deeply embedded sources detectable in the MIR were up to three orders of magnitude brighter at 100~\micron\ than 10~\micron \citep{lada87}. Three of our sources are more than three orders of magnitude fainter at $\lambda\lesssim10$~\micron\ than at $\lambda\sim100$~\micron: N12, N53, and N63. Thus, we argue that these should be at least considered possible Class 0 sources.

The average and standard deviation of $p$ for our eight sources alone are -1.1 and 0.3, respectively. That makes our sub-sample statistically inconsistent with the theoretical slope of -1.5 for isolated sources in free fall, even as all the sources we could find in the literature put together have a mean that is statistically consistent with -1.5 within their dispersion. However, as the three-color \emph{Herschel} images in Fig.~\ref{fig:cutouts} show, our sources are not isolated. There are at least four potential ways that the modelled envelope profiles could be influenced physically or observationally by their surroundings, any or all of which could be in effect:
\begin{enumerate}
    \item \textbf{Geometric.} The shallower slopes are an artefact of spherically averaging over asymmetrical features, like bipolar outflows, core fragments, and spiral arms.
    \item \textbf{Observational.} The density profiles of most PILS-Cygnus sources, like their fluxes, are convolved with those of the filaments in which they reside.
    \item \textbf{Physical.} The envelope may have filaments feeding material into the clump.
    \item \textbf{FUV.} The clumps' envelopes may be inflated by a moderately strong ISRF. 
\end{enumerate}

The geometric factors encapsulate a great number of circumstances that vary from source to source. The PILS-Cygnus sample, though small, is quite diverse in morphology, and asymmetries are readily visible in the continuum images at both submm (Fig.~\ref{fig:smaimgs}) and FIR (Fig.~\ref{fig:cutouts}) wavelengths. Several follow-up studies have examined members of the \citetalias{motte_earliest_2007} sample in search of fragmentation, with special attention paid to the clumps that were IR-quiet (weak to non-existent emission in the $1\lesssim\lambda\lesssim30$~\micron\ range, roughly). Studies by \citealt{bontemps_fragmentation_2010} and \citealt{duarte_cabral_2013} looked at substructure in N12, N48, N53, and N63 with millimetre continuum and CO, and found all of them had two or more components. In fact, the only sources in the PILS-Cygnus survey not known to contain multiple cores are N38, N51, and N54. S26 looks monolithic, but given its apparent distance and luminosity, it is highly likely to be a protocluster rather than a single protostar; if it is a single protostar, it is likely to have an outflow pointed directly down our line of sight \citep{trinidad_2003}, enhancing its apparent luminosity.

All of the sources have at least one pair of outflows (Skretas et al. submitted), and some appear to originate from a location offset from the continuum peak---strong indicators that both the geometric and physical factors are at work since outflows are indirect tracers of accretion \citep[see for example][]{blandford82,pelletier92,stahler94,hennebelle2008}. N30, N51, S8, and S26 are known to be associated with compact H\textsc{ii} regions: N30 with W75N(B) \citep{w75nb1981}, especially components VLA 1--3 \citep{vla1997}; N51 with MSX6C G081.7522+00.5906 \citep{urquhart_msx_2011}; S8 with LBN~77.61+01.59 \citep{lbn1965}; and S26 with RAFGL~2591 VLA~3 \citep{rafgl1983}. It is perhaps noteworthy that the only protostellar envelope with $|p|>1.5$ is a part of this group. Conversely, given the possibility of association between N48 and 2MASS J20390285+4222001, and the identification of the latter as an H\textsc{ii} region by \citealt{maud2015}, it is unclear whether or not N48 ($|p|\lesssim0.5$), should also be in this group. If N48 is associated with an H\textsc{ii} region, unless the effects of multiplicity can be disentangled and wholly account for the extremely flat density profile, our limited statistics would no longer support a correlation of $p$ with later-stage protostellar evolution.

It is also worth noting that including a warm inner disk and treating it as part of the spherical envelope is expected to steepen the emission profile, ergo $p$. This is a likely explanation for the handful of sources with $|p|>2$ from \citealt{mueller02}, \citealt{hatch03}, and \citealt{crimier_2010}, although all have sufficiently large uncertainties in $p$ that they are still consistent with $|p|=2$. Since we mostly observe flattened profiles among our 8 sources, flattening effects on the density profile (for example core fragmentation) for these sources appear to totally overwhelm any countervailing steepening effects.

The observational and physical factors are especially relevant for N30, N48, N51, and N53, as they are embedded in the larger filamentary structure known as DR21. At the FIR wavelengths covered by \emph{Herschel}, N53 is mostly likely at least somewhat blended with N54, and N48 is blended with at least two other objects in DR21 South. N30 has a large filament bisecting it from equatorial north-east to south-west, plus at least one additional spur extending towards the east, the latter of which was identified in \citetalias{motte_earliest_2007} as N31. As discussed in \S\ref{sec:meth}, we tried to account for blending using Gaussian decomposition of the sources; however, because we could not effectively model the embedding filament, the filamentary contribution will affect the density profile, especially at large radii. 

We are skeptical that FUV irradiation can have much of a flattening effect on the density profile; the observed impact of the ISRF on the surrounding filaments, where visible, looks more like compression than expansion. In Fig.~\ref{fig:cutouts}, N12 sits in the head of pillar that appears to protrude into a bubble bounded on the equatorial west side by DR17. The head of this pillar shows a slight increase in opacity that seems indicative of dust and gas being compressed between N12 and radiation from the vicinity of DR17. N12 also has the smallest projected distance from Cygnus-OB2, at about 30 pc compared to $\gtrsim40$~pc for most other sources, caveats listed in \S\ref{sec:meth} notwithstanding. Yet the shapes of the SED and density profile of N12 are most similar to possibly the most deeply embedded source in our sample, N63. In any case, none of the SED-fitting and \tP-modelling parameters for N12 or N63 are outliers for this sample, which tentatively suggests that the effects of external irradiation are either degenerate with or relatively insignificant compared to more intrinsic variations (for example mass and core multiplicity).

\subsection{Connecting Low- and High-Mass Envelopes}\label{ssec:discall}
The fits to the aggregate of our data with literature discussed in \S\ref{sssec:mcmc} and shown in Fig.~\ref{fig:crim} suggest that, within the measurement uncertainties, high-mass protostellar envelopes only differ from their low-mass counterparts in scale. Our eight sources sit in the upper intermediate to high mass range and appear to follow established trends, where any are visible: envelope mass, outer envelope radius, and gas number density at 1000~AU all seem to increase as power laws in \lbol. Only the gas density power law index appears to be more or less independent of luminosity, with an uncertainty-weighted mean value of 1.4 and sample standard deviation of 0.4 for all sources put together. This mean and standard deviation is consistent with the theoretical $|p|=-1.5$ for sources in free-fall. Recall, however, that our eight sources alone had a mean $|p|$ and $\sigma_p$ of $1.1\pm0.3$, which are marginally inconsistent with the theoretical $|p|=-1.5$. This means that, while our 8 sources do help fill in the intermediate luminosity gap, they are not a very representative sample in $p$, at least given current observational limitations.

If the relative importance of, say, turbulence or magnetic fields as compared to gravity differed between protostellar mass regimes, one would expect that to show up as a systematic difference in typical values of $|p|$, which should propagate to the slopes of \nkau\ and/or \menv\ versus \lbol. A cursory glance at Fig.~\ref{fig:crim}, without the benefit of visible error bars on every point, does seem to vaguely suggest a transition or break at \lbol\ between a few tens and a few hundred $L_{\odot}$ for all relationships except the trend in \rout\ with \lbol. However, the uncertainties are so large that only a first order trend can be established with any confidence; any apparent change of trend with increasing \lbol\ is simply not statistically significant. More data are needed on new sources, and measurements for existing sources need to be refined further.

Fits to \menv, \rout, and \nkau\ as functions of luminosity seem primarily driven by just the three most extensive data sets (after excluding all older measurements of the same objects): \citealt{mueller02} and \citealt{williams05} at the high-mass, high-luminosity end, and \citealt{wish2} at the low-mass, low-luminosity end. At the high-luminosity end, much of the data from \citealt{williams05} stands noticeably apart from most other data sets in the same luminosity regime. For \rout\ and \nkau, the uncertainties on these data are so large that the fitting routine does not give them much weight, but for the \menv-\lbol\ relationship, these data seem to skew the fit towards a steeper slope than the rest of the data appear to warrant. We repeated the fits of single and broken power laws to this relationship excluding the data from \citealt{williams05} to see if the two functions then became statistically distinguishable. They did not, likely because some data from \citealt{hatch03} also have large scatter towards higher \menv, even though they are not outliers in the trends of \rout\ and \nkau\ with \lbol.

For the upper two panels of Fig.~\ref{fig:crim}, we caution that envelope mass depends strongly on outer envelope radius, and methods for measuring the outer envelope radii varied from study to study. For most of the low-mass sources, particularly from \citealt{jes_2002} and \citealt{wish2}, \rout\ was defined as the radius where temperature dropped to 10~K. For high-mass sources, radii were typically measured in submm or short mm-wave images, either as an average radius of a contour above a background threshold or as some multiple of a Gaussian FWHM or $\sigma$; often the definition of \rout\ was unclear or not defined in terms of Gaussian parameters. 
For most high mass sources in general, it does not make much sense to try to calculate a 10~K radius because high-mass star-forming regions tend to be highly irradiated environments where only the heavily shielded centres of prestellar cores get that cold. The majority of our sources in particular are within 300 pc of Cygnus-OB2 or other OB clusters within Cygnus-X, so the ambient medium should be warmer, and indeed, the map of dust temperatures around DR21 by \citealt{henneman2012} shows dust temperatures mostly between 15 and 22~K. These systematic differences in how \rout\ is measured in different mass and luminosity regimes may be at least partially responsible for how the mass-luminosity and radius-luminosity trends visually appear slightly flatter for high-mass sources than low-mass sources. As of now, this visual change of trend between luminosity regimes is not statistically significant, but if more data make it so, future astronomers will have to start trying both methods on the same sources wherever feasible to quantify the difference and see if there is a typical fractional discrepancy that can be applied as a correction to older data. 

Additionally, the methods typically used for measuring the radii of high-mass sources yield somewhat wavelength dependent results, with less emission being recovered at mm wavelengths than closer to the wavelength of the peak of the dust SED. Furthermore, ground-based observations are limited by sky background, which will necessarily decrease the measured radius. For our sources, we again caution that the method we used to fit the envelope radii yields large and sometimes poorly-defined uncertainties due to the degeneracy with the density power law index and---in the case of N48 and S8---complex, radially asymmetric morphology (see the discussion of Table~\ref{tab:tptab} in \ref{sec:res}).



\section{Conclusions} \label{sec:conc}
In this study, we have deconvolved and fit greybody SEDs to archival IR data from 1.2 to 850~\micron\ on eight intermediate- to high-mass Class 0-I protostellar sources. We used the derived luminosities and total mass estimates from SED fitting, and measurements of the sources' sizes at 450~\micron, to perform 1D radiative transfer modelling using \tP, with which we derived our sources' envelope masses, radial density structures, and temperature profiles. Finally, we combined these results with similar data in the literature for as many distinct sources as we could find, and fit the aggregated data to determine if low- and high-mass stars are statistically distinguishable. Our results are as follows:
\begin{enumerate}
    \item Most of our eight sources have envelope masses between a few tens and a couple hundred $M_{\odot}$, and luminosities of several hundred to several thousand $L_{\odot}$, placing them in an intermediate mass and luminosity range that is otherwise poorly sampled in the literature.
    \item The power law indices of our eight sources' radial density profiles cluster around $|p|=1.1$ with a dispersion of 0.3, slightly shallower than predicted for an isolated, gravitationally collapsing source ($|p|=1.5$). We attribute the flattening to some combination of active accretion, spherical averaging of non-spherical features (for example disks and outflow cavities), and convolution with surrounding cloud structures. We briefly discuss a fourth possibility, inflation by a strong IRSF, but find it unconvincing.
    \item When our sources are considered with other sources in the literature, we find the scatter is too large in envelope mass, outer radius, density power-law index, and gas density at 1000~AU to definitively establish if trends differ for low and high mass-sources. Low- and high-mass sources appear to differ significantly only in scale.
    \item We establish to first-order that log(\menv)$\;\propto0.79^{+0.01}_{-0.02}$log(\lbol), log(\rout)$\;\propto0.23\pm-0.02$log(\lbol), and log(\nkau)$\;\propto0.2^{+0.3}_{-0.2}$log(\lbol) over seven decades of \lbol, with a scatter that is consistently about an order of magnitude above and below these fits.
    \item For the aggregate of sources in the literature plus our sources, we determine that the distribution of density power law indices is consistent with a constant $1.4\pm0.4$. With this much scatter, if low- and high-mass protostars get different relative amounts of turbulent or magnetic field support against gravity, we do not have the precision to detect the difference at a statistically significant level. It also implies that our eight sources were not a representative sample in this parameter.
\end{enumerate}
Our inability to distinguish trends in the envelope parameters of low- and high-mass protostars (\menv, \rout, \nkau, $p$) clearly warrants further investigation, especially more and larger surveys that focus on protostars in the \lbol\ range of 50--500 $L_{\odot}$. Nonetheless, the results presented here suggest that the differences between low- and high-mass sources do not arise from variations in large-scale physical properties.

\section*{Acknowledgements}
The research of RLP and LEK is supported by a research grant (19127) from VILLUM FONDEN. JKJ acknowledges support from the Independent Research Fund Denmark (grant number DFF0135-00123B). This research has made use of the NASA/IPAC Infrared Science Archive, which is operated by the Jet Propulsion Laboratory, California Institute of Technology, under contract with the National Aeronautics and Space Administration (NASA).
This research was made possible with the use of the Submillimeter Array. The Submillimeter Array is a joint project between the Smithsonian Astrophysical Observatory and the Academia Sinica Institute of Astronomy and Astrophysics and is funded by the Smithsonian Institution and the Academia Sinica.
This research used archival data from Herschel, an ESA space observatory with science instruments provided by European-led Principal Investigator consortia and with important participation from NASA.
This work is based in part on archival data obtained with the Spitzer Space Telescope, which was operated by the Jet Propulsion Laboratory, California Institute of Technology under a contract with NASA. Support for this work was provided by an award issued by JPL/Caltech. This research also made use of data products from the Midcourse Space Experiment. Processing of the data was funded by the Ballistic Missile Defense Organisation with additional support from NASA Office of Space Science. This publication makes use of data products from the Two Micron All Sky Survey, which is a joint project of the University of Massachusetts and the Infrared Processing and Analysis Center/California Institute of Technology, funded by the NASA and the National Science Foundation. 
This research also used the facilities of the Canadian Astronomy Data Centre (CADC) operated by the National Research Council of Canada with the support of the Canadian Space Agency.
This research made use of Astropy,\footnote{http://www.astropy.org} a community-developed core Python package for Astronomy \citep{astropy2013, astropy2018}.

\section*{Data availability}
The majority of the continuum data used for this project are publicly available through IRSA, or in the case of the SCUBA-2 data, through the Canadian Astronomy Data Centre. 
 



\bibliographystyle{aa}
\bibliography{pilscygxlib} 




\appendix
\onecolumn
\section{Measured and deconvolved continuum fluxes}\label{app:flux}
\begin{table*}[hbt]
\small
    \centering \caption{Measured or deconvolved fluxes of PILS-Cygnus sources.}\label{tab:fluxtab}
    \begin{threeparttable}
    \begin{tabular*}{\linewidth}{lllllll} \hline
Filter & $\lambda$ ($\mu$m) &  $F_\lambda$(N12) &  $F_\lambda$(N30) &  $F_\lambda$(N38)\tnote{a} &  $F_\lambda$(N48) &  $F_\lambda$(N51)\\ \hline
2MASS-J  &  1.22  &  $\cdots$  &  $<1\times10^{-3}$  &  $\cdots$  &  $\cdots$  &  $<1\times10^{-3}$\\
2MASS-H  &  1.63  &  $<3\times10^{-3}$  &  $8\pm3\times10^{-3}$ &  $\cdots$  &  $\cdots$  & $<3\times10^{-3}$\\
2MASS-Ks  &  2.19  &  $4\pm2\times10^{-3}$ &  $4\pm2\times10^{-3}$ &  $\cdots$  &  $\cdots$  & $3\pm2\times10^{-3}$\\
IRAC-1	&  3.6	&  $1.243\pm.002\times10^{-2}$	& $6.417\pm0.004\times10^{-1}$ & $2.13\pm0.07\times10^{-3}$	& $2.9\pm0.5\times10^{-4}$	& $1.352\pm0.1\times10^{-1}$\\
IRAC-2	&  4.5	&  $3.344\pm0.002\times10^{-2}$	& $1.981\pm0.001$ & $1.34\pm0.02\times10^{-2}$ & $4.87\pm0.02\times10^{-3}$ & $6.692\pm0.004\times10^{-1}$\\
IRAC-3 &  5.8 &	$3.915\pm0.007\times10^{-2}$ &	$3.902\pm0.002$ & $1.97\pm0.03\times10^{-2}$  &	$1.14\pm0.03\times10^{-2}$	&  $1.617\pm0.001$ \\
IRAC-4	&  8.0	&  $2.911\pm0.006\times10^{-2}$ &  $5.361\pm0.003$	&  $1.22\pm0.03\times10^{-2}$  & $2.42\pm0.06\times10^{-2}$ &  $2.495\pm0.003$\\
MSX-A  &  8.28  &  $\cdots$  &  $8\pm4$  &  $\cdots$  &  $\cdots$  &  $1.77\pm0.07$\\
MSX-C  &  12.13  &  $\cdots$  &  $14\pm5$  &  $\cdots$  &  $\cdots$  &  $3.6\pm0.2$\\
MSX-D  &  14.65  &  $\cdots$  &  $19\pm6$  &  $\cdots$  &  $\cdots$  &  $6.8\pm0.4$\\
MSX-E  &  21.4  &  $\cdots$  &  $54\pm6$  &  $\cdots$  &  $\lesssim28$  &  $27\pm 2$\\
MIPS  &  24  &  $1.28\pm0.03$  &  $\cdots$  &  $\cdots$  &  $\cdots$  &  $\gtrsim17$\\
PACS-B  &  70  &  $130\pm20$  &  $4800\pm300$  &  $\cdots$  &  $280\pm130$  &  $180\pm50$\\
PACS-R  &  160  &  $230\pm10$  &  $4000\pm200$  &  $\cdots$  &  $370\pm110$  &  $330\pm30$\\
SPIRE-B  &  250  &  $163\pm9$  &  $2000\pm100$  &  $\cdots$  &  $150\pm80$  &  $240\pm20$\\
SPIRE-G  &  350  &  $99\pm8$  &  $660\pm60$  &  $\cdots$  &  $\cdots$  &  $130\pm20$\\
SCUBA-B  &  450  &  $80\pm30$  &  $360\pm60$  &  $230\pm110$  &  $100\pm50$  &  $140\pm70$\\
SPIRE-R  &  500  &  $53\pm4$  &  $320\pm20$  &  $\cdots$  &  $\cdots$  &  $60\pm30$\\
SCUBA-R  &  850  &  $9\pm2$  &  $60\pm20$  &  $24\pm6$  &  $15\pm5$  &  $10\pm5$\\ \hline\\
 \hline
Filter & $\lambda$ ($\mu$m) &  $F_\lambda$(N53) &  $F_\lambda$(N54)\tnote{b} &  $F_\lambda$(N63) &  $F_\lambda$(S8)\tnote{c} &  $F_\lambda$(S26)\tnote{c}\\ \hline
2MASS-J  &  1.22 & $\cdots$  &  $2\pm1\times10^{-3}$  &  $\cdots$  &  $2.0\pm0.1\times10^{-2}$  &  $5\pm1\times10^{-3}$\\
2MASS-H  &  1.63  &  $\cdots$  &  $2\pm2\times10^{-3}$  &  $\cdots$  &  $9.0\pm0.2\times10^{-2}$  &  $7.1\pm0.2\times10^{-2}$\\
2MASS-Ks  &  2.19  &  $\cdots$  &  $4\pm2\times10^{-3}$  &  $\cdots$  &  $2.58\pm0.06\times10^{-1}$  &  $1.29\pm0.03$\\
IRAC-1  &  3.6  & $8.56\pm0.05\times10^{-3}$ &	$1.596\pm0.004\times10^{-2}$ &	$1.77\pm0.01\times10^{-3}$ &  $8.9\pm0.2\times10^{-1}$ & $73.0\pm0.6$\\
IRAC-2  &  4.5  &  $2.97\pm0.01\times10^{-2}$ &	$6.913\pm0.006\times10^{-2}$ &	$9.32\pm0.02\times10^{-3}$ & $1.2\pm0.1$ & $176\pm4$\\
IRAC-3  &  5.8  &  $2.66\pm0.03\times10^{-2}$ &	$1.992\pm0.004\times10^{-1}$ &	$1.033\pm0.006\times10^{-2}$ &  $3.2\pm0.2$&  $324\pm2$\\
IRAC-4  &  8.0 &  $4\pm2\times10^{-3}$ & $3.18\pm0.02\times10^{-1}$ & $5.76\pm0.07\times10^{-3}$ &  $5.4\pm0.4$  &  $\gtrsim190$  \\
MSX-A  &  8.28  &  $\cdots$  &  $2.8\pm0.1$  &  $\cdots$  &  $4.4\pm0.2$  &  $310\pm30$\\
MSX-C  &  12.13  &  $\cdots$  &  $3.3\pm0.2$  &  $\cdots$  &  $14.6\pm0.7$  &  $570\pm30$\\
MSX-D  &  14.65  &  $\cdots$  &  $3.4\pm0.2$  &  $\cdots$  &  $31\pm2$  &  $800\pm50$\\
MSX-E  &  21.4  &  $\cdots$  &  $11\pm1$  &  $\cdots$  &  $145\pm9$  &  $1020\pm60$\\
MIPS  &  24  &  $0.52\pm0.04$  &  $\cdots$  &  $0.29\pm0.01$  &  $\lesssim241$  &  $\cdots$\\
IRAS-25  &  25 &  $\cdots$  &  $\cdots$  &  $\cdots$  &  $\lesssim240$  &  $\lesssim1100$\\
IRAS-60  &  60  &  $\cdots$  &  $\cdots$  &  $\cdots$  &  $\lesssim1050$  &  $\lesssim5300$\\
PACS-B  &  70  &  $50\pm50$  &  $\cdots$  &  $57\pm6$  &  $1140\pm60$  &  $3600\pm200$\\
PACS-R  &  160  &  $140\pm30$  &  $\cdots$  & $139\pm8$  &  $1040\pm50$  &  $1910\pm90$\\
SPIRE-B  &  250  &  $140\pm20$  &  $\cdots$  &  $106\pm5$  &  $390\pm20$  &  $580\pm30$\\
SPIRE-G  &  350  &  $90\pm20$  &  $\cdots$  &  $66\pm4$  &  $220\pm20$  &  $270\pm10$\\
SCUBA-B  &  450 &  $110\pm50$  &  $30\pm40$  &  $45\pm5$  &  $160\pm50$  &  $220\pm110$\\
SPIRE-R  &  500 &  $60\pm10$  &  $\cdots$  &  $37\pm3$  &  $117\pm7$  &  $107\pm7$\\
SCUBA-R  &  850  &  $12\pm6$  &  $12\pm6$  &  $8\pm3$  &  $30\pm10$  &  $24\pm5$\\ \hline
    \end{tabular*}
    \begin{tablenotes}
        \item[a]In most filters, N38 could not be identified because the centroid of the nearest source was almost as far away as N48. The few measurements given should be treated with extreme caution.
        \item[b]In most filters, N54 could not be deblended from another object, N52, which was included in the \citetalias{motte_earliest_2007} survey but not later works.
        \item[c]Note that upper limits from IRAS are included for these two sources in Cygnus-X South.
      \end{tablenotes}
    \end{threeparttable}
\end{table*}

\twocolumn
\section{SEDs fits}\label{app:figs1}
\begin{figure}[hbt]
    \centering
    \includegraphics[width=0.49\textwidth]{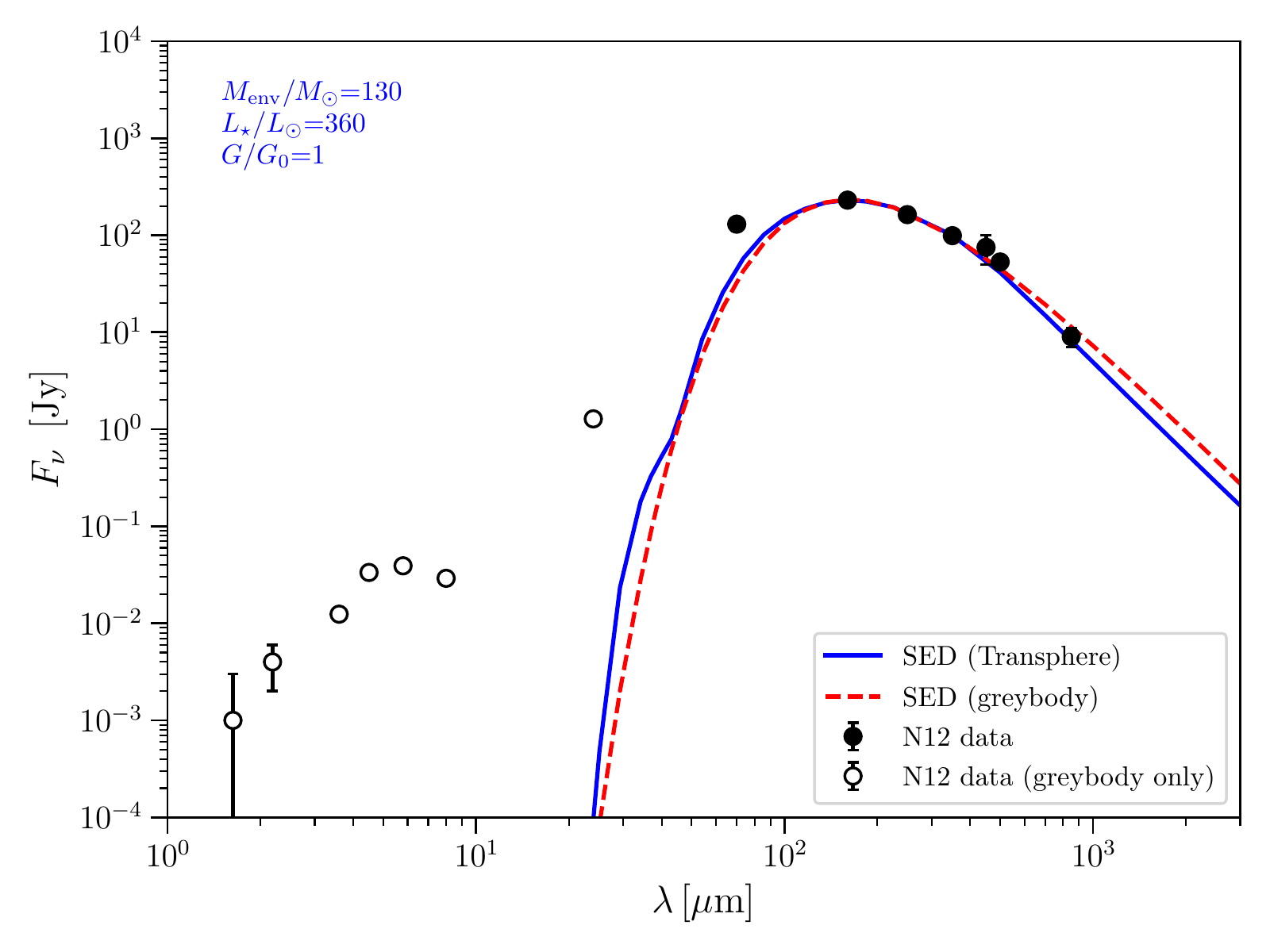}
    \caption{Same as Fig.~\ref{fig:n30sed}, but for N12. It is uncertain whether the PACS-70~$\mu$m data are part of the same temperature component as the longer-wavelength data; the fits both with and without it are within the uncertainties, but not including the 70~\micron\ data yields a more realistic $\beta$ value.}
    \label{fig:n12sed}
\end{figure}

\begin{figure}[hbt]
    \centering
    \includegraphics[width=0.49\textwidth]{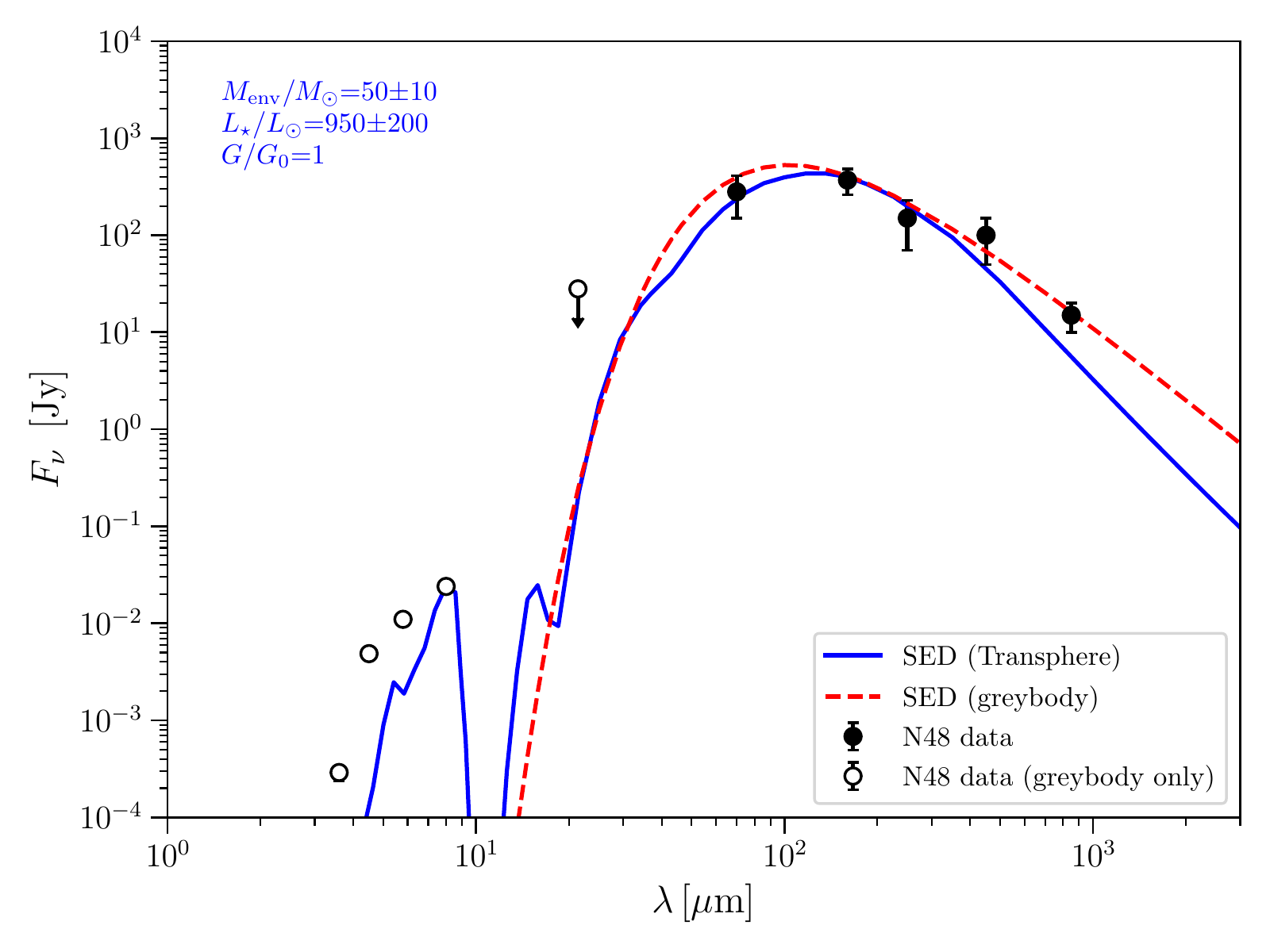}
    \caption{Same as Fig.~\ref{fig:n12sed}, but for N48. N48 is located in an especially dense part of the DR21 ridge and is blended with two other clumps identified in \protect{\citetalias{motte_earliest_2007}} at $\lambda>250\;\mu$m. We advise caution in using the derived SED parameters for this object in any precision analysis.}
    \label{fig:n48sed}
\end{figure}

\begin{figure}[hbt]
    \centering
    \includegraphics[width=0.49\textwidth]{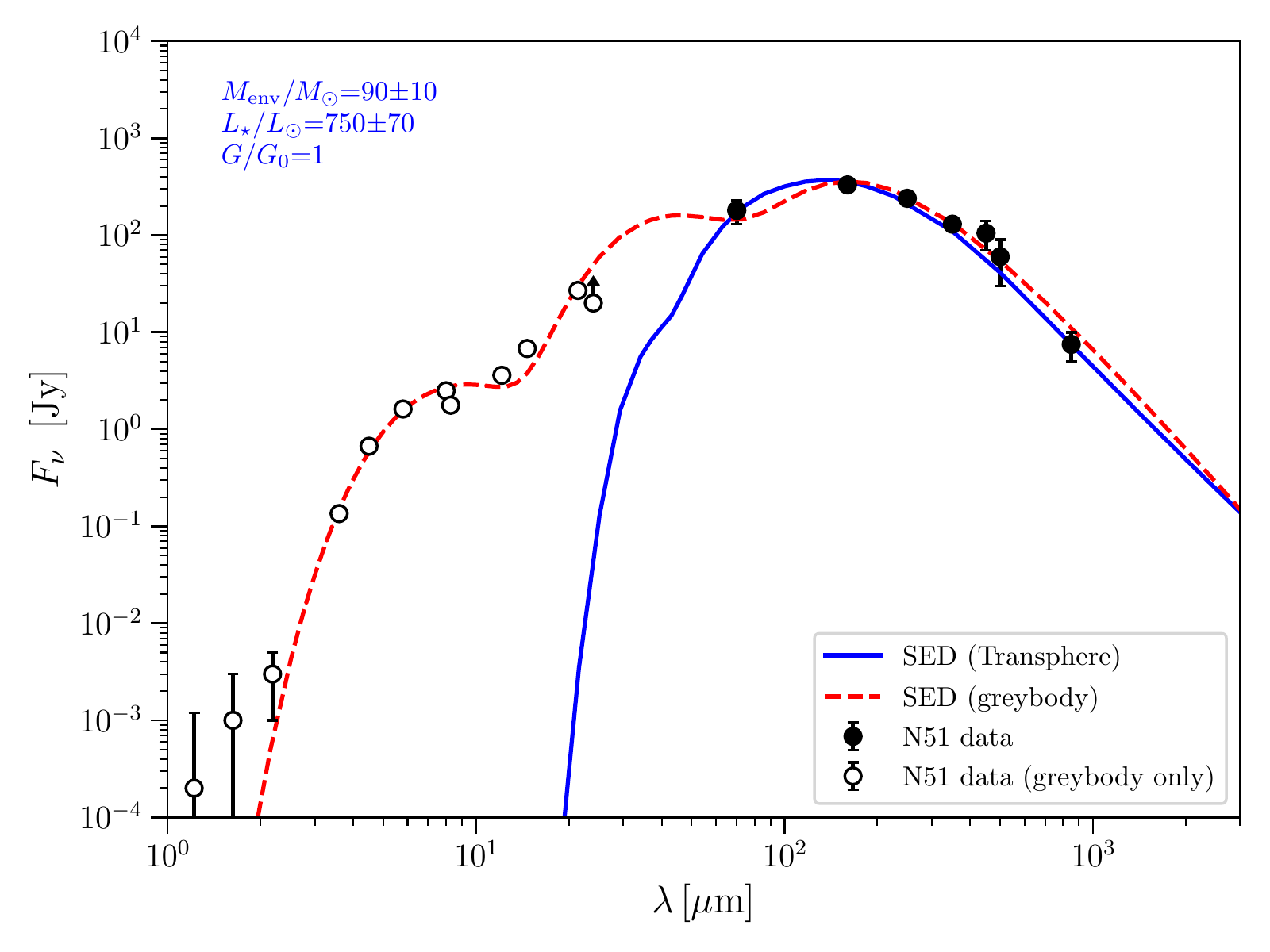}
    \caption{Same as Fig.~\ref{fig:n12sed}, but for N51.}
    \label{fig:n51sed}
\end{figure}

\begin{figure}[hbt]
    \centering
    \includegraphics[width=0.49\textwidth]{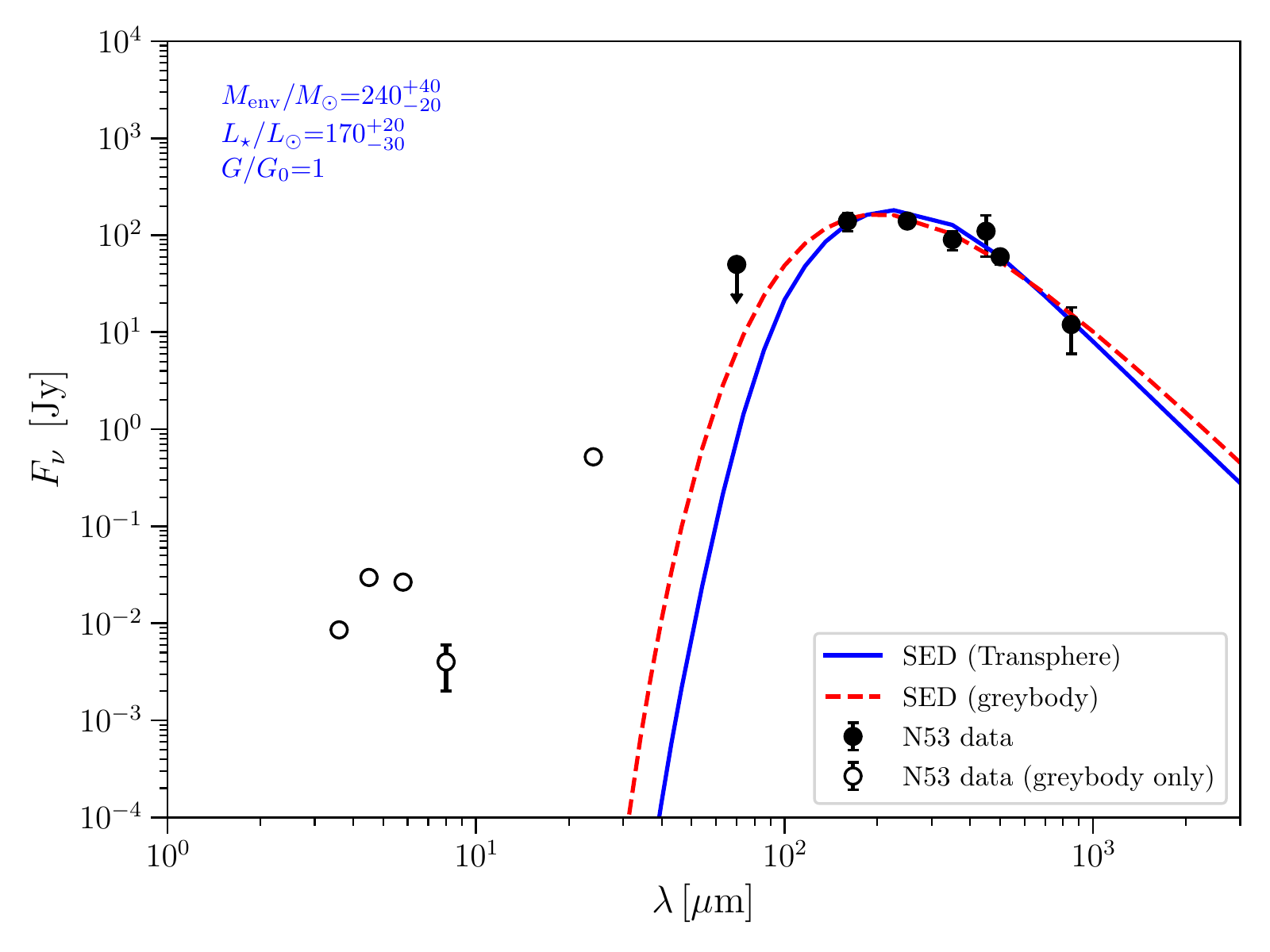}
    \caption{Same as Fig.~\ref{fig:n12sed}, but for N53.}
    \label{fig:n53sed}
\end{figure}

\begin{figure}[hbt]
    \centering
    \includegraphics[width=0.49\textwidth]{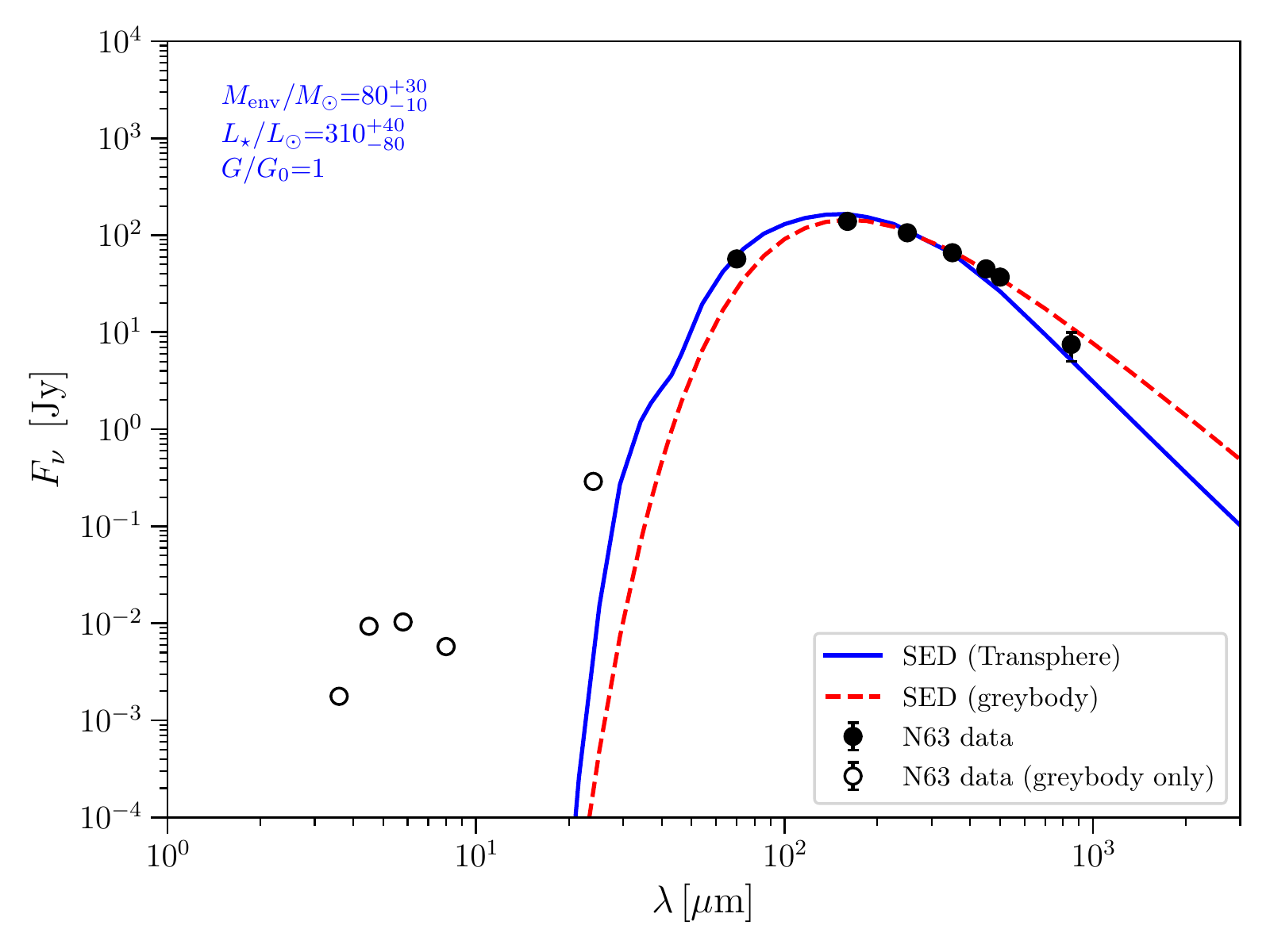}
    \caption{Same as Fig.~\ref{fig:n12sed}, but for N63.}
    \label{fig:n63sed}
\end{figure}

\begin{figure}[hbt]
    \centering
    \includegraphics[width=0.49\textwidth]{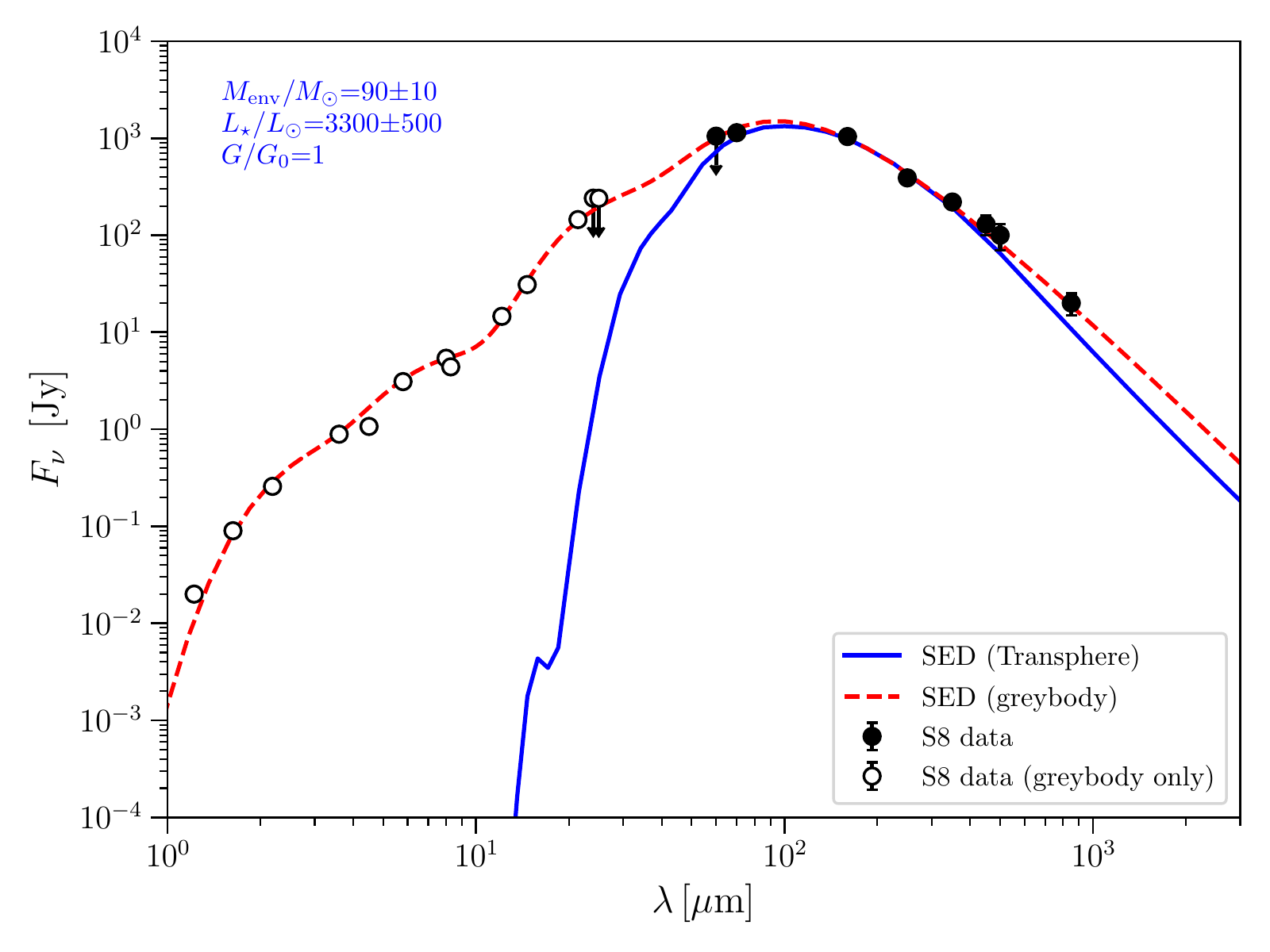}
    \caption{Same as Fig.~\ref{fig:n12sed}, but for S8. The distance to S8 is highly uncertain and it has a spiral arm feature that could only be accounted for to first order, so SED fitting parameters should be used with caution.}
    \label{fig:s8sed}
\end{figure}

\begin{figure}[hbt]
    \centering
    \includegraphics[width=0.49\textwidth]{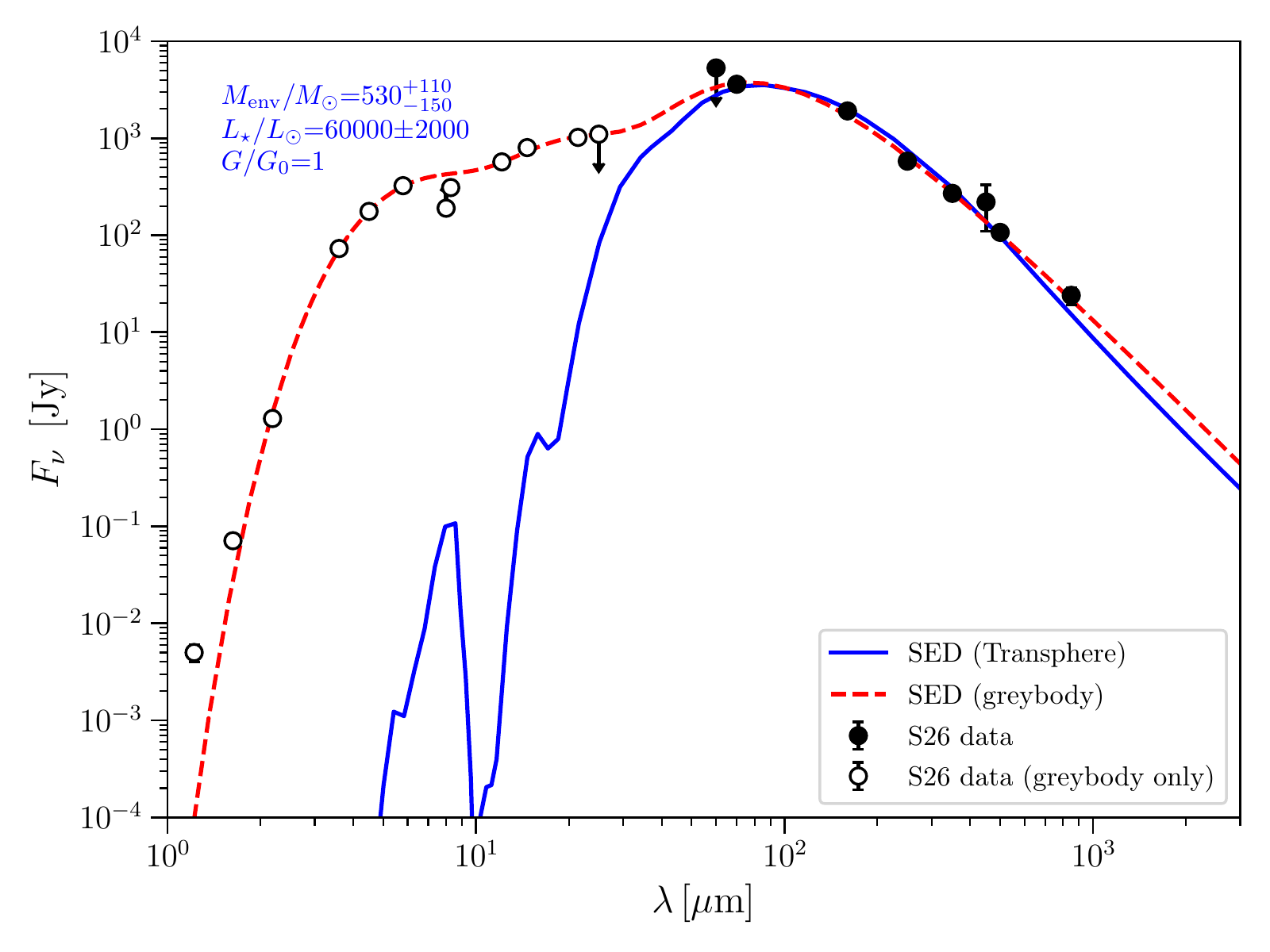}
    \caption{Same as Fig.~\ref{fig:n12sed}, but for S26. We note the change in y-axis scaling.}
    \label{fig:s26sed}
\end{figure}

Figures~\ref{fig:n12sed}-\ref{fig:s26sed} show the data (black filled and hollow circles with error bars), greybody SED fits (dashed red lines), and \tP\ SED fits (solid blue lines) for all other objects in the PILS-Cygnus survey that could be modelled, besides N30. The parameters, processes, and caveats involved in each fitting routine are discussed in \S\ref{sec:meth}. Hollow circles indicate data that were excluded from the fit with \tP, but were fit if possible with our initial greybody fitting routine.

\section{Density power-law index fits, and temperature and density profiles}\label{app:figs2}
\begin{figure*}
    \centering
    \subfloat[A.]{\includegraphics[width=0.49\textwidth]{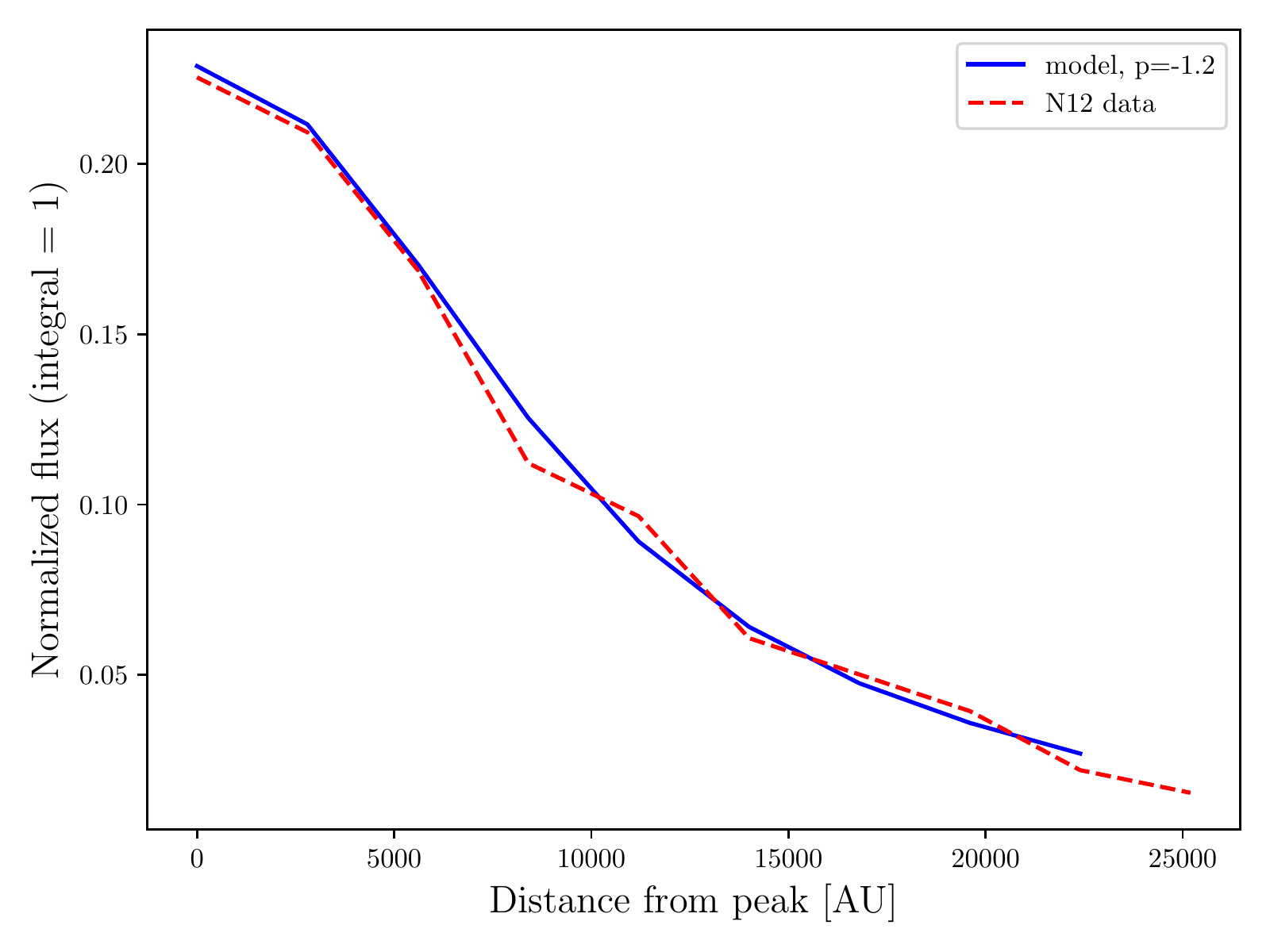}}\quad
    \subfloat[B.]{\includegraphics[width=0.49\textwidth]{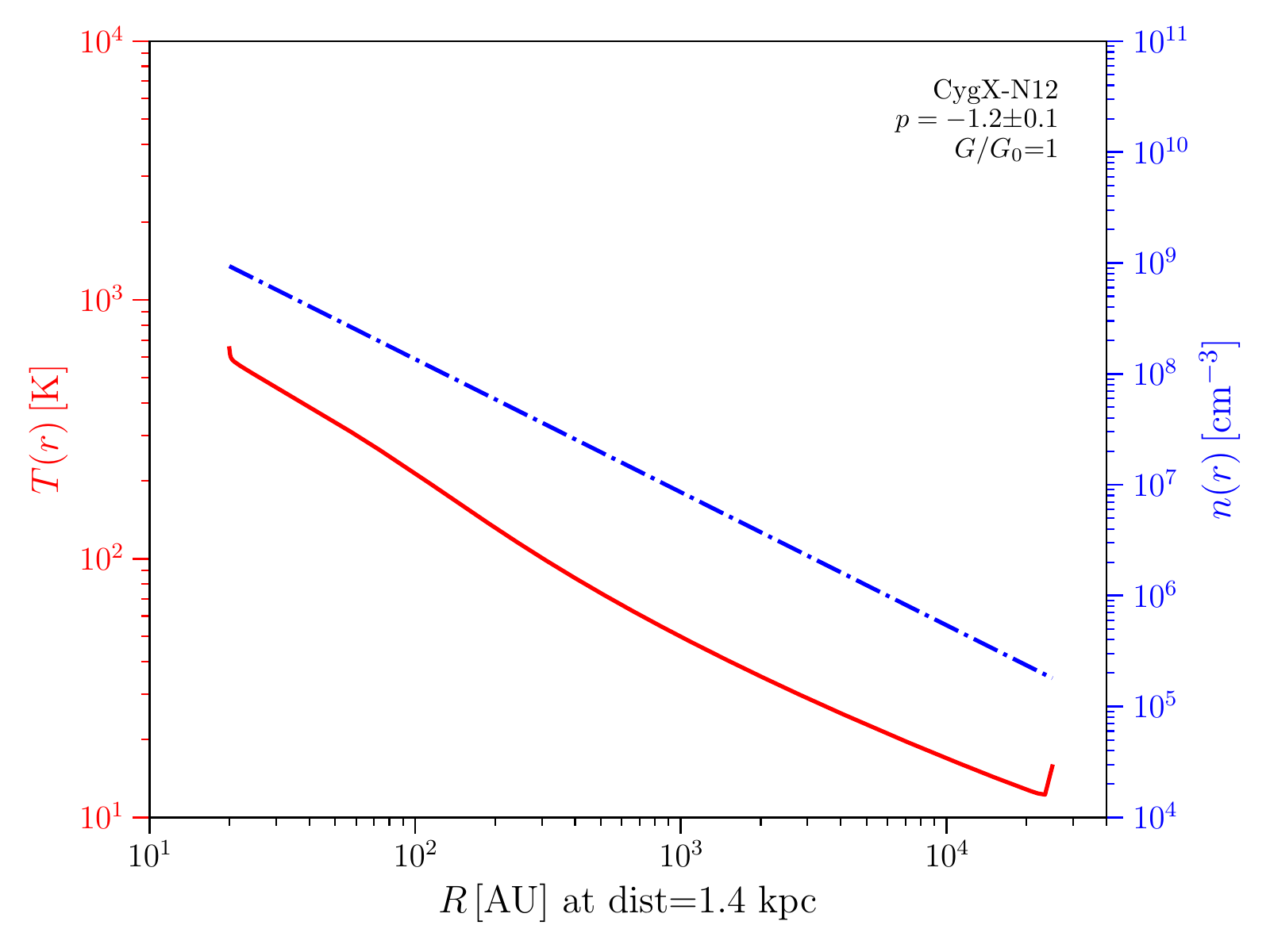}}  
    \caption{Same as Fig.~\ref{fig:n30profs}, but for N12 (using the fit that excludes the 70~\micron\ data).}
    \label{fig:n12profs}
\end{figure*}

\begin{figure*}
    \centering
    \subfloat[A.]{\includegraphics[width=0.49\textwidth]{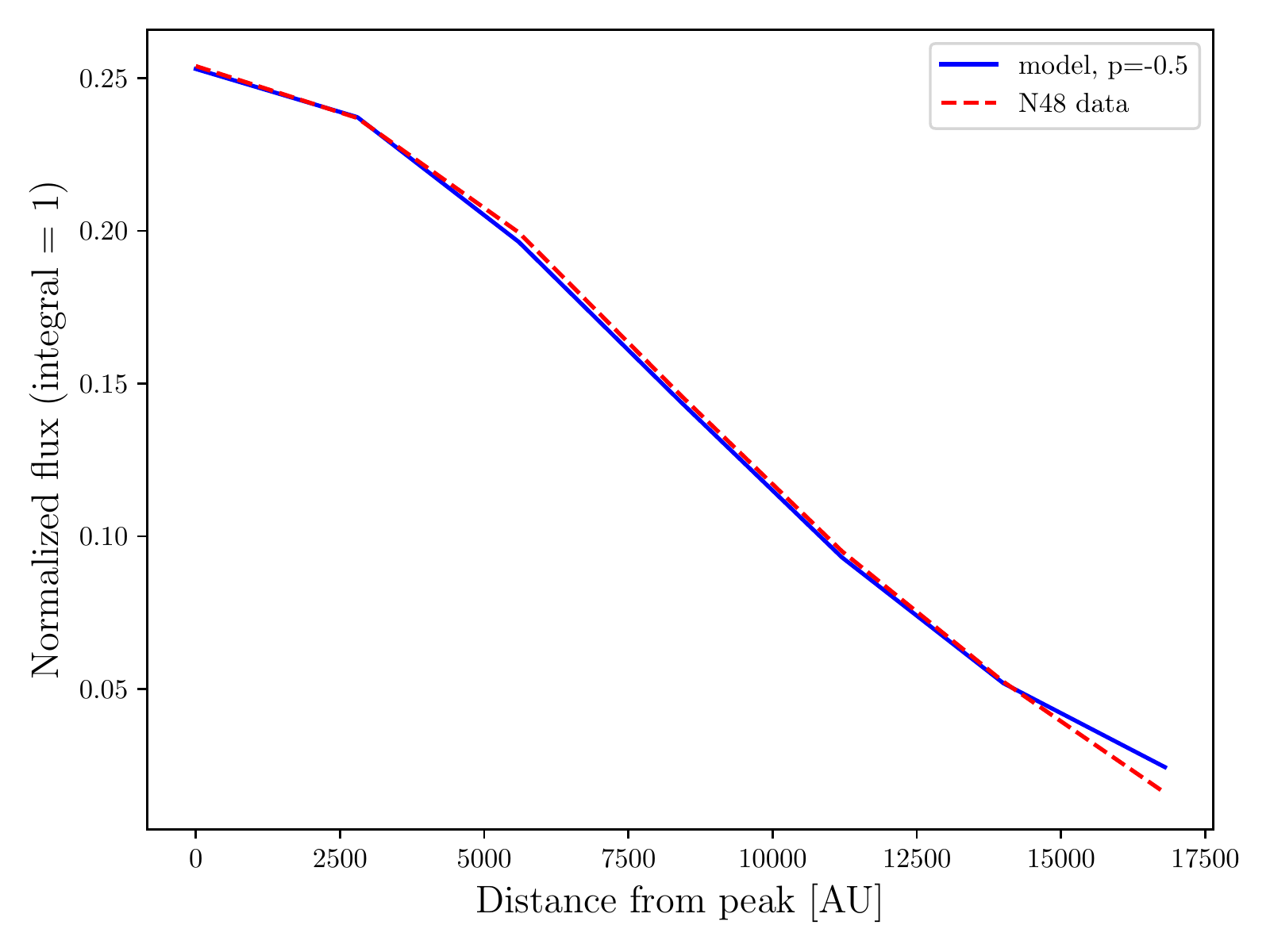}}\quad
    \subfloat[B.]{\includegraphics[width=0.49\textwidth]{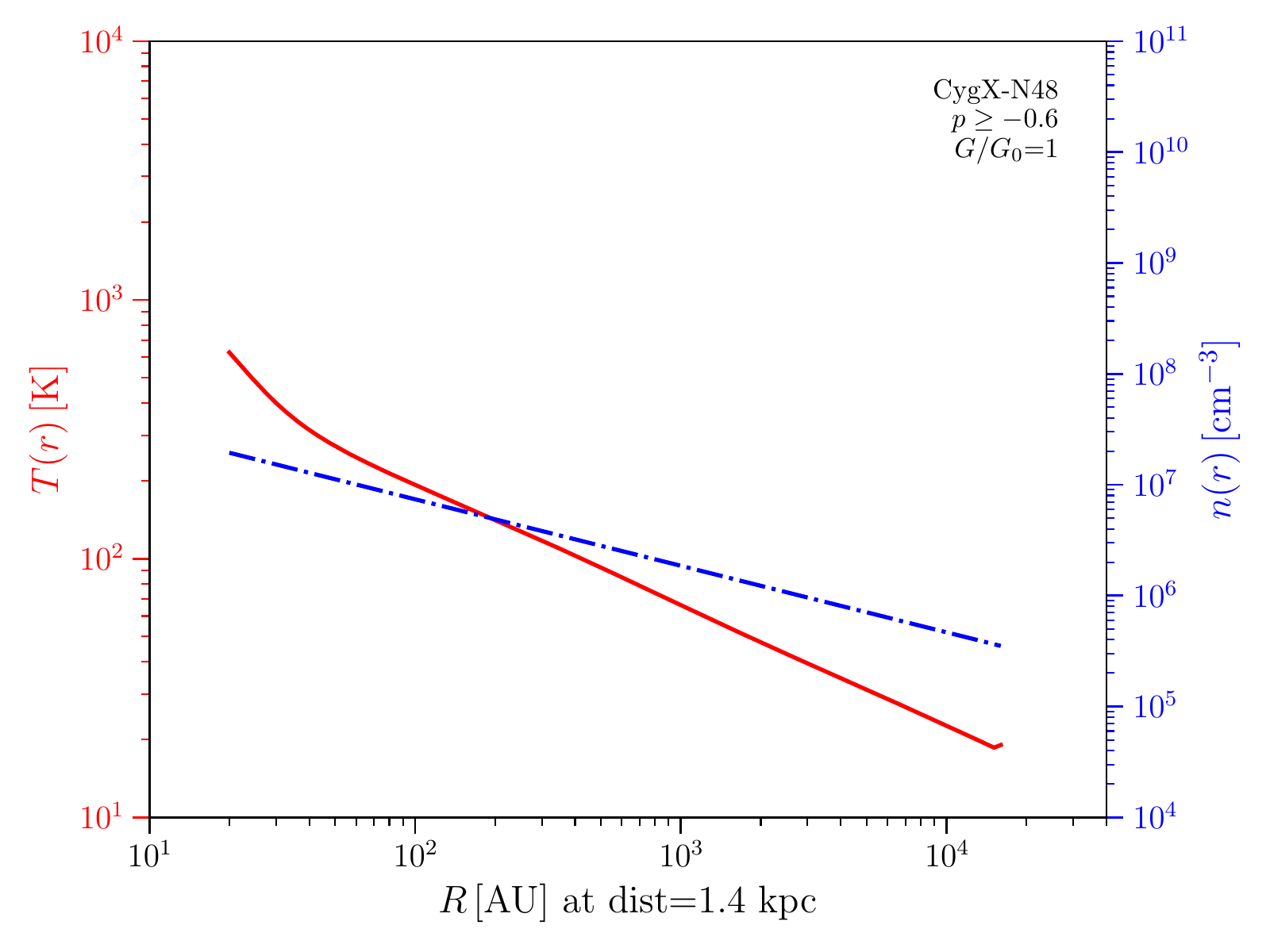}}    
    \caption{Same as Fig.~\ref{fig:n30profs}, but for N48.}
    \label{fig:n48profs}
\end{figure*}

\begin{figure*}
    \centering
    \subfloat[A.]{\includegraphics[width=0.49\textwidth]{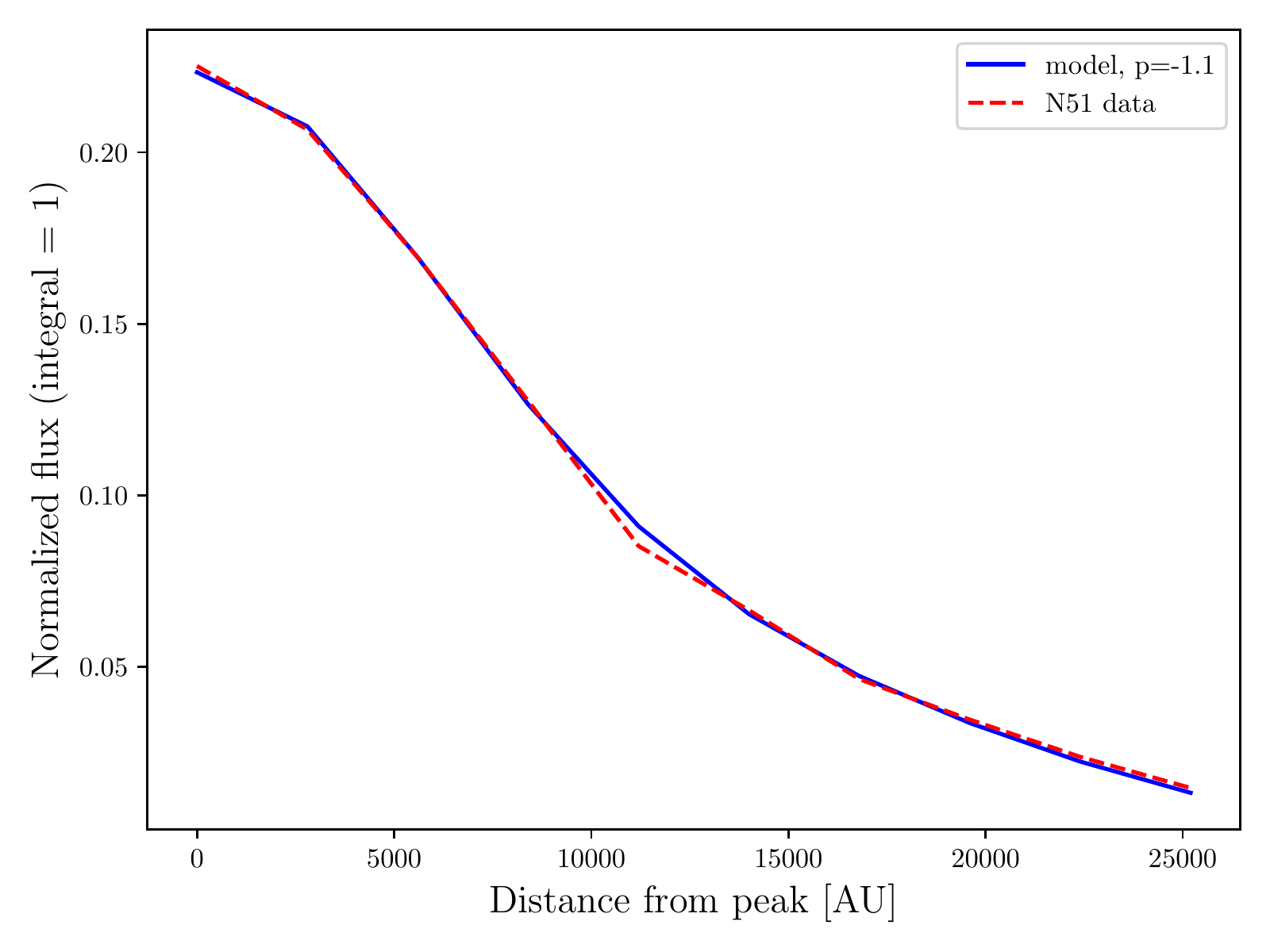}}\quad
    \subfloat[B.]{\includegraphics[width=0.49\textwidth]{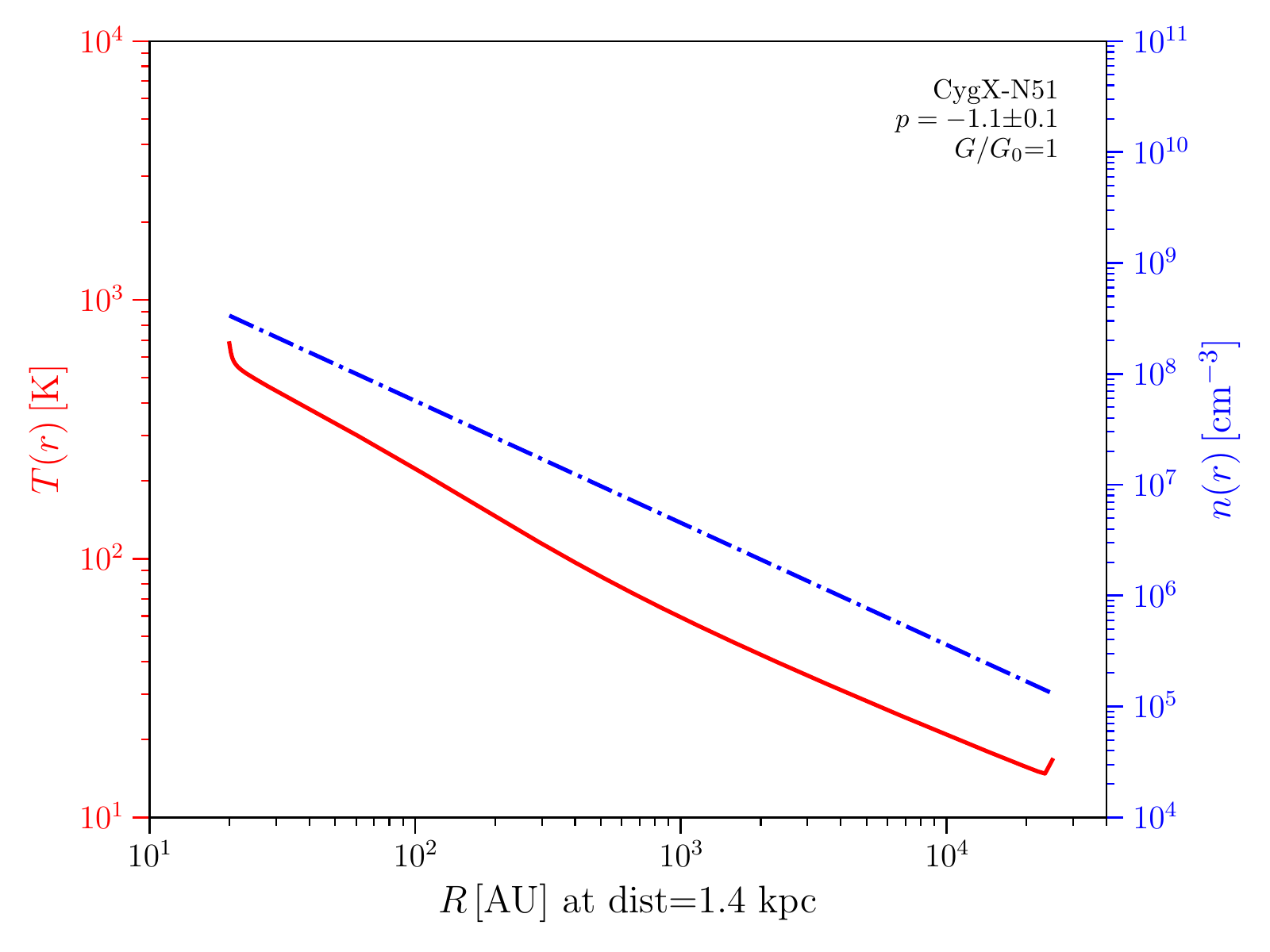}}    
    \caption{Same as Fig.~\ref{fig:n30profs}, but for N51.}
    \label{fig:n51profs}
\end{figure*}

\begin{figure*}
    \centering
    \subfloat[A.]{\includegraphics[width=0.49\textwidth]{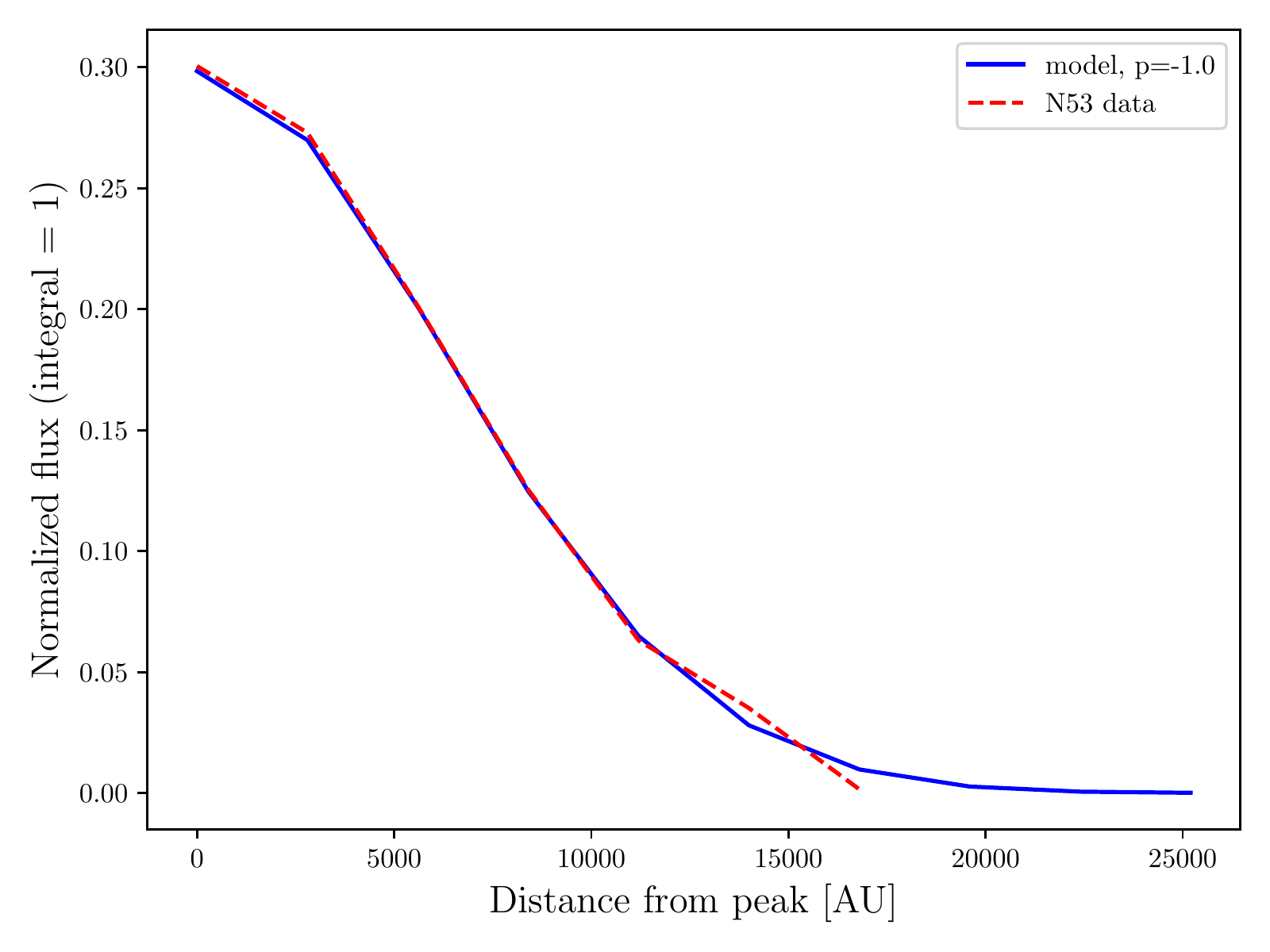}}\quad
    \subfloat[B.]{\includegraphics[width=0.49\textwidth]{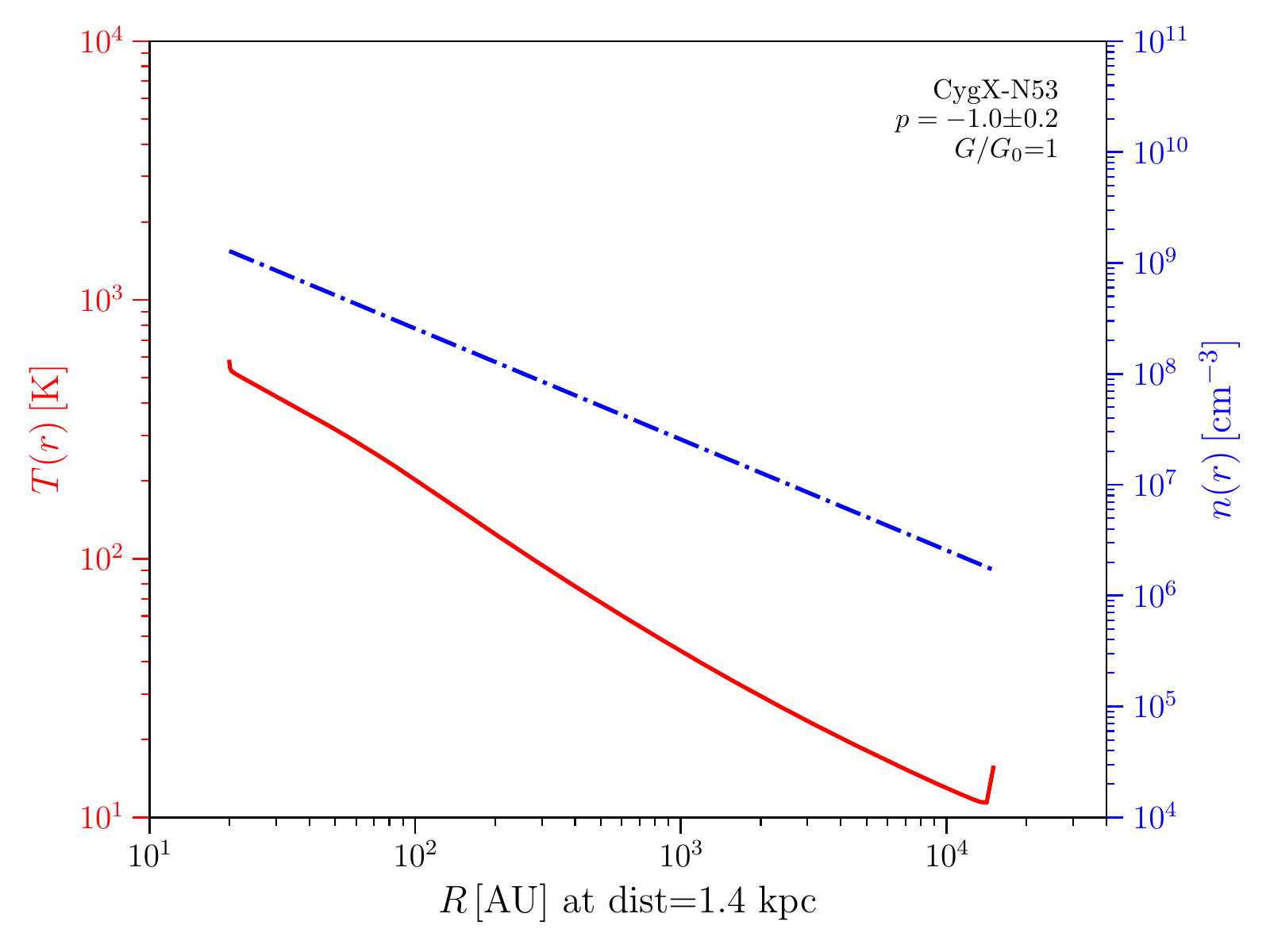}}    
    \caption{Same as Fig.~\ref{fig:n30profs}, but for N53.}
    \label{fig:n53profs}
\end{figure*}

\begin{figure*}
    \centering
    \subfloat[A.]{\includegraphics[width=0.49\textwidth]{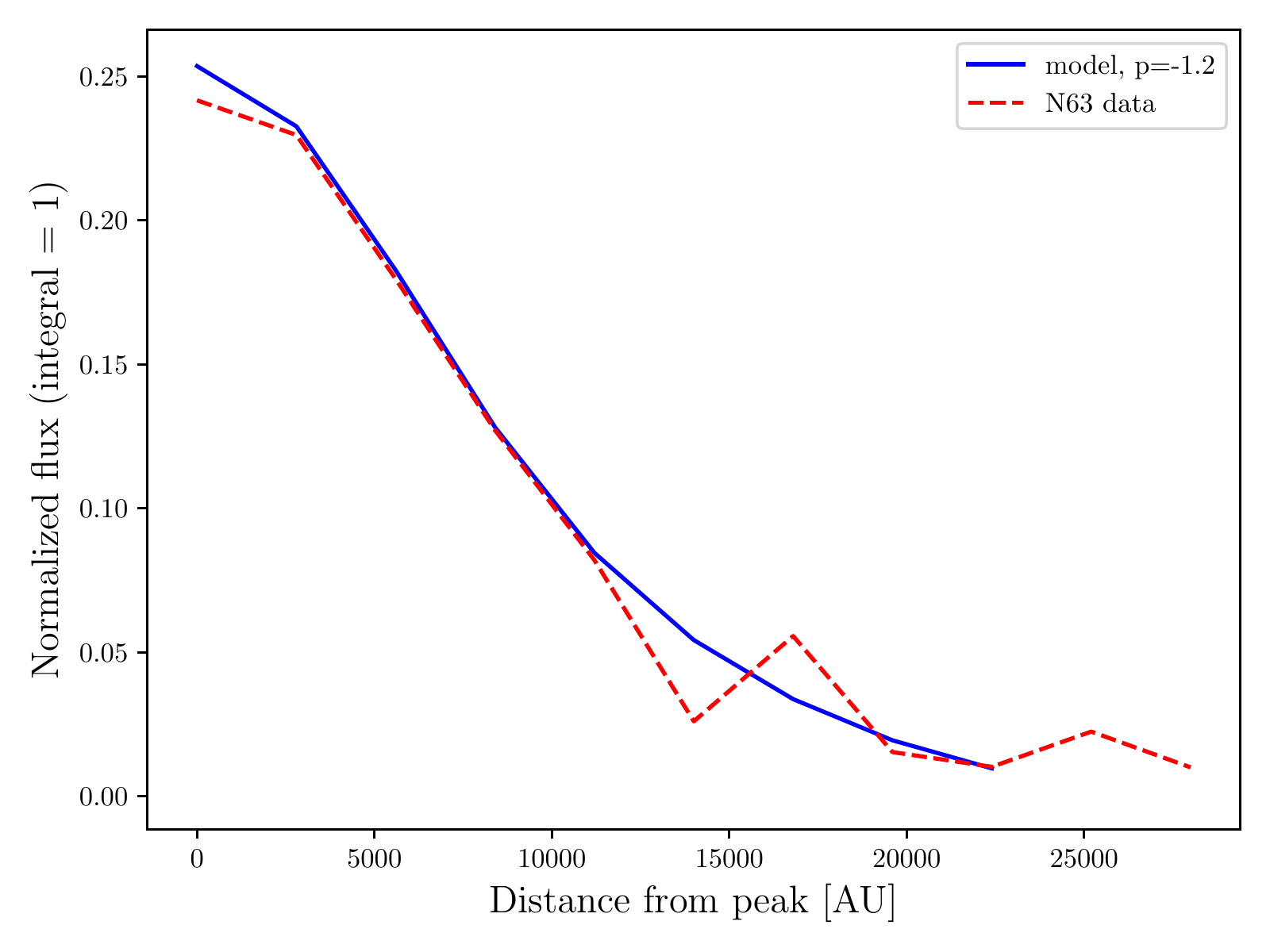}}\quad
    \subfloat[B.]{\includegraphics[width=0.49\textwidth]{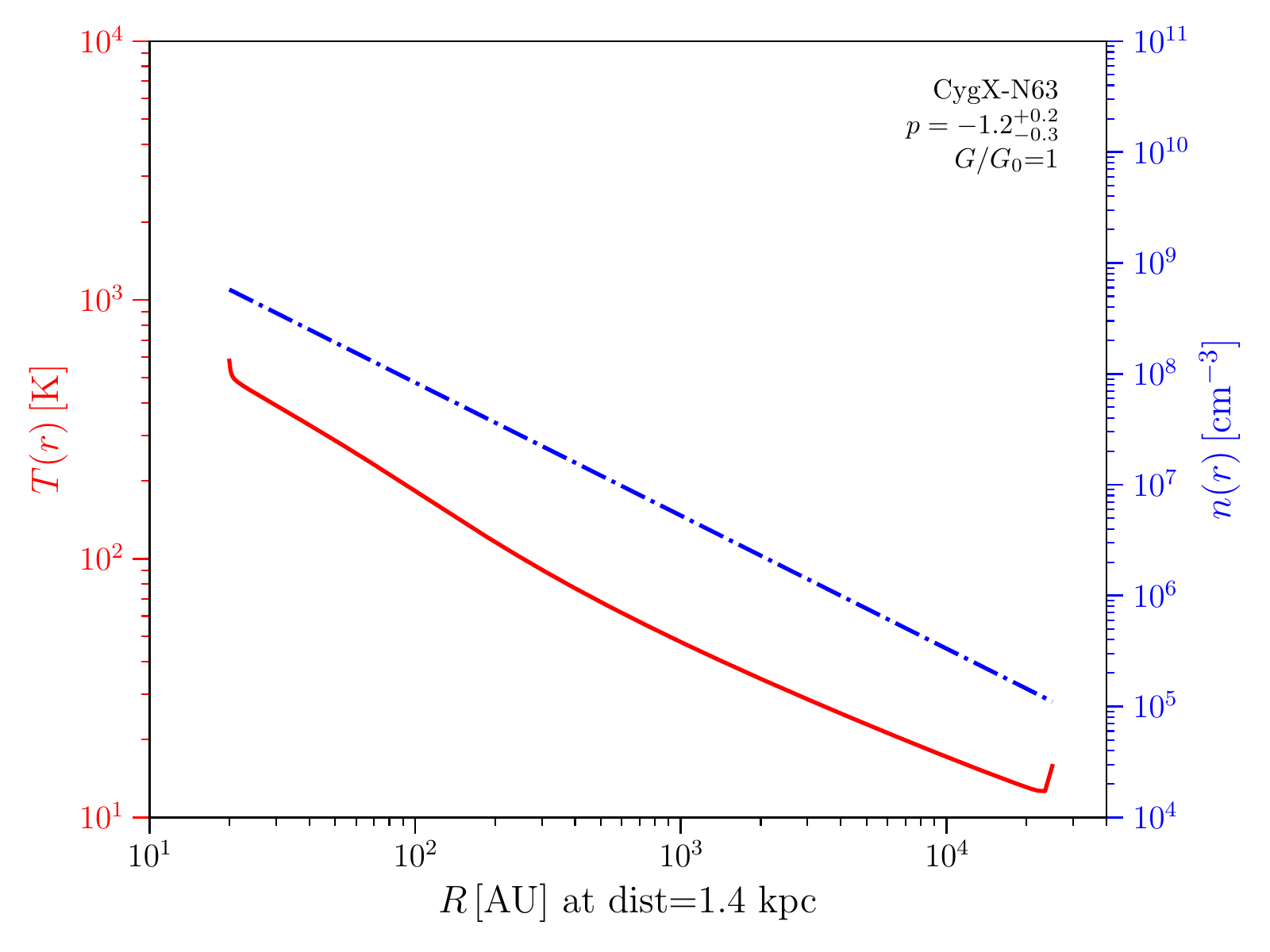}}    
    \caption{Same as Fig.~\ref{fig:n30profs}, but for N63.}
    \label{fig:n63profs}
\end{figure*}

\begin{figure*}
    \centering
    \subfloat[A.]{\includegraphics[width=0.49\textwidth]{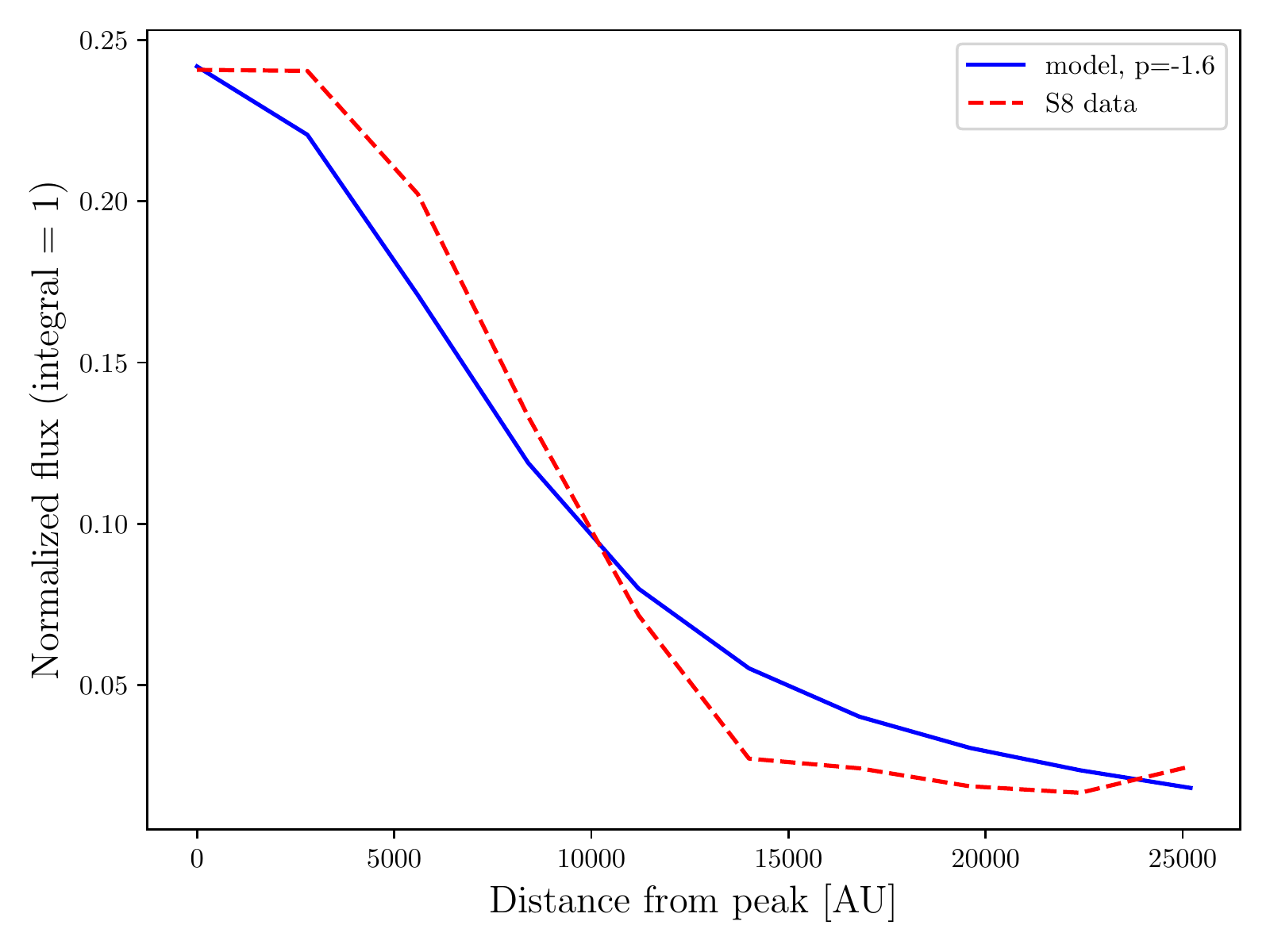}}\quad
    \subfloat[B.]{\includegraphics[width=0.49\textwidth]{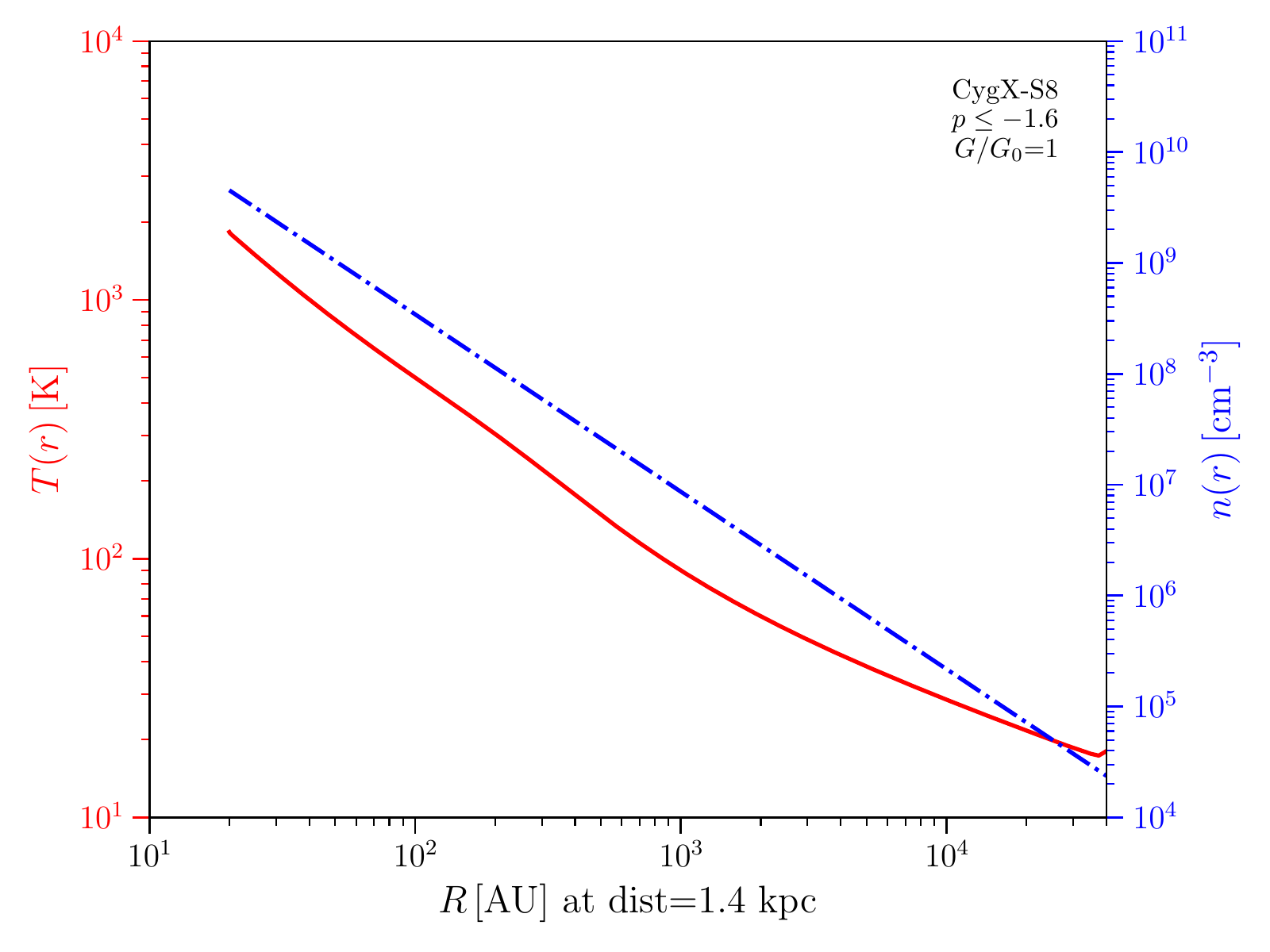}}    
    \caption{Same as Fig.~\ref{fig:n30profs}, but for S8.}
    \label{fig:s8profs}
\end{figure*}

\begin{figure*}
    \centering
    \subfloat[A.]{\includegraphics[width=0.49\textwidth]{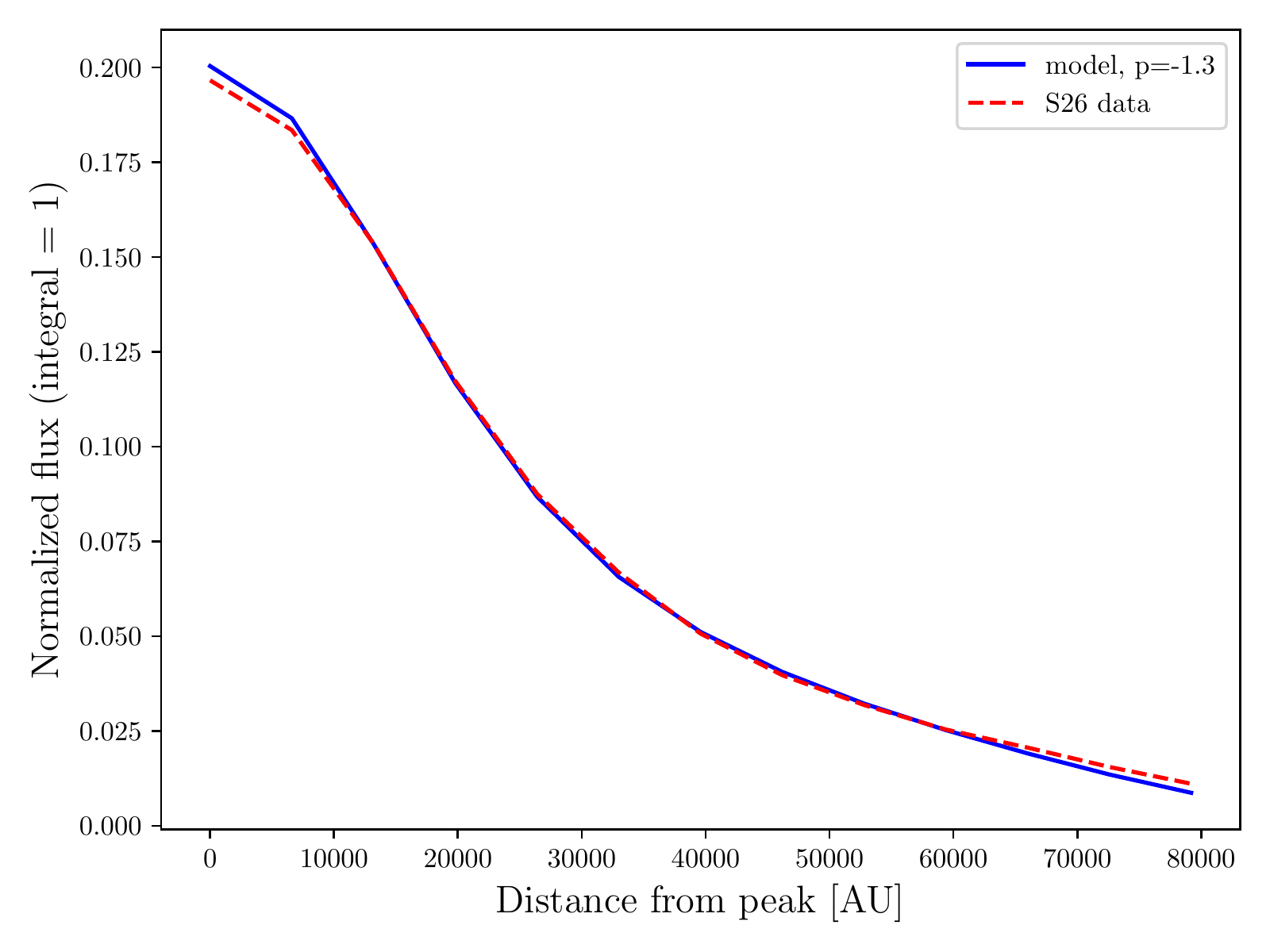}}\quad
    \subfloat[B.]{\includegraphics[width=0.49\textwidth]{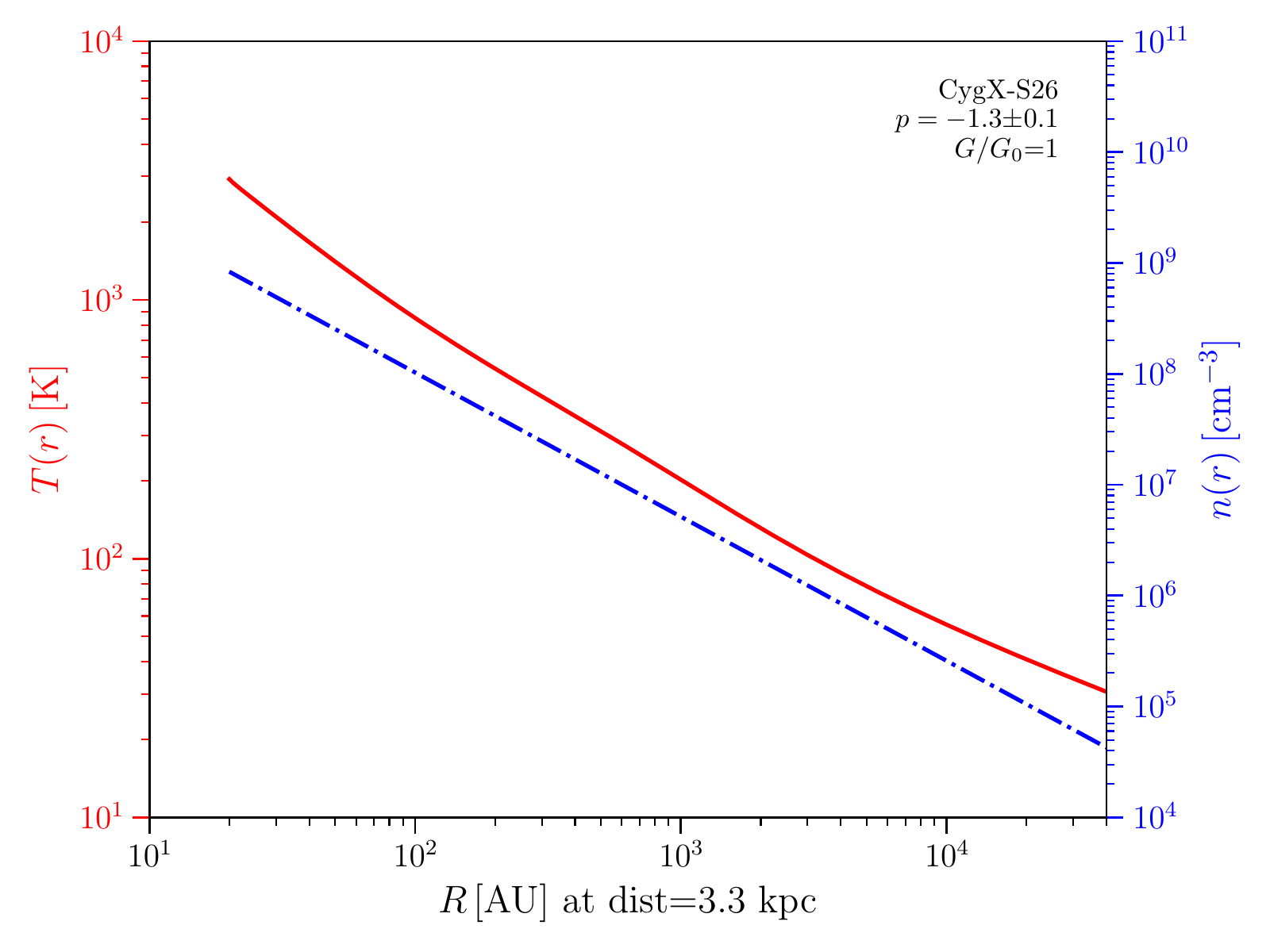}}    
    \caption{Same as Fig.~\ref{fig:n30profs}, but for S26. We note the change of x-axis scaling.}
    \label{fig:s26profs}
\end{figure*}

Figures~\ref{fig:n12profs}-\ref{fig:s26profs} are the equivalent of Fig.~\ref{fig:n30profs} for all other sources besides N30.

\section{Optimisation and posterior distributions of power-law fits to envelope parameters}\label{app:mcmc}
In Fig.~\ref{fig:crim}, we fitted power-law trends in \menv, \rout, $|p|$, and \nkau\ with \lbol\ to see if trends for high- and low-mass sources were statistically distinguishable from a single trend for all sources. The details of the process are discussed in \S\ref{sssec:mcmc}. The following figures show the posterior distributions of the parameters of each of the pair of fits to each relationship, except for the fit to $|p|$ as a function of \lbol. As we discussed in the main text, we did not find a significant correlation between $p$ and \lbol.

\begin{figure*}
    \centering
    \subfloat[Single Power Law]{\includegraphics[height=0.36\textheight]{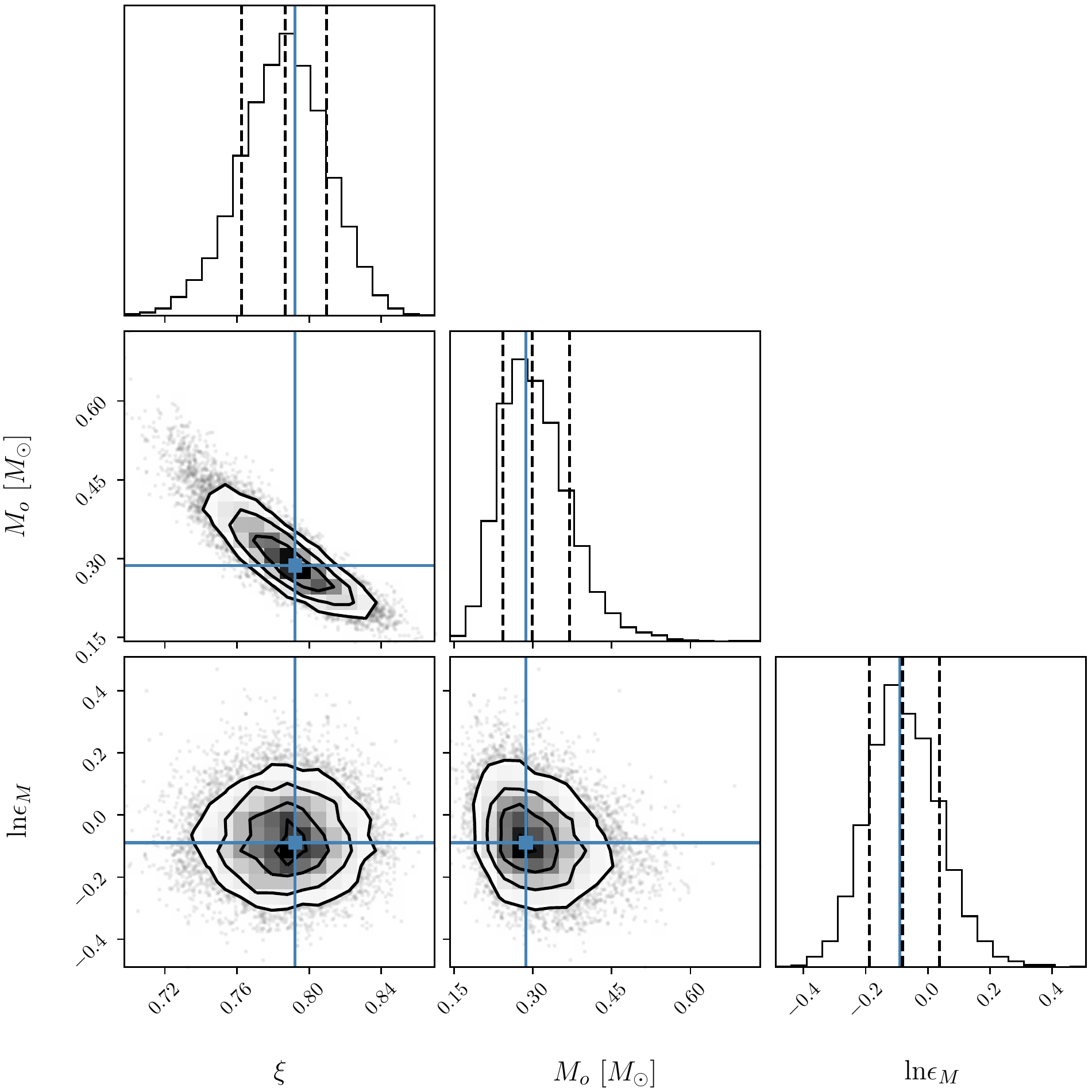}}
    
    \subfloat[Broken Power Law]{\includegraphics[height=0.48\textheight]{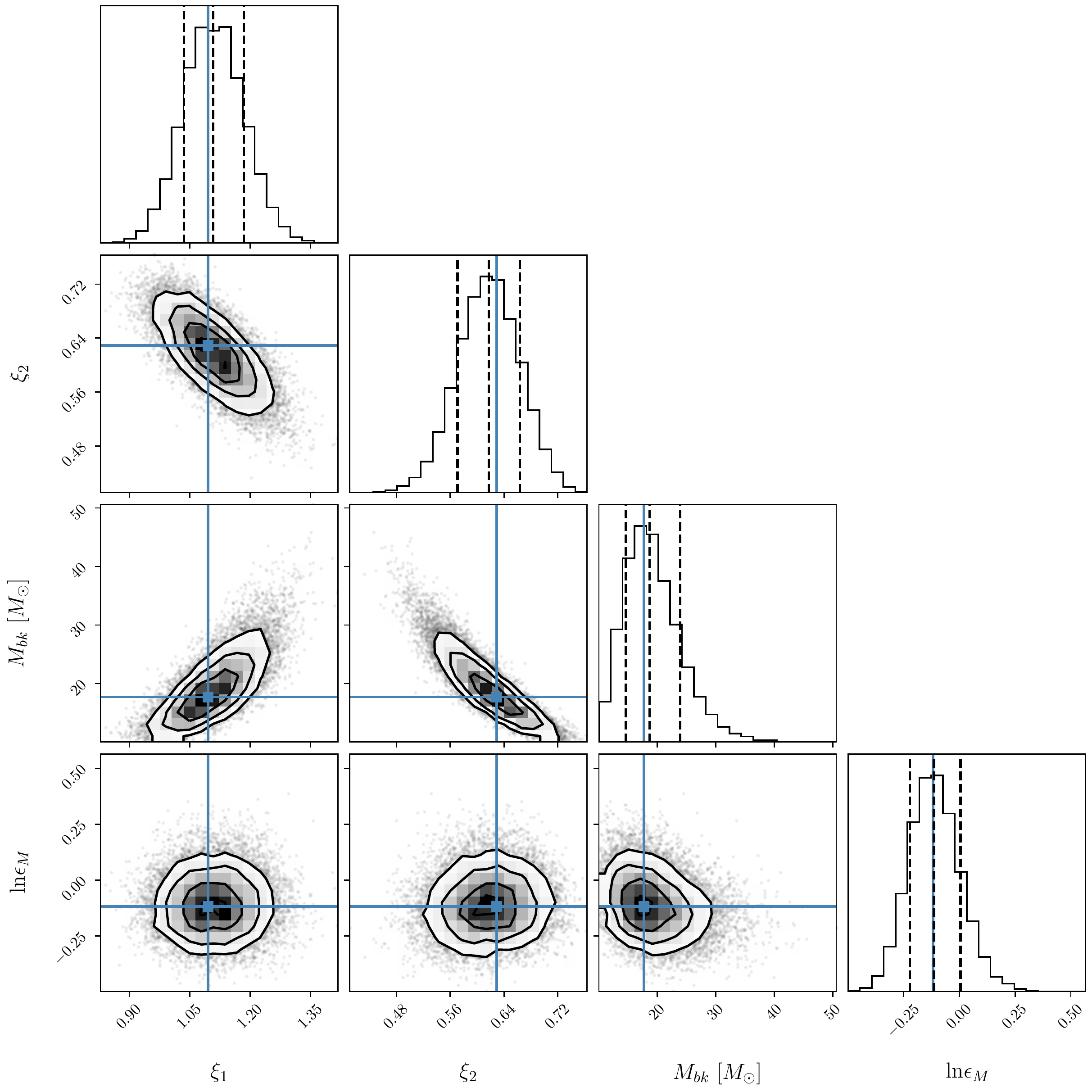}}
    \caption{\textit{(a)} Posterior distributions of the power-law exponent ($\xi$), normalisation ($M_o$), and residual variance parameter (ln$\epsilon_M$) for the single power-law fit to \menv\ vs. \lbol, shown as the solid black line in the top left panel of Fig.~\ref{fig:crim}. \textit{(b)} Posterior distributions of the power-law exponents ($\xi_1$ and $\xi_2$), normalisation ($M_{bk}$), and residual variance parameter (ln$\epsilon_M$) for the broken power-law fit to \menv\ vs. \lbol, shown as the dashed black line in the top left panel of Fig.~\ref{fig:crim}.}
    \label{fig:postm}
\end{figure*}

\begin{figure*}
    \centering
    \subfloat[Single Power Law]{\includegraphics[height=0.36\textheight]{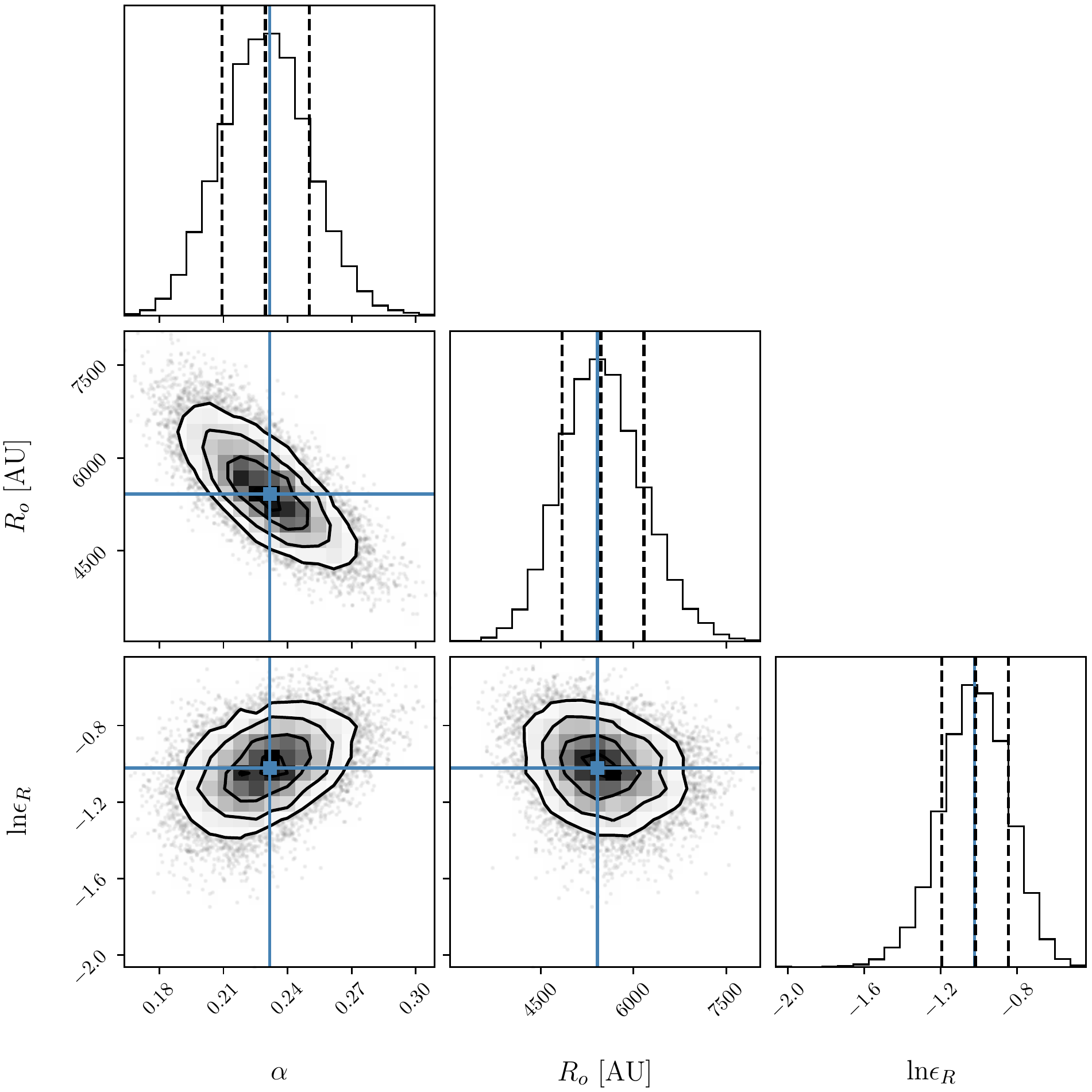}}
    
    \subfloat[Broken Power Law]{\includegraphics[height=0.48\textheight]{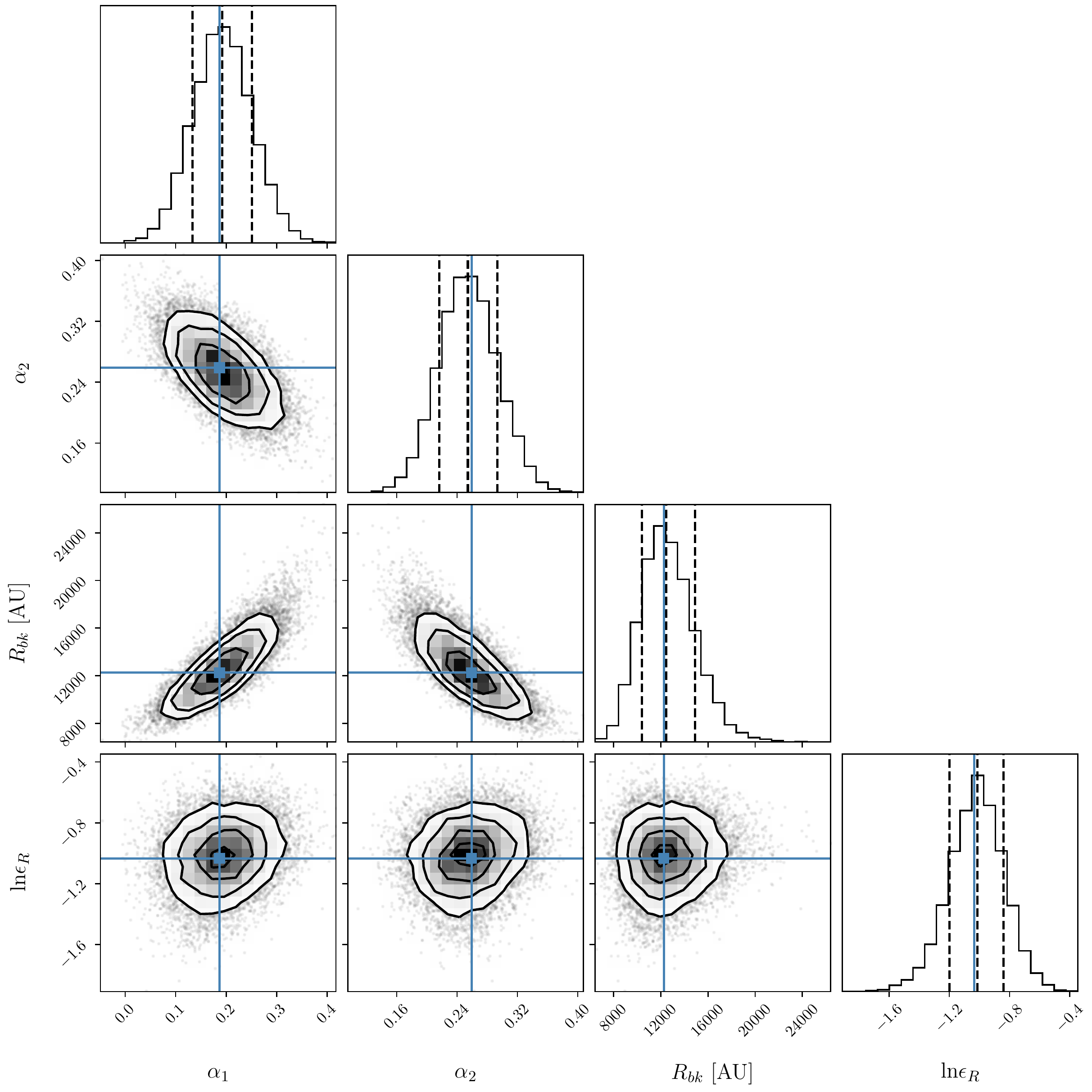}}
    \caption{\textit{(a)} Posterior distributions of the power-law exponent ($\alpha$), normalisation ($R_o$), and residual variance parameter (ln$\epsilon_R$) for the single power-law fit to \rout\ vs. \lbol, shown as the solid black line in the top right panel of Fig.~\ref{fig:crim}. \textit{(b)} Posterior distributions of the power-law exponents ($\alpha_1$ and $\alpha_2$), normalisation ($R_{bk}$), and residual variance parameter (ln$\epsilon_R$) for the broken power-law fit to \rout\ vs. \lbol, shown as the dashed black line in the top right panel of Fig.~\ref{fig:crim}.}
    \label{fig:postr}
\end{figure*}

\begin{figure*}
    \centering
    \subfloat[Single Power Law]{\includegraphics[height=0.30\textheight]{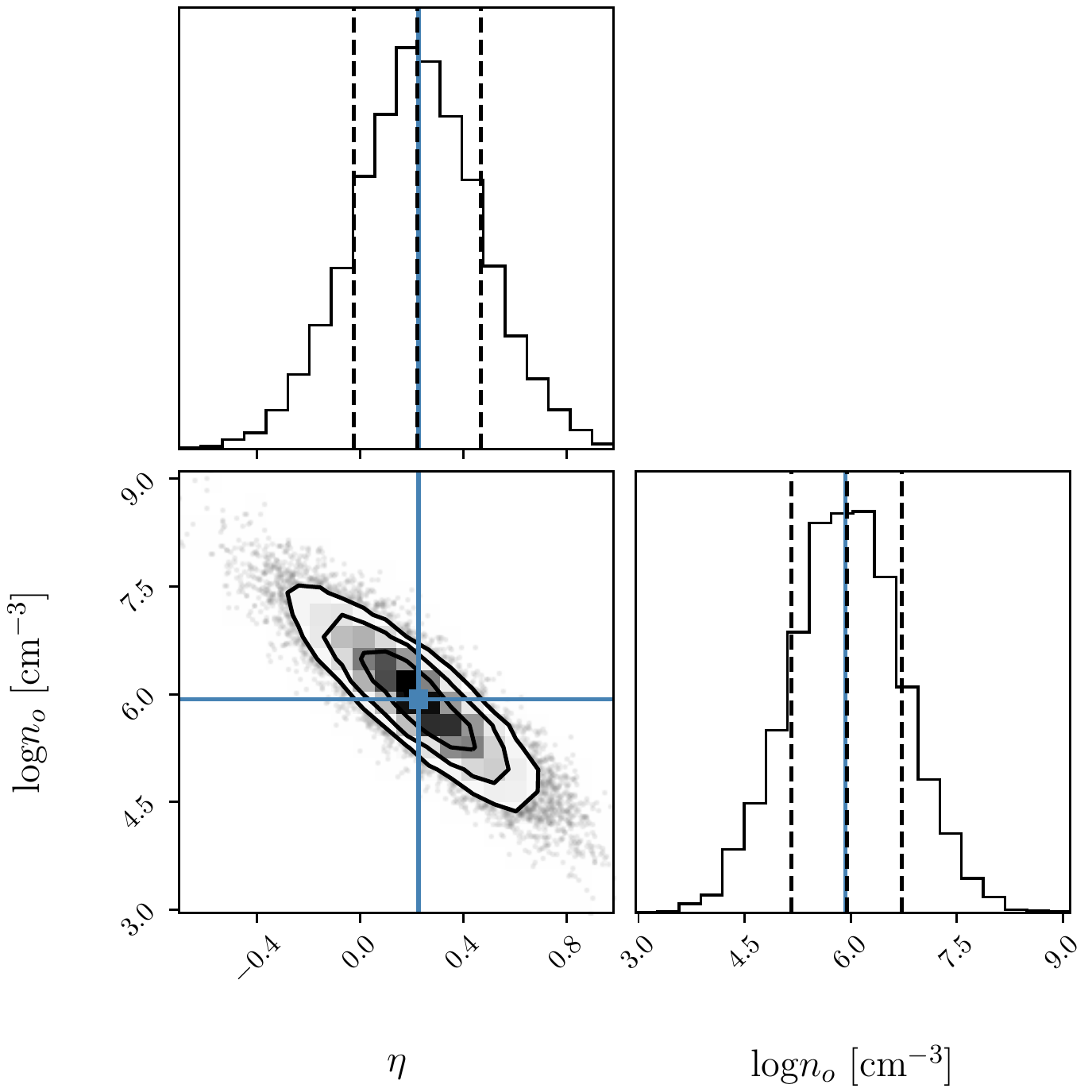}}
    
    \subfloat[Broken Power Law]{\includegraphics[height=0.56\textheight]{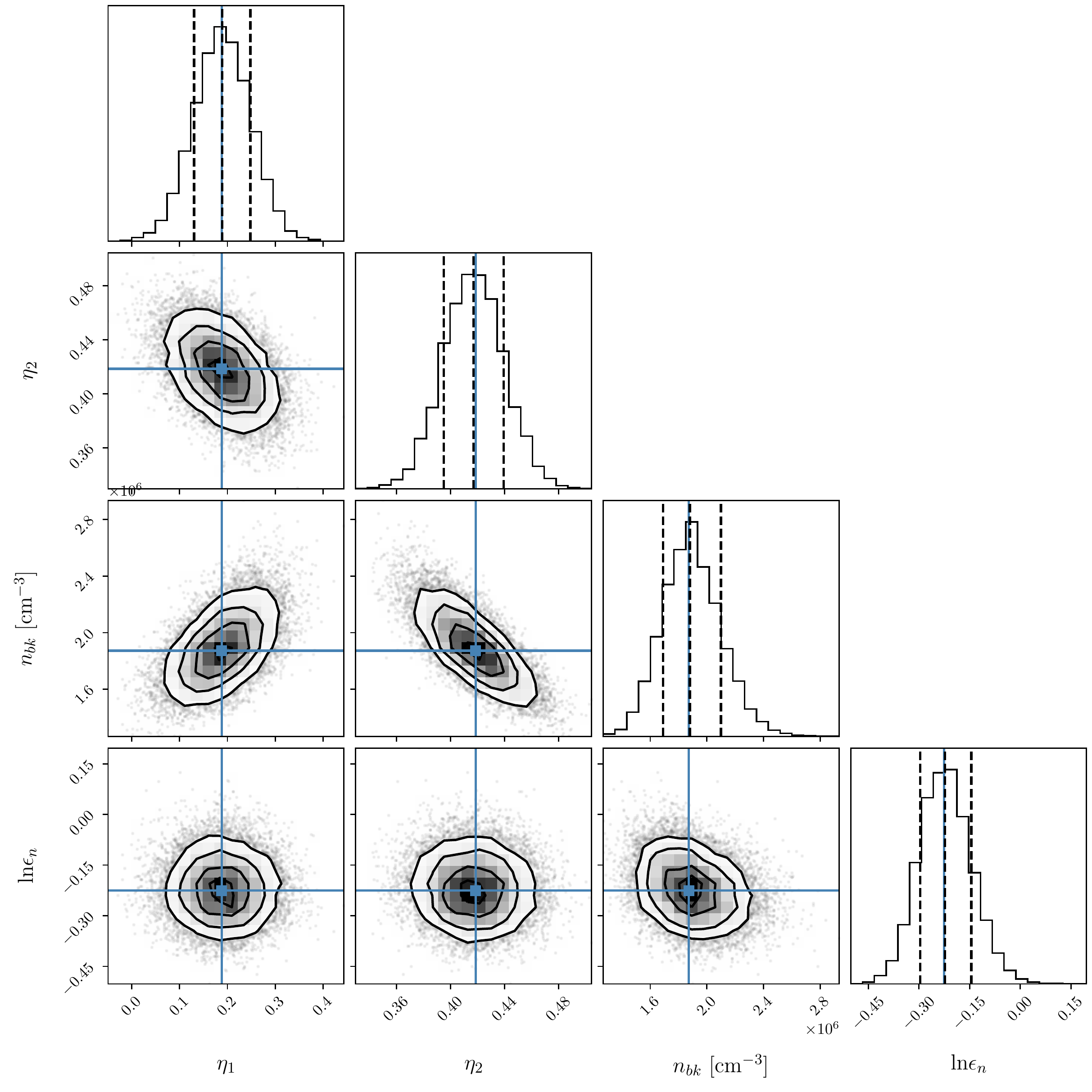}}
    \caption{\textit{(a)} Posterior distributions of the power-law exponent ($\eta$) and normalisation ($n_o$) for the single power-law fit to \nkau\ vs. \lbol, shown as the solid black line the bottom right panel of in Fig.~\ref{fig:crim}. \textit{(b)} Posterior distributions of the power-law exponents ($\eta_1$ and $\eta_2$), normalisation ($n_{bk}$), and residual variance parameter (ln$\epsilon_n$) for the broken power-law fit to \nkau\ vs. \lbol, shown as the dashed black line in the bottom right panel of Fig.~\ref{fig:crim}.}
    \label{fig:postn}
\end{figure*}

\label{lastpage}
\end{document}